\documentclass[final,onefignum,onetabnum]{siamonline171218}

\usepackage{lipsum}
\usepackage{amsfonts}
\usepackage{graphicx}
\usepackage{epstopdf}
\usepackage{algorithmic}
\ifpdf
  \DeclareGraphicsExtensions{.eps,.pdf,.png,.jpg}
\else
  \DeclareGraphicsExtensions{.eps}
\fi

\usepackage{enumitem}
\setlist[enumerate]{leftmargin=.5in}
\setlist[itemize]{leftmargin=.5in}

\newsiamremark{remark}{Remark}
\newsiamremark{hypothesis}{Hypothesis}
\crefname{hypothesis}{Hypothesis}{Hypotheses}
\newsiamthm{claim}{Claim}

\headers{Known Boundary Emulation of Complex Computer Models}{I. Vernon, S. Jackson, and J. A. Cumming}

\title{Known Boundary Emulation of Complex Computer Models
}

\author{Ian Vernon\thanks{Department of Mathematical Sciences, Durham University, Stockton Road, DH1 3LE Durham, UK 
  (\email{i.r.vernon@durham.ac.uk}, \email{samuel.jackson@durham.ac.uk}, \email{j.a.cumming@durham.ac.uk})}
\and Samuel E. Jackson\footnotemark[1] 
\and Jonathan A. Cumming\footnotemark[1]}

\usepackage{amsopn}

\usepackage{lscape,epsfig,amsmath,setspace}
\usepackage{xypic}

\usepackage{acronym}
\usepackage{color}
\usepackage{multirow}
\usepackage{subcaption}

\usepackage{amsmath,amsfonts,amssymb}
\usepackage[all]{xy}
\usepackage{psfrag}

\usepackage{fullpage} 
\usepackage{amsmath,amsfonts,amssymb}
\usepackage{psfrag}
\usepackage{epsfig}
\usepackage[all]{xy}
\usepackage[linewidth=1pt]{mdframed}

\newcommand{\e}[1]{\ensuremath{{\rm E}\left[#1\right]}}
\newcommand{\ed}[2]{\ensuremath{{\rm E}_{#1}\left[#2\right]}}
\newcommand{\var}[1]{\ensuremath{{\rm Var}\left[#1\right]}}

\newcommand{\vard}[2]{\ensuremath{{\rm Var}_{#1}\left[#2\right]}}
\newcommand{\rvard}[2]{\ensuremath{{\rm RVar}_{#1}(#2)}}

\newcommand{\cov}[2]{\ensuremath{{\rm Cov}\left[#1,#2\right]}}
\newcommand{\covd}[3]{\ensuremath{{\rm Cov}_{#1}\left[#2,#3\right]}}
\newcommand{\covb}[2]{\ensuremath{{\rm Cov}[#1,#2]}}

\newcommand{\be}{\begin{equation}}
\newcommand{\ee}{\end{equation}}
\newcommand{\ba}{\begin{eqnarray}}
\newcommand{\ea}{\end{eqnarray}}
\newcommand{\bi}{\begin{itemize}}
\newcommand{\ei}{\end{itemize}}
\newcommand{\bn}{\begin{enumerate}}
\newcommand{\en}{\end{enumerate}}
\newcommand{\bfi}{\begin{figure}}
\newcommand{\efi}{\end{figure}}

\newcommand{\mx}{\mathcal{X}}
\newcommand{\mk}{\mathcal{K}}
\newcommand{\ml}{\mathcal{L}}

\newcommand{\qq}{\quad \quad}

\ifpdf
\hypersetup{
  pdftitle={An Example Article},
  pdfauthor={D. Doe, P. T. Frank, and J. E. Smith}
}
\fi

\newcommand{\red}[1]{\textcolor{black}{#1}}% IRV comments/changes		

\begin{document}

\maketitle

% REQUIRED
\begin{abstract}
Computer models are now widely used across a range of scientific disciplines to describe various complex physical systems, however to perform 
full uncertainty quantification we often need to employ emulators.
An emulator is a fast statistical construct that mimics the complex computer model, and greatly aids the vastly more computationally intensive
uncertainty quantification calculations that a serious scientific analysis often requires.
In some cases, the complex model can be solved far more efficiently for certain parameter settings, leading to boundaries or hyperplanes in the input parameter space where the model is essentially known.
We show that for a large class of Gaussian process style emulators, multiple boundaries can be formally incorporated into the 
emulation process, by Bayesian updating of the emulators with respect to the boundaries, for trivial computational cost. The resulting 
updated emulator equations are given analytically. This leads to emulators that possess increased accuracy across large portions of the input parameter 
space. We also describe how a user can incorporate such boundaries within standard black box GP emulation packages that are currently 
available, without altering the core code.
Appropriate designs of model runs in the presence of known boundaries are then analysed, with two kinds of general purpose 
designs proposed. We then apply the improved emulation and 
design methodology to an important systems biology model of hormonal crosstalk in Arabidopsis Thaliana. 

\end{abstract}

% REQUIRED

% REQUIRED
%\begin{AMS}
%  68Q25, 68R10, 68U05
%\end{AMS}

\section{Introduction}

The use of mathematical models to describe complex physical systems is now commonplace in a wide variety of scientific disciplines.
We refer to such models as {\it simulators}. Often they possess high numbers of input and/or output dimensions, 
and are sufficiently complex that they may require substantial time for the completion of a single evaluation. 
The simulator may have been developed to aid understanding of the real world system in question, or to be
compared to observed data, necessitating a high-dimensional parameter search or model calibration, 
or to make predictions of future system behaviour, possibly with the goal of aiding a future decision process. 
The responsible use of simulators in all of the above contexts usually requires a full (Bayesian) uncertainty analysis~\cite{bernardo2006bayesian}, which will aim to incorporate 
all the major relevant sources of uncertainty, for example, parametric uncertainty on the inputs to the model, observation uncertainties on the data, 
structural model discrepancy that represents the uncertain differences between the simulator and the real system, both for past and future and the relation between them, and also all the many uncertainties related to the decision process~\cite{Vernon10_CS}.

However, such an uncertainty analysis, which represents a critically important part of any serious scientific study, usually requires
a vast number of simulator evaluations. 
For complex simulators with even a modest runtime, this means that such an analysis is utterly infeasible. 
A solution to this problem is found through the use of emulators: an emulator is a statistical construct that seeks to mimic the 
behaviour of the simulator over its input space, but which is several orders of magnitude faster to evaluate.
Early uses of Gaussian process emulators for computer models were given by \cite{SWMW89_DACE,Currin91_BayesDACE} with a more detailed account given 
in \cite{Santner03_DACE}.
A vital feature of an emulator is that it gives both an expectation of the simulator's outputs at an unexplored input location as well as an 
uncertainty statement about the emulator's accuracy at this point, an attribute that elevates emulation above interpolation 
or other simple proxy modelling approaches. 
Therefore, emulators fit naturally within a Bayesian approach, and help 
facilitate the full uncertainty analysis described above. For an early example of an uncertainty analysis using multilevel emulation combined with structural discrepancy modelling in a Bayesian history matching context see~\cite{Craig96_Pressure,Craig97_Pressure}, and for a fully Bayesian calibration of a complex 
nuclear radiation model, again incorporating structural model discrepancy, see~\cite{Kennedy01_Calibration}.

Emulators have now been successfully employed across several scientific disciplines, including 
cosmology~\cite{Vernon10_CS,Vernon10_CS_rej,vernon_astro,PhysRevD.78.063529,Higdon09_Coyote2,galf_stat_sci,Vernon:2016aa}, climate 
modelling~\cite{Williamson:2013aa,Rougier:2009aa,S:2015aa,Holden:aa} (the later employing emulation to increase the efficiency of an Approximate 
Bayesian Computation algorithm),  
epidemiology~\cite{Yiannis_HIV_1,Yiannis_HIV_2,Yiannis_HIV_3,McKinley:2017aa,McCreesh2017}, systems biology~\cite{Vernon_sysbio_hm_2016,Jackson1}, oil reservoir 
modelling~\cite{JAC_Handbook,JAC_sma_samp},
environmental science~\cite{asses_mod}, traffic modelling~\cite{Boukouvalas:2014aa}, vulcanology~\cite{Bayarri:2009aa} and even to Bayesian analysis itself~\cite{BABA_paper1}.
%\ian{list some more from other authors: OHagan, Gosling, Oakley, Wilkinson, Higdon, Berger etc}.
The development of improved emulation strategies therefore has the potential to benefit multiple scientific areas, allowing more accurate 
analyses with lower computational cost. 
If additional prior insight into the physical structure of the model is available, it is of real importance that emulator structures capable 
of incorporating such insights have been developed to fully exploit this information.  

Here we describe such an advance in emulation strategy that, when applicable, can lead to substantial improvements in 
emulator performance. 
In most cases, complex deterministic simulators have to be solved numerically for arbitrary input specifications, which leads to substantial runtimes.
However, for some simulators, there exist input parameter settings, lying possibly on boundaries or hyperplanes in the input parameter space, where the simulator can be solved far more efficiently,
either analytically in the ideal case or just significantly faster using a much more efficient and simpler solver. 
This may be due to the system in question, or at least a subset of the system outputs, behaving in a much simpler way for particular input settings, possibly due for 
example to various modules decoupling from more complex parts of the model (possibly when certain inputs are set to zero, switching some processes  off). Note that this leads to Dirichlet boundary conditions, i.e. known simulator behaviour on various hyperplanes, that impose constraints on the emulator itself, and that these are distinct from \red{Dirichlet boundary conditions imposed on the physical simulator model, that we do not require here (which are} for example, analysed approximately using KL expansions by \cite{Tan:2017aa}). 
The goal then, is to incorporate these known boundaries, situated where we essentially know the function output, into the Bayesian emulation process, 
which should lead to significantly improved emulators. We do this by formally updating the emulators by the information contained
on the known boundaries, obtaining analytic results, and show that this is possible for a large class of emulators, for multiple boundaries of various forms \red{(specifically collections of parallel or perpendicular hyperplanes)}, and most importantly, for trivial extra computational cost. 
\red{We note that \cite{Tan:2018aa} have examined this problem, however they used an approach which requires multiple extra emulator parameters that have to be estimated, as they essentially included substantial extra modelling to ensure both the mean and the variance of the emulator were consistent with the known boundary a priori. In contrast, our approach includes no extra modelling, and zero additional parameters, instead updating the Gaussian process style emulator with the boundary information in a natural way.}
We also detail how users can include known boundaries when using standard black box Gaussian Process software (without altering the core code), although this method is
less powerful than implementing the fully updated emulators that we develop here.
We then analyse the design problem of how to choose an efficient 
set of runs of the full simulator, given that we are aware of the existence of one or more known boundaries.
Finally we apply this approach to a model of hormonal crosstalk in Arabidopsis, an important model in systems biology, which possesses these features.

The article is organised as follows. In section~\ref{sec_emulCCM} we describe a standard emulation approach for deterministic 
simulators. In section~\ref{sec_KBE} we develop the full known boundary emulation (KBE) methodology, explicitly constructing emulators that have been updated by one or two perpendicular (or parallel) known boundaries. In section~\ref{sec_design_KB} we discuss the design problem of how to choose efficient 
sets of simulator runs. In section~\ref{sec_arabid} we apply both the KBE and the design techniques to the systems biology Arabidopsis model, before concluding in section~\ref{sec_conc}.

\section{Emulation of Complex Computer Models}\label{sec_emulCCM}

We consider a complex computer model represented as a function $f(x)$, where $x\in\mathcal{X}$ denotes a $d$-dimensional vector 
containing the 
computer model's input parameters, and $\mathcal{X}\subset \mathbb{R}^d$ is a pre-specified input parameter space of interest. We imagine that 
a single evaluation of the computer model takes a substantial amount of time to complete, and hence we will only be able evaluate it at a small 
number of locations. Here we assume $f(x)$ is univariate, but the results we present should directly generalise to the corresponding multivariate 
case.

We represent our beliefs about the unknown $f(x)$ at unevaluated input $x$ via an emulator. For now, we assume that the form of the emulator is that of a pure Gaussian process (or in a less fully specified version, a weakly second order stationary stochastic process):
\be\label{eq_GPem}
f(x) \;=\; u(x)
\ee  
We make the judgement, consistent with most of the computer model literature, that the $u(x)$ have a product correlation structure:
\be \label{eq_cor_struc}
\cov{u(x)}{u(x')} \;=\; \sigma^2 r(x-x') \;=\; \sigma^2 \prod_{i=1}^d r_i(x_i - x'_i)
\ee
with $r_i(0)=1$, corresponding to deterministic $f(x)$.  Product correlation structures are very common, with the most popular being the Gaussian correlation structure given 
by:
\be\label{eq_prodgausscor}
 r(x-x') \;=\; \exp\{-|x-x'|^2/\theta^2\} \;=\;   \prod_{i=1}^d \exp\{-|x_i-x_i'|^2/\theta^2\}
\ee
which can be generalised to have different $\theta_i$ in each direction, while maintaining the product structure.  As usual, we will also assume stationarity, but the following derivations do not require this assumption.

If we perform a set of runs at locations $x_D=(x^{(1)},\dots,x^{(n)})$ over the input space of interest $\mathcal{X}$, giving computer model outputs 
as the column vector $D = (f(x^{(1)}),\dots,f(x^{(n)}))^T$, then we can update our beliefs about the computer model $f(x)$ in light of $D$.
This can be done either using Bayes theorem (if $u(x)$ is assumed to be a Gaussian process) or using the Bayes linear update formulae
(which following DeFinetti \cite{DeFinetti1}, treats expectation as primitive, and requires only a second order specification~\cite{Goldstein_99,Goldstein07_BayesLinearBook}):
\ba
\ed{D}{f(x)} &=& \e{f(x)} + \cov{f(x)}{D} \var{D} ^{-1}(D- \e{D}) \label{eq_BLm}\\
\vard{D}{f(x)} &=& \var{f(x)} - \cov{f(x)}{D} \var{D}^{-1}\cov{D}{f(x)} \label{eq_BLv} \\
\covd{D}{f(x)}{f(x')} &=& \cov{f(x)}{f(x')} - \cov{f(x)}{D} \var{D}^{-1} \cov{D}{f(x')} \label{eq_BLc} 
\ea
where $\ed{D}{f(x)}$, $\vard{D}{f(x)}$ and $\covd{D}{f(x)}{f(x')}$ are the expectation, variance and covariance of $f(x)$ adjusted by $D$~\cite{Goldstein_99,Goldstein07_BayesLinearBook}. \red{Equation~(\ref{eq_BLv}) is obviously a special case of equation~(\ref{eq_BLc}), but we include it explicitly to aid clarity in subsequent derivations.}
The fully Bayesian calculation, using Bayes theorem, would yield similar update formulae for the analogous posterior quantities. 
Although we will work within the 
Bayes linear formalism, the derived results will apply directly to the fully Bayesian case, were one willing to make the additional assumption of full normality that use of a Gaussian process entails. In this case all Bayes linear adjusted quantities can be directly mapped to the corresponding posterior versions e.g. $\ed{D}{f(x)} \rightarrow \e{f(x)|D}$ and $\vard{D}{f(x)} \rightarrow \var{f(x)|D}$. See \cite{Goldstein_99,Goldstein07_BayesLinearBook} for discussion of the benefits of using a Bayes linear approach, and  \cite{Vernon10_CS,Vernon10_CS_rej} for its benefits within a computer model setting. 

The results presented in this article rely on the product correlation structure of the emulator.  As such, expansion of these methods to more general emulator forms requires further calculation. For example, a more advanced emulator is given by \cite{Craig97_Pressure,Vernon10_CS}:
\be\label{eq_fullem}
f(x) \;=\; \sum_{j} \beta_j g_j(x_A) + u(x_A) + v(x)
\ee
where the active inputs $x_A$ are a subset of $x$ that are strongly influential for $f(x)$,
the first term on the right hand side is a regression term containing known functions $g_j(x_A)$ and possibly unknown $\beta_j$,
$u(x_A)$ is a Gaussian process over the active inputs only, and 
$v(x)$ is an uncorrelated nugget term, representing the inactive variables.  See also \cite{JAC_Handbook} and \cite{Vernon10_CS,Vernon10_CS_rej} for 
discussions of the benefits of using an emulator structure of this kind, and see~\cite{Kennedy01_Calibration,Higdon08a_calibration} for discussions of alternative structures. 
We will discuss the generalisation of our results to equation~(\ref{eq_fullem}) in section~\ref{sec_conc}, but currently 
we note that if the regression 
parameters $\beta_j$ are assumed known, perhaps due to sufficiently large run number, and if all variables are assumed active,
%nugget vanishes only on the known boundaries, 
then equation~(\ref{eq_fullem}) reduces to the required form,
%case of a pure Gaussian process, as given by equation~(\ref{eq_GPem}),  
and all our results will apply.

\section{Known Boundary Emulation}\label{sec_KBE}

We now consider the situation where the computer model is analytically solvable on some lower dimensional boundary $\mathcal{K}$. 
Hence we can evaluate $\{ f(x):x\in \mk \}$ a vast number of times $m$ on $\mk$, and use these to supplement our standard emulator evaluations over $\mathcal{X}$ to produce an emulator that respects the functional behaviour of $f(x)$ along $\mk$.
We first examine the case of finite (but large) $m$, which can be analysed using the standard Bayes linear update, but structure our calculations so that they can be simply generalised to continuous model evaluations on $\mk$, which will require a generalised version of the Bayes linear update, as described in section~\ref{ssec_continuous_case}.

Call the corresponding length $m$ vector of model evaluations $K$. Unfortunately simply plugging these $m$ runs into the Bayes Linear update equations~(\ref{eq_BLm}), (\ref{eq_BLv}) and (\ref{eq_BLc}), replacing $D$ with $K$, would be infeasible due to the size of the $m\times m$ matrix inversion $\var{K}^{-1}$. For example, if the dimension $d_{\mk}$ of $\mk$ is not small, we may need $m$ to be extremely large (billions or trillions say) to capture all the information contained in $\mk$.
Hence a direct update of the emulator in light of the information in $K$ is non-trivial. 
Here we show from first principles that this update can be performed analytically for a wide class of emulators. We do this by exploiting a sufficiency argument briefly described in the supplementary material of~\cite{Kennedy01_Calibration}, and in \cite{Rougier:aa}, but which, to our knowledge, has not been fully explored or utilised in the context of known boundary emulation.
The emulation problem is further compounded \red{in the general case where we have both evaluations $K$ on the boundary, and in the main bulk $D$ defined as above. For this case we will develop a sequential update that first updates analytically by $K$ to obtain $\ed{K}{f(x)}$, $\vard{K}{f(x)}$ and $\covd{K}{f(x)}{f(x')}$, 
as developed in Section~\ref{ssec_singleKB}, and then subsequently updates by $D$, 
to obtain $\ed{D \cup K}{f(x)}$, $\vard{D \cup K}{f(x)}$ and $\covd{D\cup K}{f(x)}{f(x')}$, as described in Section~\ref{sssec_up_fur}}.

%
%Note also that if we have both runs $K$ on the boundary, and in the main bulk $D$ defined as above, then we would wish to update the emulator in light of both $K$ and $D$, that is to find $\ed{D \cup K}{f(x)}$, $\vard{D \cup K}{f(x)}$ and $\covd{D\cup K}{f(x)}{f(x')}$. Our analysis allows us to do this  sequentially, by first updating analytically by $K$ to obtain $\ed{K}{f(x)}$, $\vard{K}{f(x)}$ and $\covd{K}{f(x)}{f(x')}$, and then subsequently updating by $D$, as we go on to describe in section~\ref{sssec_up_fur}. \ian{Remove this paragraph?}

\subsection{A Single Known Boundary}\label{ssec_singleKB}
\begin{figure}
\begin{center}
\begin{tabular}{cc}
\includegraphics[scale=0.6]{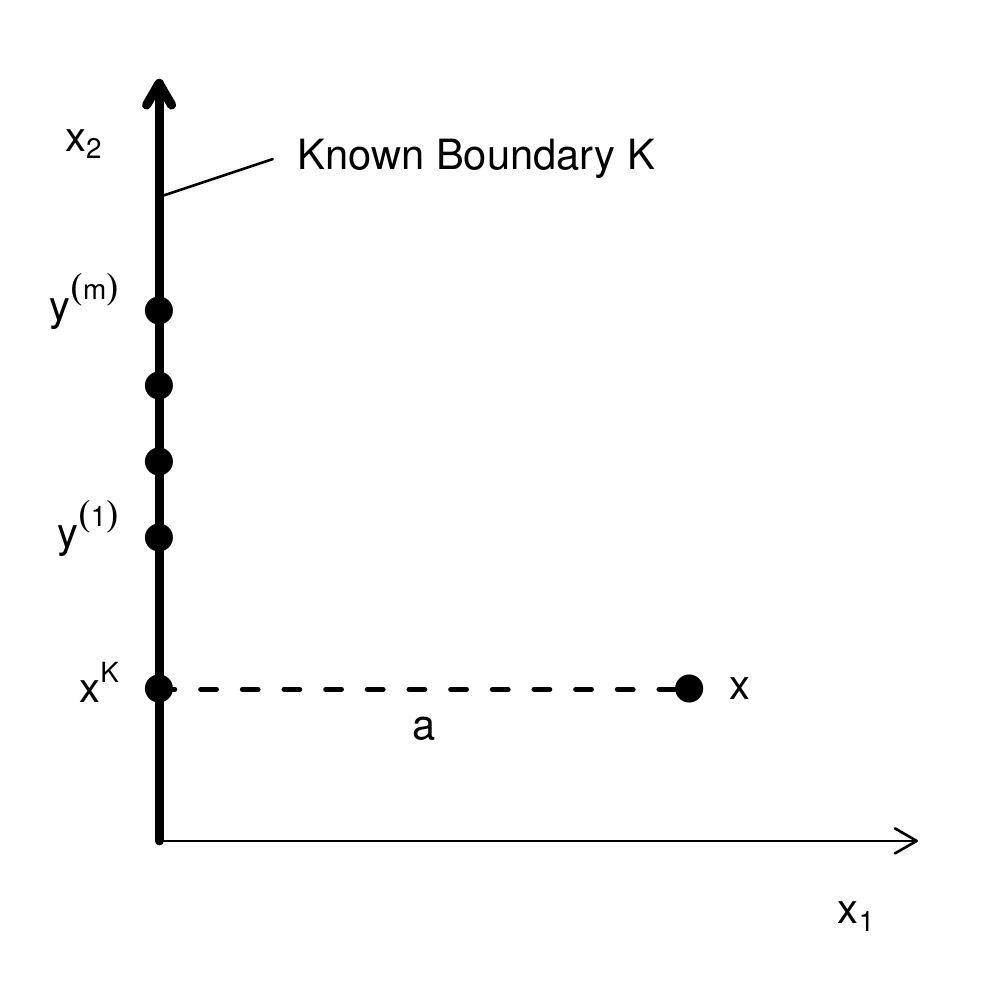} & \hspace{0.5cm}
\includegraphics[scale=0.6]{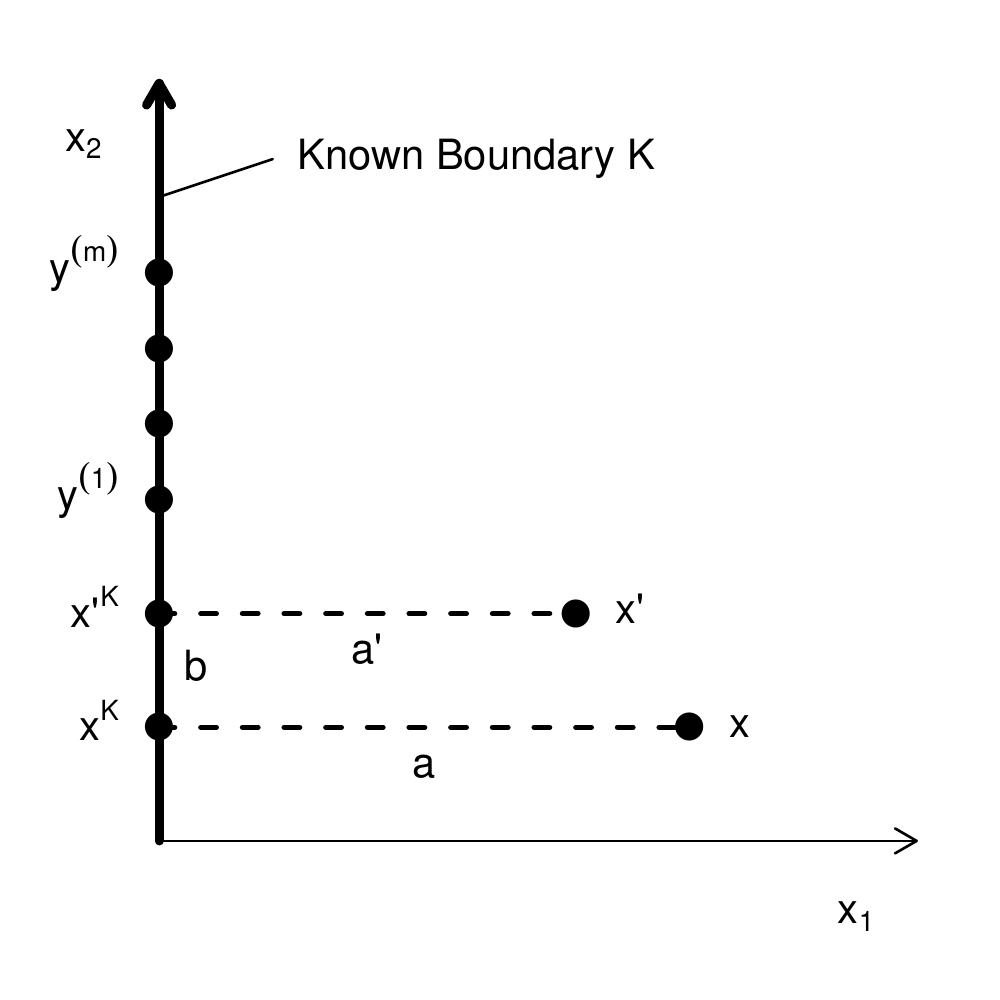} \\
\end{tabular}
\end{center}
\vspace{-1.0cm}
\caption{\footnotesize{The single known boundary case. Left panel: the points required for the $\ed{K}{f(x)}$ and $\vard{K}{f(x)}$ calculation. $x$ is the point we wish to emulate at, $x^K$ its orthogonal projection onto the known boundary $\mk$ at distance $a$. Right panel: the points required for the \covd{K}{f(x)}{f(x')} calculation. $x$ and $x'$ are points we wish to update the covariance at, while $x^K$ and $x'^K$ are their orthogonal projection onto the known boundary $\mk$, at distances $a$ and $a'$ respectively. In both panels, the $y^{(i)}$ represent a large number of points for which we can evaluate $f(y^{(i)})$ analytically (or at least very quickly).}}
\label{fig_single_bound}
\vspace{-0.5cm}
\end{figure}

We wish to update the emulator, and hence our beliefs about $f(x)$, at the input point $x\in \mathcal{X}$ in light of a single 
known boundary $\mk$, where $\mk$ is a $d-1$ dimensional hyperplane perpendicular to the $x_1$ direction 
(but we note that our results naturally extend to lower dimensional boundaries).
%i.e. whenever $x_1=c$ (for some constant $c$) the simulator reduces to a simple analytic function in the remaining $d-1$ variables. 
To capture the simulator behaviour along $\mk$, we evaluate $f(x)$ at a large number, $m$, of points on $\mk$ which we denote $y^{(1)},\dots,y^{(m)}$. We also evaluate the perpendicular projection of the point of interest, $x$, onto the boundary $\mk$, which we denote as $x^K$. We therefore extend the collection of boundary evaluations, $K$, to be the $m+1$ column vector  
\[
K=(f(x^K),f(y^{(1)}),\dots,f(y^{(m)}))^T,
\]
which is illustrated in Figure~\ref{fig_single_bound} (left panel). 
%As discussed above, for $m$ large, the calculation of the adjusted emulator expectation
We start by examining the expression for $\ed{K}{f(x)}$
\be
\ed{K}{f(x)} \;=\; \e{f(x)} + \cov{f(x)}{K} \var{K}^{-1}(K- \e{K}). \label{eq_BLmK}
\ee
As noted above, this calculation is seemingly infeasible due to the $\var{K}^{-1}$ term. However, if we evaluate it at the point $x^K$ itself, which 
lies on $\mk$, and as we have evaluated $f(x^K)$, we must find, for the emulator of a smooth deterministic function with suitably chosen correlation structure, that $\ed{K}{f(x^K)} = f(x^K)$ 
(and that $\vard{K}{f(x^K)} = 0$).
%As noted above, for large $m$ this calculation is seemingly infeasible due to the $\var{K}^{-1}$ term.
%%will be infeasible due to $m$ being large making the inversion $\var{K}^{-1}$ expensive or impossible. 
%A similar problem is encountered with the adjusted variance. To simplify this calculation and avoid direct matrix inversion, we consider the behaviour of the emulator at the perpendicular projection point $x^K$ given the boundary
%\[
%\ed{K}{f(x^K)} \;=\; \e{f(x^K)} + \cov{f(x^K)}{K} \var{K}^{-1}(K- \e{K}).
%\]
%Trivially, as $x^K$ lies on $\mk$ and has been evaluated as $f(x^K)$, we must find that, for the emulator of a smooth deterministic function with suitably chosen correlation structure, both $\ed{K}{f(x^K)} = f(x^K)$ and $\vard{K}{f(x^K)} = 0$. 
This is indeed the case as can be seen by examining the structure of the $\var{K}^{-1}$ term.
As $f(x^K)$ is included as the first element of $K$, we note that
\begin{align}
I_{(m+1)}  &=\;\var{K} \var{K}^{-1} \label{eq_inv_var}\\
	 &=\; \begin{pmatrix} 
	\covb{f(x^K)}{K} \\
	\covb{f(y^{(1)})}{K} \\
	\vdots \\
	\covb{f(y^{(m)})}{K} 
	\end{pmatrix}  \var{K}^{-1}.   \label{eq_varKrowvec}
\end{align}
where $I_{(m+1)}$ is the identity matrix of dimension $(m+1)$. Taking the first row of equation~(\ref{eq_varKrowvec}) gives
% immediately gives the result
\be
 \cov{f(x^K)}{K} \var{K}^{-1} =\; (1,0, \cdots, 0) \label{eq_covvar}
\ee
Substituting equation~(\ref{eq_covvar}) into the adjusted mean and variance naturally gives $\ed{K}{f(x^K)} = f(x^K)$ and $\vard{K}{f(x^K)}=0$ as it must. Whilst unsurprising, this simple result is of particular value when considering the behaviour at the point of interest, $x$. %,$\ed{K}{f(x)}$, and by exploiting the symmetry of the product correlation structure, we have the following. 
As we have defined $x^K$ as the perpendicular projection of $x$ onto $\mk$, we can write 
$x= x^K + (a,0,\dots,0)$,
for some constant $a$. Now we can exploit the symmetry of the product correlation structure~(\ref{eq_cor_struc}), to obtain the following covariance expressions
\ba
\covb{f(x)}{f(x^K)} &=& \sigma^2 \prod_{i=1}^d r_i(x_i - x^K_i) \;=\; \sigma^2 r_1(x_1 - x^K_1) \;=\; \sigma^2 r_1(a)  \nonumber\\
&=&   r_1(a) \,\covb{f(x^K)}{f(x^K)} \label{eq_shortcov2}
\ea
since $x_i=x_i^K$ for $i=2,\dots,d$ and $r_i(0)=1$. Furthermore,
\ba
\covb{f(x)}{f(y^{(j)})} &=& \sigma^2 \prod_{i=1}^d r_i(x_i - y^{(j)}_i) \;=\; \sigma^2 r_1(x_1 - x^K_1)  \prod_{i=2}^d r_i(x_i - y^{(j)}_i)  \nonumber \\
&=& \sigma^2 r_1(a)  \prod_{i=2}^d r_i(x^K_i - y^{(j)}_i)   \nonumber\\
&=& r_1(a) \,\covb{f(x^K)}{f(y^{(j)})} \label{eq_prod_ra_xy}
\ea
since the first component of $x^K$ and $y^{(j)}$ must be equal as they all lie on $\mk$ (i.e. $x^K_1 = y^{(j)}_1$).
%\ba
%\covb{f(x)}{f(x^K)} &=& \sigma^2 \prod_{i=1}^d r_i(x_i - x^K_i) \;=\; \sigma^2 r_1(x_1 - x^K_1)\nonumber  \\
%&=&  \sigma^2 r_1(a) \;=\;  r_1(a) \,\covb{f(x^K)}{f(x^K)}\\
%\covb{f(x)}{f(y^{(j)})} &=& \sigma^2 \prod_{i=1}^d r_i(x_i - y^{(j)}_i)  \nonumber \\
%&=& \sigma^2 r_1(x_1 - x^K_1)  \prod_{i=2}^d r_i(x_i - y^{(j)}_i)  \nonumber \\
%&=& \sigma^2 r_1(a)  \prod_{i=2}^d r_i(x^K_i - y^{(j)}_i)  \nonumber \\
%&=& r_1(a) \,\covb{f(x^K)}{f(y^{(j)})} \label{eq_prod_ra_xy}
%\ea
Combining \eqref{eq_shortcov2} and \eqref{eq_prod_ra_xy}, the covariance between point $x$ and the set of boundary evaluations is given by
\ba
\cov{f(x)}{K} &=& \left(
		\covb{f(x)}{f(x^K)}, \covb{f(x)}{f(y^{(1)})}, \cdots , \covb{f(x)}{f(y^{(m)})} 
		\right)   \nonumber \\
		&=& r_1(a) \left(
		\covb{f(x^K)}{f(x^K)}, \covb{f(x^K)}{f(y^{(1)})}, \cdots , \covb{f(x^K)}{f(y^{(m)})}  \nonumber
		\right)  \\
		&=& r_1(a) \, \cov{f(x^K)}{K},  \label{eq_rcovxK}
\ea
Equations~(\ref{eq_covvar}) and (\ref{eq_rcovxK}) are very useful results that greatly simplify the emulator calculations.
%\subsubsection{Adjusted expectation, variance and covariance}\label{sssec_Adj}
We can use them to write the adjusted emulator expectation for $f(x)$ given in \eqref{eq_BLmK} as 
\ba
\ed{K}{f(x)} &=& \e{f(x)} + \cov{f(x)}{K} \var{K}^{-1}(K- \e{K})\nonumber\\
	&=& \e{f(x)} + r_1(a) \, \cov{f(x^K)}{K}   \var{K}^{-1} (K- \e{K}) \nonumber \\
	&=& \e{f(x)} + r_1(a) (1,0, \cdots, 0) (K- \e{K}) \nonumber \\
&=&  \e{f(x)} + r_1(a) (f(x^K) -  \e{f(x^K)})  \label{eq_EK1}
\ea 
Thus we have eliminated the need to explicitly invert the large matrix $\var{K}$ entirely by exploiting the symmetric product correlation structure and the identity \eqref{eq_covvar}. Similarly, we find the adjusted variance using equations~(\ref{eq_BLv}), (\ref{eq_covvar}) and (\ref{eq_rcovxK}),
\ba
\vard{K}{f(x)} &=& \var{f(x)} - r_1(a) (1,0, \cdots, 0) \cov{K}{f(x)}  \nonumber  \\
&=&  \var{f(x)} - r_1(a) \covb{f(x^K)}{f(x)} \nonumber  \\
%&=& \sigma^2 - r_1(a) \sigma^2 r_1(a) \nonumber  \\
&=& \sigma^2(1 - r_1(a)^2) \label{eq_VK1}
\ea
Equations~(\ref{eq_EK1}) and (\ref{eq_VK1}) give the expectation and variance of the emulator at a point $x$, updated by a known boundary $\mk$. As they require only evaluations of the analytic boundary function and the correlation function they can be implemented with trivial computational cost in comparison to a direct update by $K$. Note that they critically rely on the projected point $f(x^K)$ being in $K$.

Finally, we consider the Bayes linear update for the covariance between $x$ and a second input point $x' \in \mathcal{X}$ given the boundary $\mk$. We define the orthogonal projection of $x'$ onto $\mk$ as $x'^K$, and denote its perpendicular distance from $\mk$ as $a'$, as shown in Figure~\ref{fig_single_bound} (right panel).
%\begin{figure}
%\begin{center}
%\includegraphics[scale=0.65]{plots/singleKB_2x.pdf}
%\end{center}
%\caption{The single known boundary case for the \covd{K}{f(x)}{f(x')} calculation. $x$ and $x'$ are points we wish to update the covariance at, while $x^K$ and $x'^K$ are their orthogonal projection onto the known boundary $\mk$, at distances $a$ and $a'$ respectively. The $y^{(i)}$ represent a large number of points for which we can evaluate $f(y^{(i)})$ analytically (or at least very quickly).}
%\label{fig_single_bound_cov}
%\end{figure}
Equation~\eqref{eq_BLc} now gives
%Introducing the above results into the adjusted covariance formula \eqref{eq_BLc} gives
\ba
\covd{K}{f(x)}{f(x')} &=& \cov{f(x)}{f(x')} - \cov{f(x)}{K} \var{K}^{-1} \cov{K}{f(x')} \nonumber \\
&=&  \cov{f(x)}{f(x')} -  r_1(a) (1,0, \cdots, 0) \cov{K}{f(x')}  \nonumber \\
&=& \cov{f(x)}{f(x')} -  r_1(a) \cov{f(x^K)}{f(x')}  \nonumber \\
&=& \cov{f(x)}{f(x')} -  r_1(a) \cov{f(x^K)}{f(x'^K)} r_1(a')   \label{eq_covK1}
\ea
where in the final line we used the equivalent result to equation~(\ref{eq_prod_ra_xy}), rewritten for $x'$. Noting that we can also write
$x' = x'^K + (a',0,\dots,0)$, and that $x^K_1 = x'^K_1$, equation~(\ref{eq_covK1}) becomes 
\ba
\covd{K}{f(x)}{f(x')} &=& \sigma^2 \prod_{i=1}^d r_i(x_i - x'_i)-   r_1(a) r_1(a') \sigma^2 \prod_{i=1}^d r_i(x^K_i-x'^K_i)  \nonumber\\
&=& \sigma^2 r_1(a-a')\prod_{i=2}^d r_i(x_i - x'_i)-  \sigma^2 r_1(a) r_1(a')  r_1(0) \prod_{i=2}^d r_i(x^K_i - x'^K_i)  \nonumber \\
&=& \sigma^2 r_1(a-a')\prod_{i=2}^d r_i(x^K_i - x'^K_i)-  \sigma^2 r_1(a) r_1(a') \prod_{i=2}^d r_i(x^K_i - x'^K_i)   \nonumber\\
&=& \sigma^2 \left( r_1(a-a')-   r_1(a) r_1(a') \right) r_{-1}(x^K-x'^K)  \nonumber\\
&=& \sigma^2 R_1(a,a') \, r_{-1}(x^K-x'^K)  \label{eq_covK2}
\ea
where we have defined the correlation function of the projection of $x$ and $x'$ onto $\mk$ 
%(which therefore does not have any dependance on the $x_1$ direction) 
as 
\[
r_{-1}(x^K-x'^K) \;=\;  \prod_{i=2}^d r_i(x^K_i - x'^K_i)  \; = \; \cov{f(x^K)}{f(x'^K)}
\]
and defined the `updated correlation component' in the $x_1$ direction as 
\be
R_1(a,a') \;=\; r_1(a-a')-   r_1(a) r_1(a')  \label{eq:UpdCorrComp}
\ee
\red{We see of course that equation~(\ref{eq_VK1}) is a special case of equation~(\ref{eq_covK2}), with $x=x'$.}

%% Begin rearrange
These expressions for the expectation and (co)variance updated by the information at the simulator boundary provide several insights: 
\begin{enumerate}[label=(\emph{\alph*}),leftmargin=0.6cm]
\item \emph{Sufficiency:} for the updating of our beliefs about the emulator \red{at point $x$}, we see that $f(x^K)$ is sufficient for $K$. Hence, only the evaluation $K=f(x^K)$ is required and the evaluations $y^{(i)}$ are redundant (note that under an assumption of an underlying Gaussian process, this result corresponds to a conditional independence statement discussed in the supplementary material to~\cite{Kennedy01_Calibration}). 
This has important ramifications for users of black box GP packages, as we discuss in section~\ref{sssec_up_fur}. \red{It also implies that if we are interested in emulating at 
any point $x\in \mathcal{X}$, we only require the known boundary $\mk$ to contain the projection of $\mathcal{X}$. 
So, for example, $\mk$ could be only a bounded subset of a hyperplane, provided $\mathcal{X}$ is similarly bounded. }

%(such as
%BACCO~\cite{Hankin:2005aa} or GPfit~\cite{JSSv064i12} in R, or GPy~\cite{gpy2014} for Python), 
%which perhaps cannot be easily recoded to use the more sophisticated formula of equations~(\ref{eq_EK1}) and (\ref{eq_VK1}), but for which appropriate extra points on $\mk$, i.e. the projections on to $\mk$ of any points of interest, could easily be included. %\ian{Should this point actually be moved until after the cov calculation, as only then can we claim full sufficiency has been proved?}

\item \emph{The correlation structure is now no longer stationary:} the contribution to the correlation function from dimensions 2 to $d$, denoted $r_{-1}(x^K-x'^K)$, is unchanged by the update (as we would expect from symmetry arguments), however the contribution in the $x_1$ direction depends on the distance to the boundary $\mk$, through $R_1(a,a')$, breaking stationarity.

\item \emph{The correlation structure is still in product form:}
Critically, as the correlation structure has maintained its product from, this suggests that we can update by further known boundaries, either perpendicular to any of the remaining inputs $x_i$, with $i=2,\dots,d$, hence perpendicular to $\mk$ or indeed by 
a second boundary parallel to $\mk$. We perform these updates in sections~\ref{ssec_two_perp_bound} and \ref{ssec_two_para_bound}.

%\item The updated variance depends only on the distance from the point of interest, $x$, to the boundary $\mk$ in the form of $r_1(a)$.
\item \emph{Intuitive limiting behaviour}: As we move $x$ towards $\mk$, the emulator tends towards the known boundary function, and as we move away from $\mk$ the emulator reverts to its prior form, as expected:
\ba
\lim_{a\to 0}  \ed{K}{f(x)} \;=& f(x^K), \qq \qq  \lim_{a\to 0}  \vard{K}{f(x)} &=\; 0, \\
\lim_{a\to \infty}  \ed{K}{f(x)} \;=& \e{f(x)}, \qq \qq  \lim_{a\to \infty}  \vard{K}{f(x)} &=\; \var{f(x)},
\ea
as $\lim_{a\to \infty} r_1(a) = 0$.
%Similarly, the behaviour of 
%$\covd{K}{f(x)}{f(x')}$ is as expected, tending to its prior form as both $a$ and $a'$ tend to infinity with their difference remaining finite, and tending to 0 as either $a$ and $a'$ tend to zero:
Similarly, the behaviour of 
$\covd{K}{f(x)}{f(x')}$ is as expected, tending to its prior form far from the boundary (with $a-a'$ finite), 
and to zero as either $a$ and $a'$ tend to zero:
\ba
\lim_{a\to 0} \covd{K}{f(x)}{f(x')} &=&  \lim_{a'\to 0} \covd{K}{f(x)}{f(x')} \;=\; 0 \\
\lim_{a,a' \to \infty}  \covd{K}{f(x)}{f(x')} &=& \sigma^2 r(x-x') \;=\; \cov{f(x)}{f(x')}, \quad  a-a' \text{ finite} \quad
\ea

\end{enumerate}

\subsection{Application to a 2-Dimensional Model}\label{sssec_2dmodel}
\begin{figure}[t]
\begin{center}
\begin{subfigure}{0.4\columnwidth}
\centering
\includegraphics[scale=0.38]{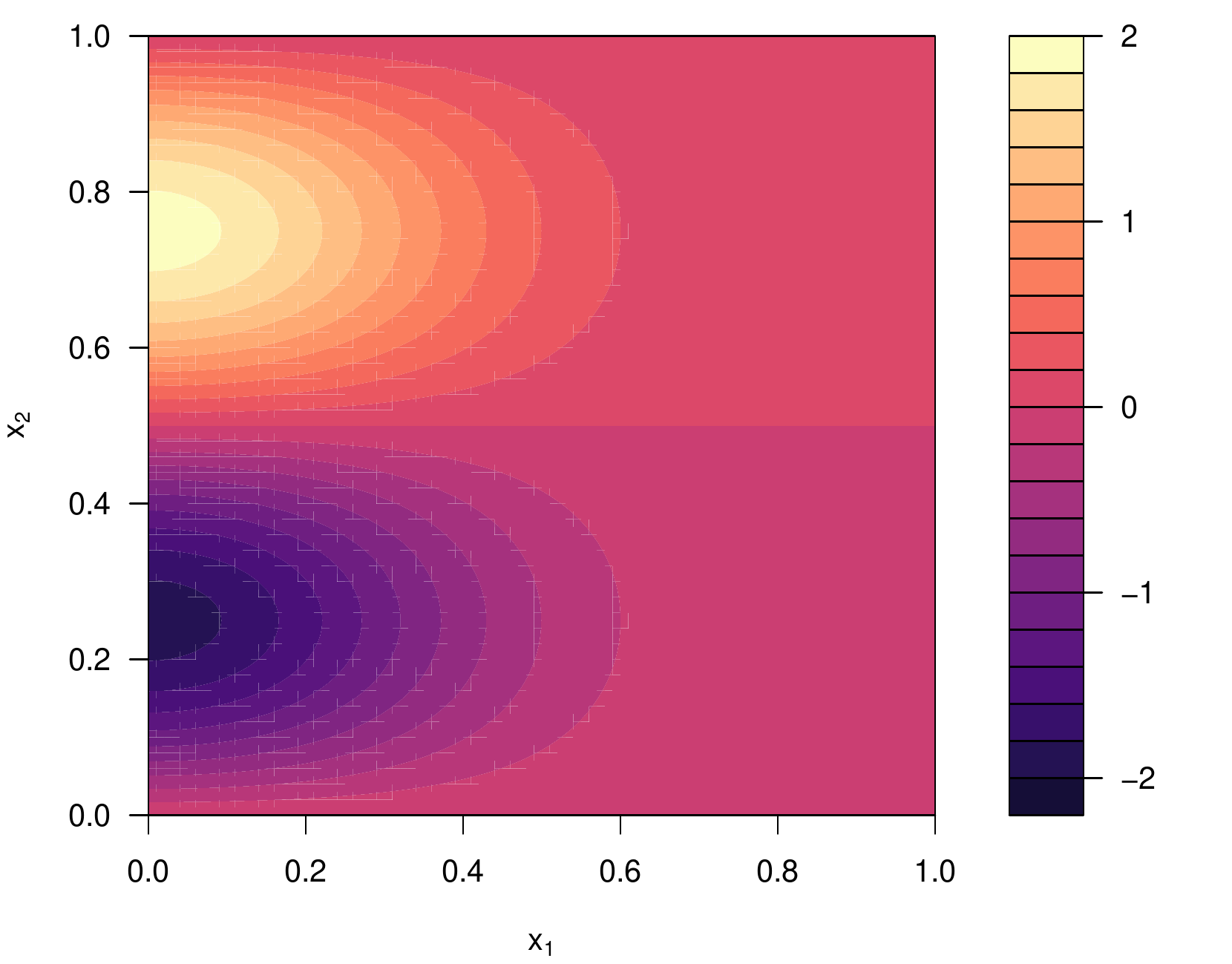}
\vspace{-0.1cm}
\caption{\footnotesize{The emulator expectation $\ed{K}{f(x)}$.}}
\label{fig_toymod1_a}
\end{subfigure}
\begin{subfigure}{0.4\columnwidth}
\centering
\includegraphics[scale=0.38]{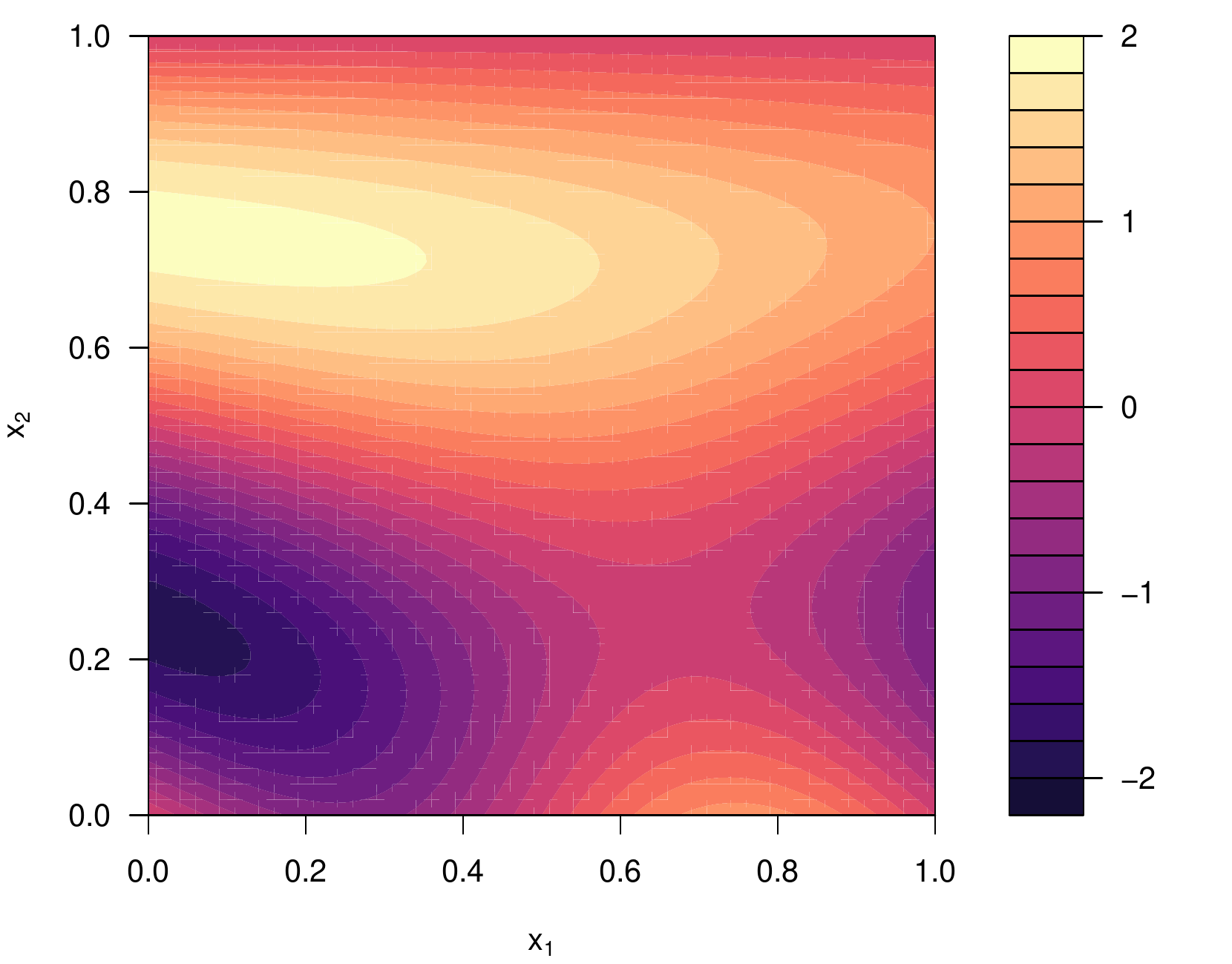}
\vspace{-0.1cm}
\caption{\footnotesize{The true 2-dimensional function $f(x)$}.}
\label{fig_toymod1_b}
\end{subfigure}
\begin{subfigure}{0.4\columnwidth}
\centering
\vspace{-0.2cm}
\includegraphics[scale=0.38]{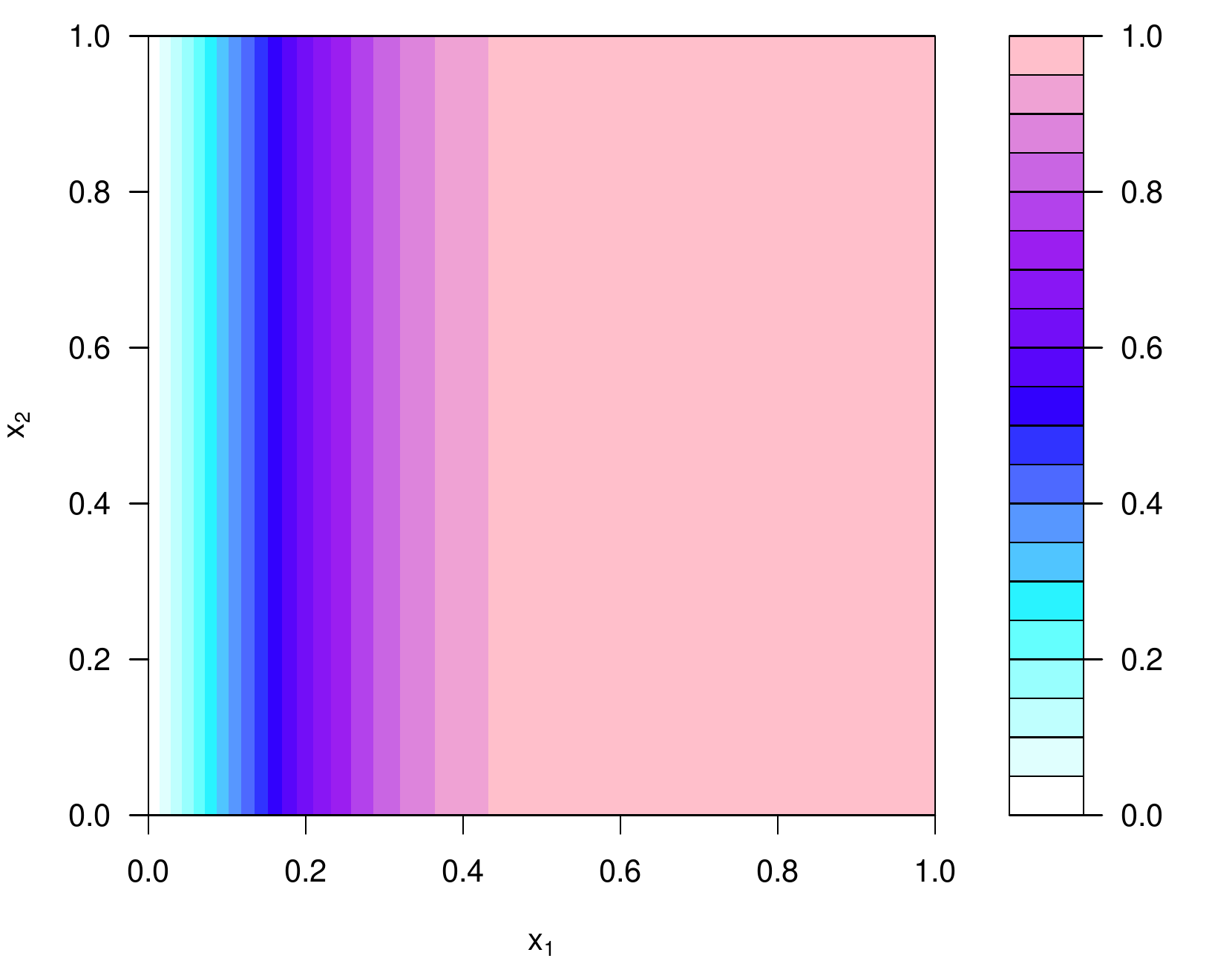}
\vspace{-0.1cm}
\caption{\footnotesize{The emulator stan. dev. 
$\sqrt{\vard{K}{f(x)}}$.}}
\label{fig_toymod1_c}
\end{subfigure}
\begin{subfigure}{0.4\columnwidth}
\centering
\vspace{-0.2cm}
\includegraphics[scale=0.38]{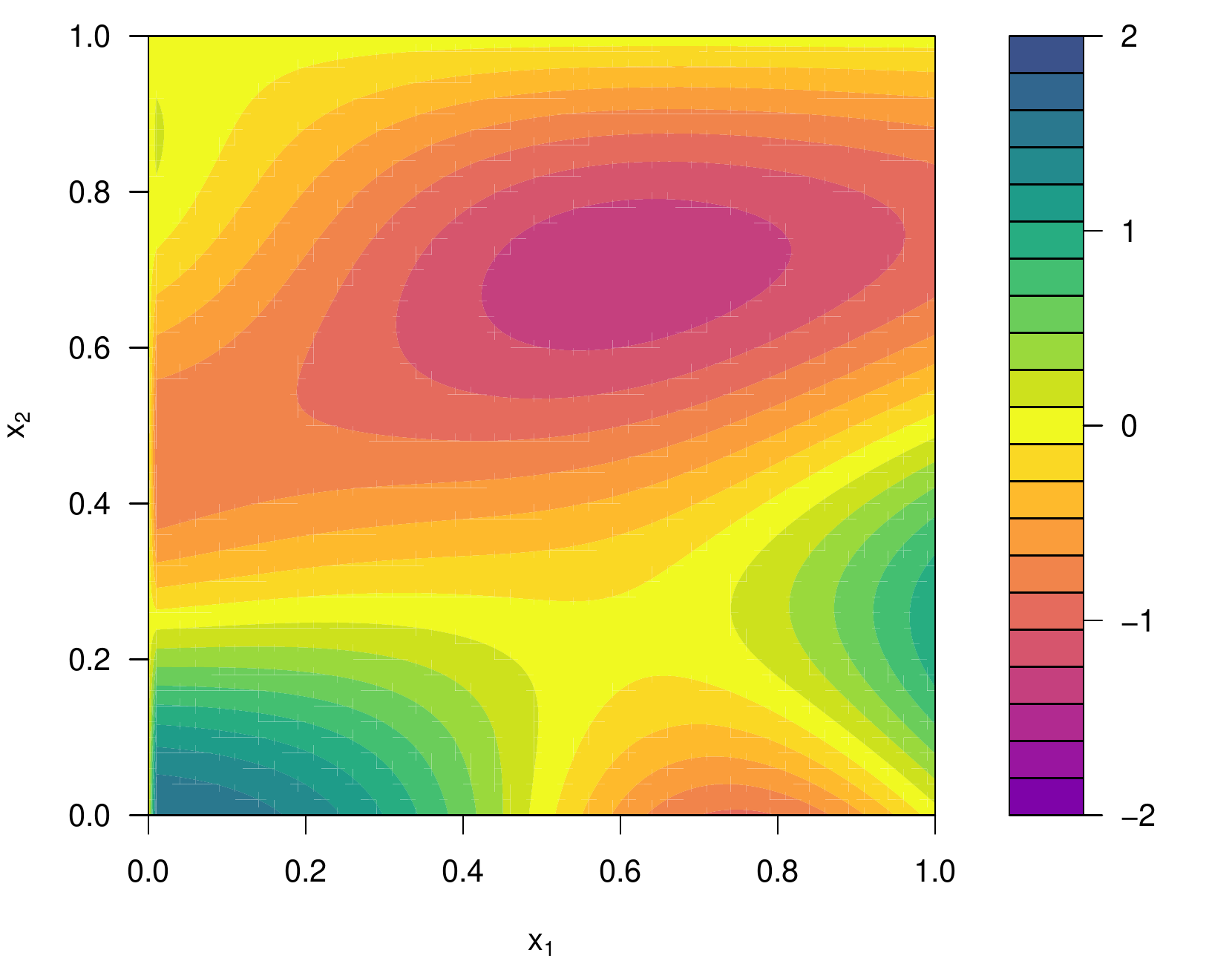}
\vspace{-0.1cm}
\caption{\footnotesize{Emulator diagnostics $S_K(x)$.}}
\label{fig_toymod1_d}
\end{subfigure}
\end{center}
%\begin{tabular}{ccc}
%\hspace{-1.cm} \includegraphics[scale=0.4]{plots/SingleKBemul_1.pdf} &
%\hspace{-0.5cm} \includegraphics[scale=0.4]{plots/toymodel.pdf} \\
%\hspace{-1cm}  \includegraphics[scale=0.4]{plots/SingleKBemul_2.pdf} &
%\hspace{-0.5cm}  \includegraphics[scale=0.4]{plots/SingleKBemul_3.pdf} \\
%\end{tabular}
%\vspace*{-0.5em}
\vspace{-0.4cm}
\caption{{Updating by a single known boundary $\mk$ at $x_1=0$.}}
\label{fig_toymod1}
\vspace{-0.4cm}
\end{figure}

For illustration, we consider the problem of emulating the 2-dimensional function 
\be
f(x) \;=\; -\sin\left(2 \pi x_2\right) + 0.9\sin\left(2 \pi (1-x_1)(1-x_2)\right)	
\ee
defined over the region $\mx$ given by $0<x_1<1$, $0<x_2 < 1$, where we assume a known boundary $\mk$ at $x_1=0$, and hence 
have that $f(x^K) = f(0,x_2) = -1.9 \sin\left(2 \pi x_2\right)$. The true output of $f(x)$ over $\mx$ is given in Figure~\ref{fig_toymod1_b} for reference.
Using a prior expectation $\e{f(x)}=0$, and a product Gaussian covariance structure with parameters $\theta=0.4$ and $\sigma=1$, we apply the expectation and variance updates (\ref{eq_EK1}) and (\ref{eq_VK1}) given the boundary $\mk$ at $x_1=0$ and find that
%\begin{align*}
%\ed{K}{f(x)} &=& -1.9 \exp\{-x_1^2/\theta^2\}  \sin(2 \pi x_2)   \\
%\vard{K}{f(x)} &=& \;\;\; 1 - \exp\{-2x_1^2/\theta^2\}
%\end{align*}
\begin{eqnarray*}
\ed{K}{f(x)} \; &=& \;  -1.9 \exp\{-x_1^2/\theta^2\}  \sin(2 \pi x_2)   \\
\vard{K}{f(x)} \; &=& \;\;\;\;1 - \exp\{-2x_1^2/\theta^2\}
\end{eqnarray*}
Figure~\ref{fig_toymod1_a} shows the adjusted expectation $\ed{K}{f(x)}$ over $\mx$, clearly illustrating how the expectation surface has been changed in the vicinity of $\mk$ to agree with the simulator behaviour. Figure~\ref{fig_toymod1_c} shows the adjusted emulator standard deviation $\sqrt{\vard{K}{f(x)}}$ and demonstrates the significant reduction in emulator uncertainty near $\mk$. Finally, Figure~\ref{fig_toymod1_d} shows simple emulator diagnostics over $\mx$ of the form of the standardised values
%\be
%S_K(x) = \frac{\ed{K}{f(x)} - f(x)}{\sqrt{\vard{K}{f(x)}}},
%\ee
$S_K(x) = (\ed{K}{f(x)} - f(x))/\sqrt{\vard{K}{f(x)}}$.
Thus any values of $x$ for which $S_K(x)$ was far from $0$ (a typical choice being $|S_K(x)| > 3$) would indicate 
%that not only our expectation for the emulator at that point was far from the simulator evaluation but also that the uncertainty on the emulator was small, which would suggest 
a conflict between emulator and simulator (see \cite{Tony_EmDiag} for details).
%Conservative limits on the behaviour of $S_K(x)$ suggest that any values outside $[-3,3]$ would indicate problems. 
For our boundary-adjusted emulator, the standardised diagnostics all maintain modest values lying well within $\pm 1.5$ standard deviations giving no cause for concern. 
\red{We specify $\theta$ and $\sigma$ a priori here, according to the Bayesian paradigm and also mainly for simplicity, but note that the known boundary approach that we describe can be used in combination with various methods of assessing such covariance function parameters (and indeed covariance functions). For example, if one wished to use maximum likelihood, the likelihood calculated given $D$ could also be reduced to tractable form using similar sufficiency arguments, by employing equation~(\ref{eq_covK2}).}

\subsection{Updating by further model evaluations}\label{sssec_up_fur}

Most importantly, as we have analytic expressions for $\ed{K}{f(x)}$, $\vard{K}{f(x)}$ and $\covd{K}{f(x)}{f(x')}$ we are now able to include additional simulator evaluations into the emulation process. 
%Having incorporated the information available on the emulator's boundary behaviour, we now wish to focus on the more traditional emulation problem of updating our emulator to represent the simulator's behaviour across all of $\mx$,  not just in the vicinity of $\mk$. 
To do this, we perform $n$ (expensive) evaluations, $D$, of the full simulator across $\mx$, and use these to supplement the evaluations, $K$, available on the boundary. We want to update the emulator by the union of the evaluations $D$ and $K$, that is to find $\ed{D \cup K}{f(x)}$, $\vard{D \cup K}{f(x)}$ and $\covd{D\cup K}{f(x)}{f(x')}$. This can be achieved via a sequential Bayes Linear update:
%\ba
%\ed{D \cup K}{f(x)} &=& \ed{K}{f(x)} + \covd{K}{f(x)}{D} \vard{K}{D}^{-1}(D- \ed{K}{D}) \label{eq_BLmDK}\\
%\vard{D \cup K}{f(x)} &=& \vard{K}{f(x)} - \covd{K}{f(x)}{D} \vard{K}{D}^{-1}\covd{K}{D}{f(x)} \label{eq_BLvDK}\\
%\covd{D\cup K}{f(x)}{f(x')} &=& \covd{K}{f(x)}{f(x')} \nonumber\\
%&&~~~- \covd{K}{f(x)}{D} \vard{K}{D}^{-1} \covd{K}{D}{f(x')} \label{eq_BLcDK}  
%\ea
\ba
 \ed{D \cup K}{f(x)}  &=&  \ed{K}{f(x)} + \covd{K}{f(x)}{D} \vard{K}{D}^{-1}(D- \ed{K}{D}) \label{eq_BLmDK}\\
\vard{D \cup K}{f(x)}  &=&  \vard{K}{f(x)} - \covd{K}{f(x)}{D} \vard{K}{D}^{-1}\covd{K}{D}{f(x)} \label{eq_BLvDK}\\
\covd{D\cup K}{f(x)}{f(x')}  &=&  \covd{K}{f(x)}{f(x')} - \covd{K}{f(x)}{D} \vard{K}{D}^{-1} \covd{K}{D}{f(x')} \label{eq_BLcDK}  
\ea
where we first update our emulator analytically by $K$, and subsequently update these quantities 
by the evaluations $D$~\cite{Goldstein07_BayesLinearBook}. 
As typically $n$ is small due to the relative expense of evaluating the full simulator, 
these calculations will remain tractable, as $\vard{K}{D}^{-1}$ will be feasible for modest $n$. 

Not only will the known boundary $\mk$ improve the accuracy of the emulator compared to just updating by $D$, for only trivial computational cost, it will also allow us to design a more informative set of runs that constitute $D$. We discuss appropriate designs for this scenario in section~\ref{sec_design_KB}.

\subsubsection{Incorporating Known Boundaries into Black Box Emulation Packages}

Consideration of the form of the sequential update given by equations~(\ref{eq_BLmDK})-(\ref{eq_BLcDK}), combined with the sufficiency 
argument presented in section~\ref{ssec_singleKB}, shows that for the full joint update by $D\cup K$, a sufficient set of points  is composed of 
a) the $n$ points in $D$,
b) the $n$ points formed from the projection of $D$ onto the boundary $\mk$, and 
c) the projection $x^K$ of the point of interest $x$, 
giving a total of $2n+1$ points. This has ramifications for users of black box Gaussian process emulation packages (such as
BACCO~\cite{Hankin:2005aa} or GPfit~\cite{JSSv064i12} in R, or GPy~\cite{gpy2014} for Python), 
which perhaps cannot be easily recoded to use the more sophisticated analytic emulation formula of equations~(\ref{eq_EK1}) and (\ref{eq_VK1}), but for which the inclusion of extra simulator evaluations is trivial. Hence such a user simply has to add the 
extra $(n+1)$ projected points on $\mk$ to their usual set of $n$ runs, and their black box GP package will produce results that 
precisely match equations~(\ref{eq_BLmDK})-(\ref{eq_BLcDK}). This however will require inverting a matrix of size $(2n+1)$ 
and hence will be slower than directly using the above analytic results, which only require inverting a matrix of size $n$.

% \emph{Sufficiency:} for the updating of our beliefs about the emulator, we see that $f(x^K)$ is sufficient for $K$. Hence, only the evaluation $K=f(x^K)$ is required and the evaluations $y^{(i)}$ are redundant \todo{conditionally independent? ``Under an assumption of an underlying Gaussian process, this result corresponds to a form of conditional independence statement.''}. This has ramifications for users of black box Gaussian process packages (such as
%BACCO~\cite{Hankin:2005aa} or GPfit~\cite{JSSv064i12} in R, or GPy~\cite{gpy2014} for Python), 
%which perhaps cannot be easily recoded to use the more sophisticated formula of equations~(\ref{eq_EK1}) and (\ref{eq_VK1}), but for which appropriate extra points on $\mk$, i.e. the projections on to $\mk$ of any points of interest, could easily be included. 

%
%Given the above results, in the next two subsections we proceed to update the emulator by a further known boundary, $\ml$. 
%In the first case the boundary is perpendicular to $\mk$, and in the second case the boundary is parallel to $\mk$.
%

\subsection{Updating by Two Perpendicular Known Boundaries}\label{ssec_two_perp_bound}
Given the above results, we now proceed to discuss the update of the emulator by a second known boundary, $\ml$. In the first case,
discussed here, $\ml$ is assumed perpendicular to $\mk$, and in the second case, discussed in section~\ref{ssec_two_para_bound}, $\ml$ is parallel to $\mk$. Detailed derivations of the results presented can be found in appendices~\ref{app_two_perp} and \ref{app_two_para}, and  the key results for single and dual-boundaries are summarised in Table~\ref{tab_sum_results}, for ease of comparison.

\begin{figure}
\begin{center}
\begin{tabular}{cc}
\includegraphics[scale=0.6]{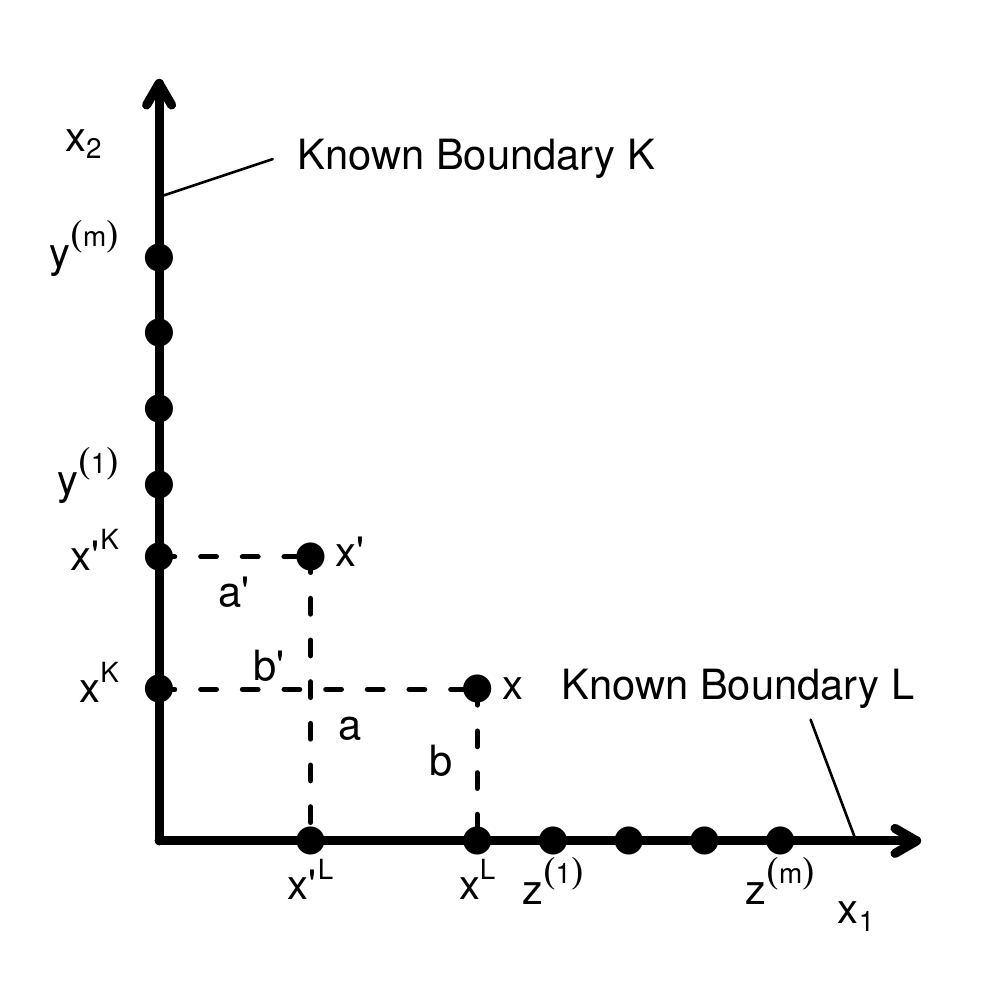} & \hspace{0.5cm}  \includegraphics[scale=0.6]{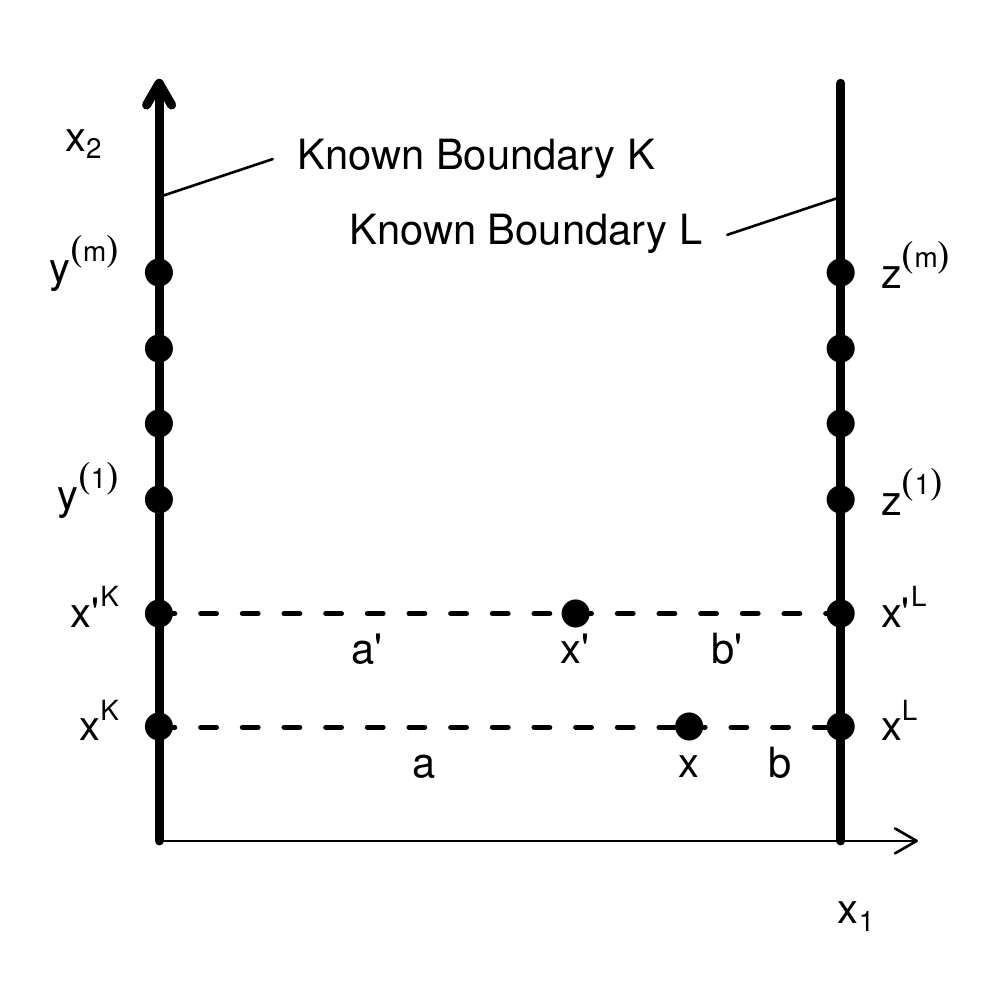} \\
\end{tabular}
\end{center}
\vspace{-1.0cm}
\caption{\footnotesize{Left panel: two perpendicular known boundaries. Right panel: two parallel known boundaries. In both cases $x$ and $x'$ are the points of interest for the emulation calculation, while $x^K$ and $x'^K$ are their orthogonal projection onto the known boundary $\mk$, and $x^L$ and $x'^L$ their orthogonal projection onto the known boundary $\ml$. The $y^{(i)}$ and $z^{(i)}$ represent a large number of points on the boundaries $\mk$ and $\ml$ respectively for which we can evaluate $f(y^{(i)})$ and $f(z^{(i)})$ analytically.}}
\label{fig_double_perp}
\vspace{-0.5cm}
\end{figure}

First, we assume the second known boundary $\ml$ is a $d-1$ dimensional hyperplane, perpendicular to the $x_2$ direction, as illustrated in Figure~\ref{fig_double_perp} (left panel). 
%Our goal is to update the emulator for $f(x)$, $x\in \mx$ by our knowledge of the function's behaviour on both boundaries $\mk$ and $\ml$, and by a set of runs $D$ within $\mx$. Thus we must find $\ed{D \cup L \cup K}{f(x)}$ and 
%$\vard{D \cup L \cup K}{f(x)}$. 
%We will show that since the updated correlation structure \eqref{eq_covK2} maintains its product from, the update by both boundaries can be achieved analytically, thus allowing us to quickly update the emulator by $K$ and then by $L$, before numerically updating by the evaluations $D$.
Our goal is to update the emulator for $f(x)$, $x\in \mx$ by our knowledge of the function's behaviour on both boundaries $\mk$ and $\ml$, and subsequently by a set of runs $D$ within $\mx$. Thus we must find $\ed{D \cup L \cup K}{f(x)}$ and 
$\vard{D \cup L \cup K}{f(x)}$. We do this sequentially by analytically updating by $K$ followed by $L$, then numerically by $D$.

As before, assume that the $f(x)$ is analytically solvable and hence inexpensive to evaluate along $\mathcal{L}$, permitting a large but finite number, $m$, of evaluations on $\ml$, denoted $z^{(1)},\dots,z^{(m)}$. As in Section~\ref{ssec_singleKB}, we define the corresponding length $m+1$ vector of boundary values $L$ as
\be \label{eq_defL}
L\;=\;\left(f(x^L),f(z^{(1)}),\dots,f(z^{(m)})\right)^T, 
\ee
which includes the projection $x^L$ of $x$ onto $\ml$.
%We now derive the analytic expressions for $\ed{L \cup K}{f(x)}$, $\vard{L \cup K}{f(x)}$ and $\covd{L \cup K}{f(x)}{f(x')}$.
%We perform the update by $K$ using the results of the previous section. We then use an analogous proof to that 
%of equation~(\ref{eq_covvar}), but
An analogous proof to that of equation~(\ref{eq_covvar}) gives%but now applied to $L$ gives
\be
 \covd{K}{f(x^L)}{L} \vard{K}{L}^{-1} \;=\; (1,0, \cdots, 0) \label{eq_covvar_K}  
\ee
while as the product correlation structure is not disturbed by the update by $K$, we also have 
\be
\covd{K}{f(x)}{L} \;=\; r_2(b) \, \covd{K}{f(x^L)}{L}  \label{eq_rcovxL} 
\ee
%Exploiting the result used in \eqref{eq_varKrowvec} and \eqref{eq_covvar} and as the product correlation structure is not disturbed by the boundary update by $K$, we can apply \eqref{eq_rcovxK} to find:
%\footnote{We are assuming there are no problems here due to the non-empty $\mk \cap \ml$. In fact the full Bayes linear update would instead use the generalised inverse if $L$ contains points on $\mk$ (which would possess zero variance), but equation~(\ref{eq_covvar_K}) will stay the same. \ian{Should we just explicitly put the generalised inverse in, or is it a little confusing?}} 
%\ba
%% \vard{K}{L}  \vard{K}{L}^{-1} &=& I_{(m+1)} \nonumber \\
%%\Rightarrow \quad 
%\covd{K}{f(x^L)}{L} \vard{K}{L}^{-1} &=& (1,0, \cdots, 0) \label{eq_covvar_K}.
%\ea
%Further, as the product correlation structure is not disturbed by the boundary update by $K$, we can apply \eqref{eq_rcovxK} to give
%\be
%\covd{K}{f(x)}{L}\vard{K}{L}^{-1} \;=\; r_2(b) \, (1,0, \cdots, 0)
%%\covd{K}{f(x^L)}{L}
%,  \label{eq_rcovxL} 
%\ee
where $b$ is the perpendicular distance from $x$ to $\ml$ and $r_2(\cdot)$ is the correlation function in the perpendicular direction to $\ml$, as shown in Figure~\ref{fig_double_perp} (left panel).
Using (\ref{eq_covvar_K}) and (\ref{eq_rcovxL}) the expectation of $f(x)$ adjusted by $K$ then $L$ can now be calculated  using the sequential update \eqref{eq_BLmDK} giving
\ba
 \ed{L \cup K}{f(x)} %&=& \ed{K}{f(x)} + \covd{K}{f(x)}{L} \vard{K}{L}^{-1}(L- \ed{K}{L})  \nonumber \\
  \;&=&\;  \ed{K}{f(x)} +  r_2(b) (1,0, \cdots, 0) (L- \ed{K}{L}) \nonumber \\
  \;&=&\;  \ed{K}{f(x)} +  r_2(b) (f(x^L) - \ed{K}{f(x^L)} )   \label{eq_expKL1_half}\\ 
 \;&=&\;   \e{f(x)} + r_1(a) (f(x^K) -  \e{f(x^K)})  \;+\; r_2(b) f(x^L)  \nonumber \\
 && \quad \;-\;  r_2(b) ( \e{f(x^L)} + r_1(a) (f(x^{LK}) -  \e{f(x^{LK})}) )  \nonumber \\
% &=&  \e{f(x)} + r_1(a) (f(x^K) -  \e{f(x^K)})  +  r_2(b) (f(x^L)  -  \e{f(x^L)}) - r_1(a) r_2(b) (f(x^{LK}) -  \e{f(x^{LK})}) \nonumber \\
 \;&=&\;  \e{f(x)} + r_1(a) \Delta f(x^K)  +  r_2(b) \Delta f(x^L) -  r_1(a) r_2(b) \Delta f(x^{LK})    \label{eq_expKL1}
\ea
where we have also used equation~(\ref{eq_EK1}) for $ \ed{K}{f(x)}$, defined $\Delta f(.) \equiv f(.) - \e{f(.)}$ and denoted the projection of
$x^L$ onto $\mk$ as $x^{LK}$, which is just the perpendicular projection of $x$ onto $\ml \cap \mk$.
%\vard{D \cup K}{f(x)} &=& \vard{K}{f(x)} - \covd{K}{f(x)}{D} \vard{K}{D}^{-1}\covd{K}{D}{f(x)} \label{eq_BLvDK}
An expression for the covariance adjusted by $K$ then $L$ is obtained by a similar argument (see appendix~\ref{app_two_perp})
%\ba
%\covd{L \cup K}{f(x)}{f(x')} 
%%&=& \covd{K}{f(x)}{f(x')} - \covd{K}{f(x)}{L} \vard{K}{L}^{-1}\covd{K}{L}{f(x')}  \nonumber \\
%%     &=& \covd{K}{f(x)}{f(x')} - r_2(b) (1,0, \cdots, 0)\covd{K}{L}{f(x')}   \nonumber \\
%%     &=& \covd{K}{f(x)}{f(x')} - r_2(b) \covd{K}{f(x^L)}{f(x')}  \nonumber \\
%%%     &=& \covd{K}{f(x)}{f(x')} - r_2(b) \covd{K}{f(x^L)}{f(x'^L)} r_2(b')  \nonumber \\
%%%     &=& r_2(b-b') \covd{K}{f(x^L)}{f(x'^L)}  - r_2(b) \covd{K}{f(x^L)}{f(x'^L)} r_2(b') \nonumber \\
%     &=& r_2(b-b') \covd{K}{f(x^L)}{f(x'^L)}  \nonumber\\
%     &&~~~~~- r_2(b) \covd{K}{f(x^L)}{f(x'^L)} r_2(b')  \label{eq_covKL1half} \\
%%     &=&  (r_2(b-b')- r_2(b) r_2(b') ) \covd{K}{f(x^L)}{f(x'^L)} \nonumber \\
%     &=& \sigma^2 R_1(a,a') \, R_2(b,b') \, r_{-1,-2}(x^{LK}-x'^{LK})   \label{eq_covKL1}
%\ea
\ba
\;\;\;\;\;  \covd{L \cup K}{f(x)}{f(x')} 
%&=& \covd{K}{f(x)}{f(x')} - \covd{K}{f(x)}{L} \vard{K}{L}^{-1}\covd{K}{L}{f(x')}  \nonumber \\
%     &=& \covd{K}{f(x)}{f(x')} - r_2(b) (1,0, \cdots, 0)\covd{K}{L}{f(x')}   \nonumber \\
%     &=& \covd{K}{f(x)}{f(x')} - r_2(b) \covd{K}{f(x^L)}{f(x')}  \nonumber \\
%%     &=& \covd{K}{f(x)}{f(x')} - r_2(b) \covd{K}{f(x^L)}{f(x'^L)} r_2(b')  \nonumber \\
%%     &=& r_2(b-b') \covd{K}{f(x^L)}{f(x'^L)}  - r_2(b) \covd{K}{f(x^L)}{f(x'^L)} r_2(b') \nonumber \\
     \;&=&\; r_2(b-b') \covd{K}{f(x^L)}{f(x'^L)}  - r_2(b) \covd{K}{f(x^L)}{f(x'^L)} r_2(b')  \label{eq_covKL1half} \\
%     &=&  (r_2(b-b')- r_2(b) r_2(b') ) \covd{K}{f(x^L)}{f(x'^L)} \nonumber \\
     \;&=&\; \sigma^2 R_1(a,a') \, R_2(b,b') \, r_{-1,-2}(x^{LK}-x'^{LK})   \label{eq_covKL1}
\ea
where we have defined the correlation function of the projection of $x$ and $x'$ onto $\ml \cap \mk$ as
\be
r_{-1,-2}(x^{LK}-x'^{LK})   \;=\;  \prod_{i=3}^d r_i(x^{LK}_i - x'^{LK}_i)  \;=\; \cov{f(x^{LK})}{f(x'^{LK})}
\ee
The updated variance is trivially obtained by setting $x=x'$ to get
\ba
\vard{L \cup K}{f(x)}  \;& = &\; \sigma^2 R_1(a,a) \, R_2(b,b)  \nonumber \\
   \;& = &\; \sigma^2 (1- r^2_1(a))(1-r^2_2(b)) \label{eq_varKL1}
\ea

%As with the single-boundary case, we note that there are intuitive limiting behaviours as the distance $b$ from the boundary $\ml$ tends to 0 or $\infty$ (see appendix~\ref{app_two_perp}). Also, as a consistency check, we see that all three expressions (\ref{eq_expKL1}), (\ref{eq_covKL1}) 
%and (\ref{eq_varKL1}) are invariant under interchange of the two boundaries, represented as the 
%transformation $K \leftrightarrow L$ and $a \leftrightarrow b$, as they should be.

As a consistency check, we see that all three expressions (\ref{eq_expKL1}), (\ref{eq_covKL1}) 
and (\ref{eq_varKL1}) are invariant under interchange of the two boundaries, represented as the 
transformation $K \leftrightarrow L$ and $a \leftrightarrow b$, as they should be. They also exhibit intuitive limiting behaviours as the distance $b$ from the boundary $\ml$ tends to 0 or $\infty$ (see appendix~\ref{app_two_perp}). 
Again, we observe that were we to sequentially update by a further $n$ evaluations, $D$, and calculate $\ed{D \cup L \cup K}{f(x)}$ and 
$\vard{D \cup L \cup K}{f(x)}$, the only points we require for sufficiency are $D$ and
the projections of $D$ and $x$ onto $\mk$, $\ml$, and $\mk\cap\ml$. This represents only $4n+3$ points, which is far fewer than
the $2(m+1) +1+ n$ points (with $m$ extremely large) that we started with. Again, users of black box emulators can easily insert these points, 
at the cost of having to invert a matrix now of size $4n+3$, instead of a single inversion of size $n$ were they to encode the above analytic results directly.

\begin{figure}[t]
\centering
\begin{subfigure}{0.325\columnwidth}
	\includegraphics[scale=0.35,viewport= 1 1 448 365,clip]{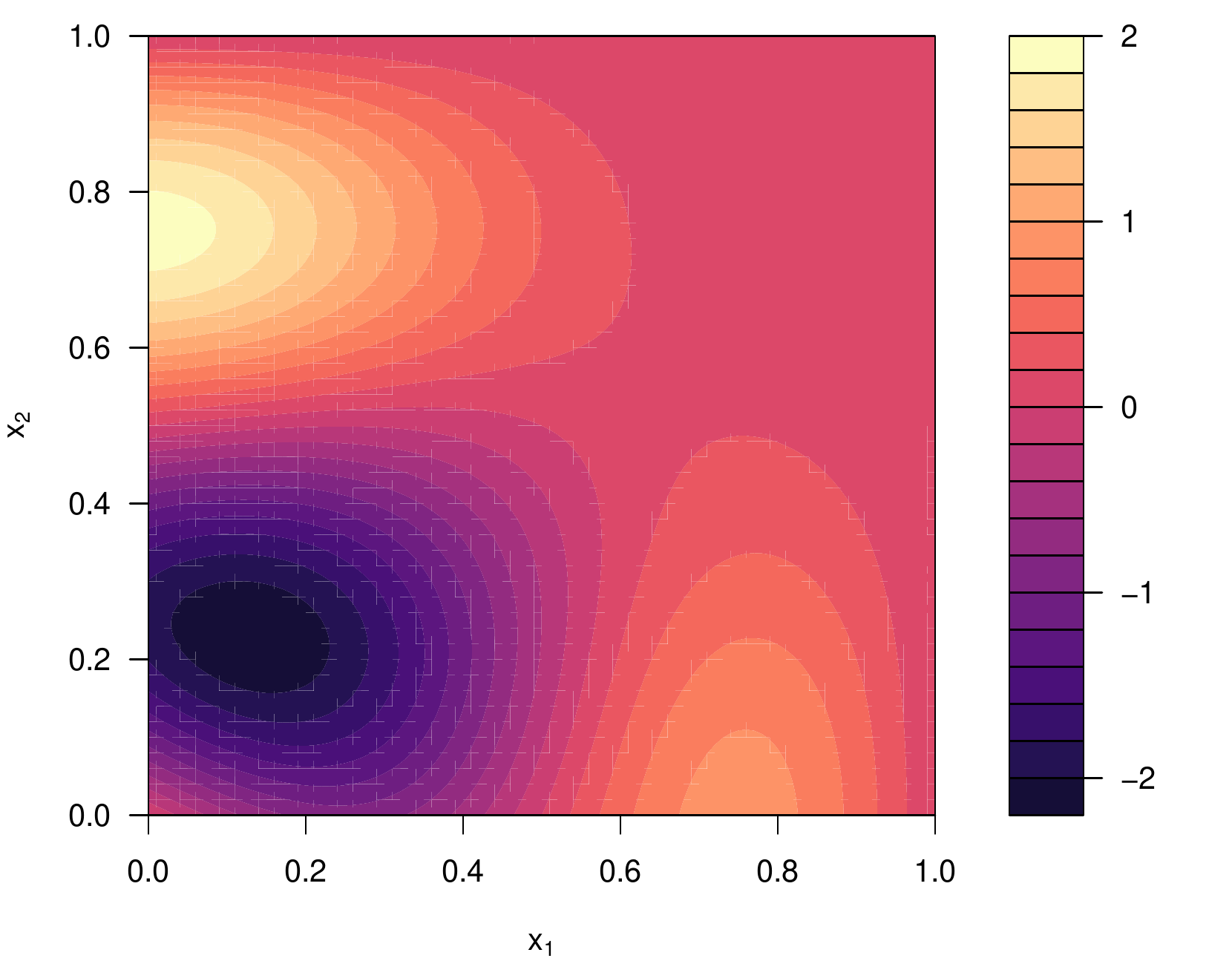}
	\vspace{-0.6cm}
	\caption{\footnotesize{$K\perp L$: $\ed{L \cup K}{f(x)}$}}
	\label{fig_toymod2_perp1}
\end{subfigure}
\begin{subfigure}{0.325\columnwidth}
	\includegraphics[scale=0.35,viewport= 1 1 454 365,clip]{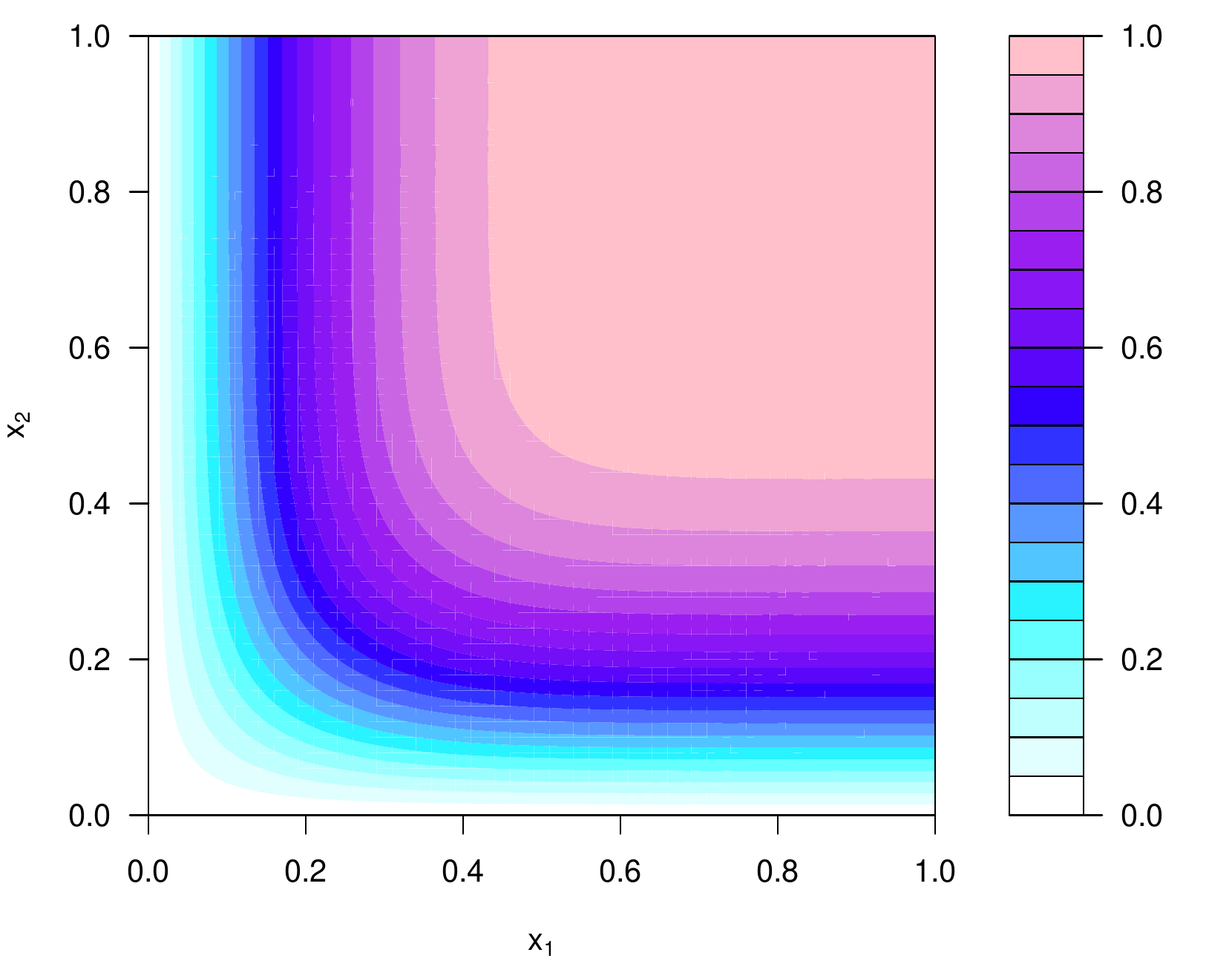}
	\vspace{-0.6cm}
	\caption{\footnotesize{$K\perp L$: $\sqrt{\vard{L \cup K}{f(x)}}$}}
	\label{fig_toymod2_perp2}
\end{subfigure}
\begin{subfigure}{0.325\columnwidth}
	\includegraphics[scale=0.35,viewport= 1 1 448 365,clip]{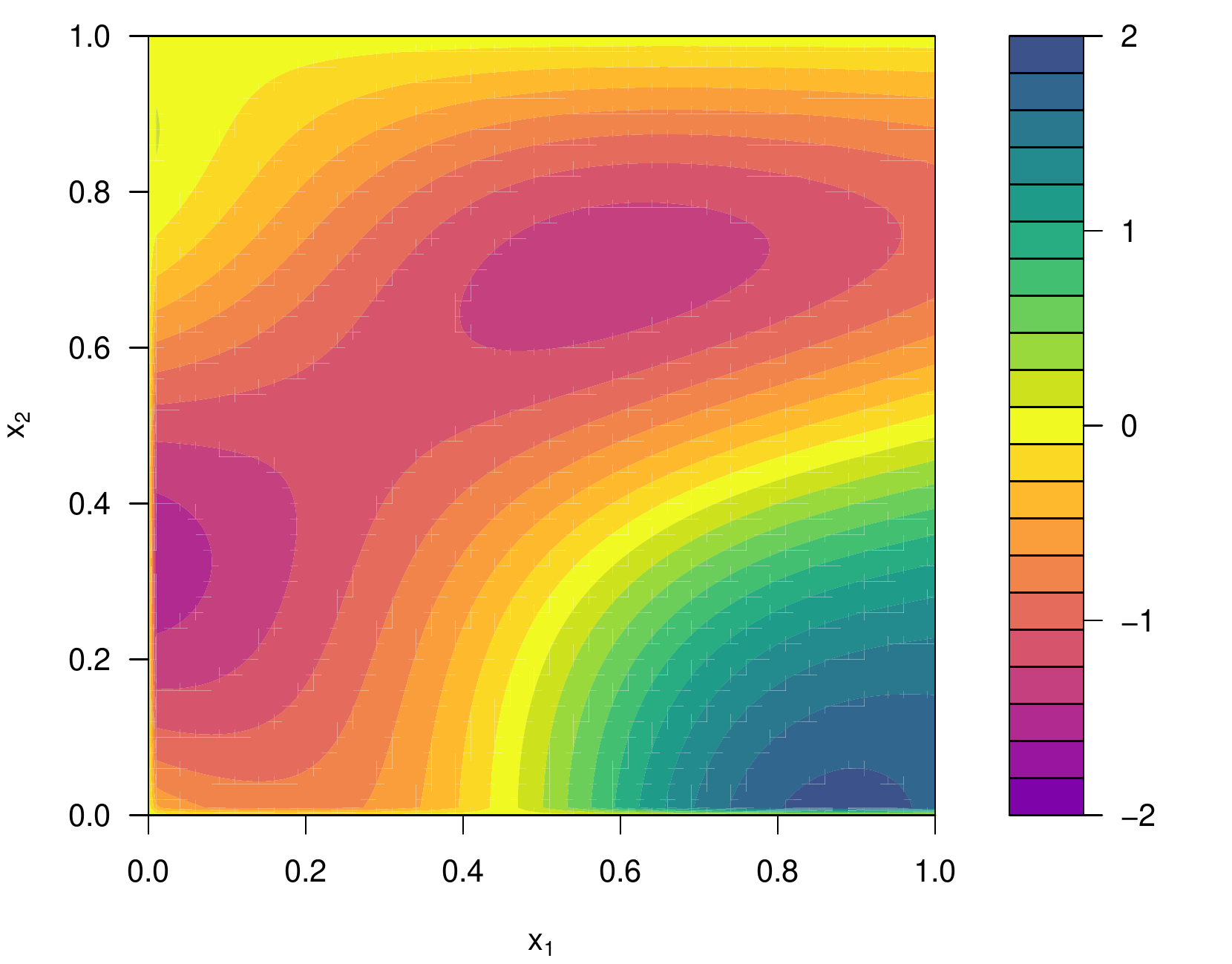}
	\vspace{-0.6cm}
	\caption{\footnotesize{$K\perp L$: Diagnostics $S_{L\cup K}(x)$}}
	\label{fig_toymod2_perp3}
\end{subfigure}

\begin{subfigure}{0.325\columnwidth}
	\vspace{-0.2cm}
	\includegraphics[scale=0.35,viewport= 1 1 448 365,clip]{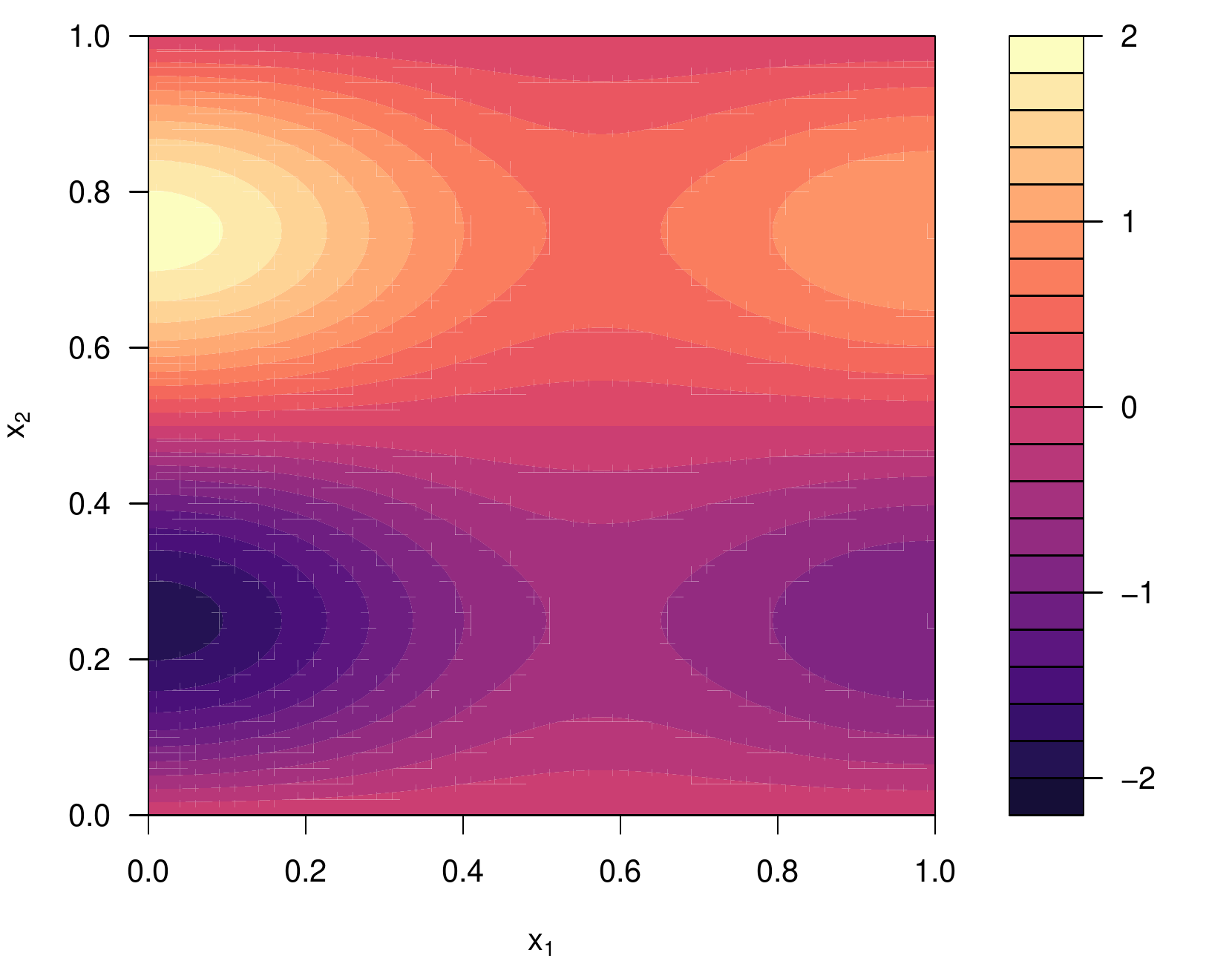}
	\vspace{-0.6cm}
	\caption{\footnotesize{$K\parallel L$: $\ed{L \cup K}{f(x)}$}}
	\label{fig_toymod2_par1}
\end{subfigure}
\begin{subfigure}{0.325\columnwidth}
	\vspace{-0.2cm}
	\includegraphics[scale=0.35,viewport= 1 1 454 365,clip]{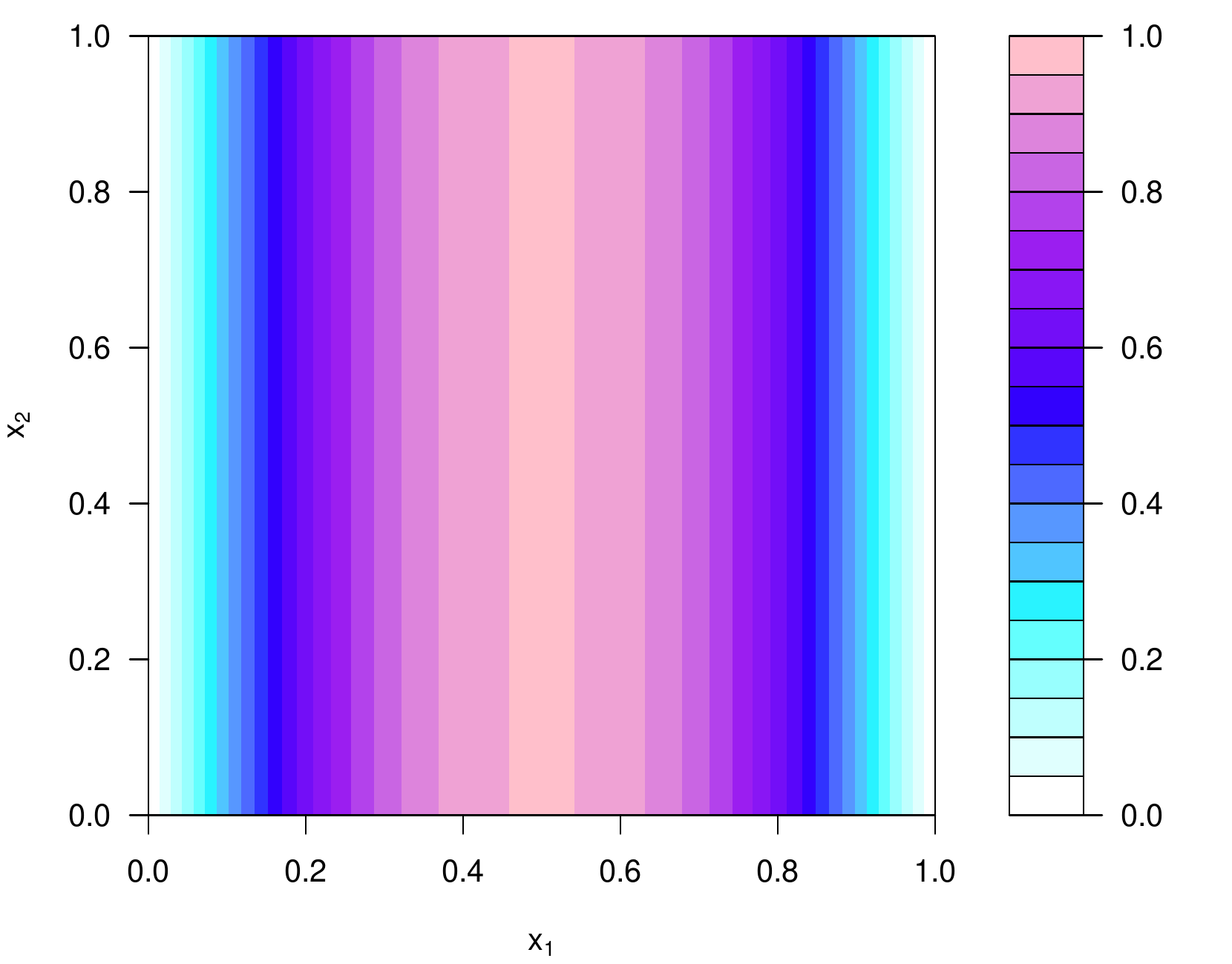}
	\vspace{-0.6cm}
	\caption{\footnotesize{$K\parallel L$: $\sqrt{\vard{L \cup K}{f(x)}}$}}
	\label{fig_toymod2_par2}
\end{subfigure}
\begin{subfigure}{0.325\columnwidth}
	\vspace{-0.2cm}
	\includegraphics[scale=0.35,viewport= 1 1 448 365,clip]{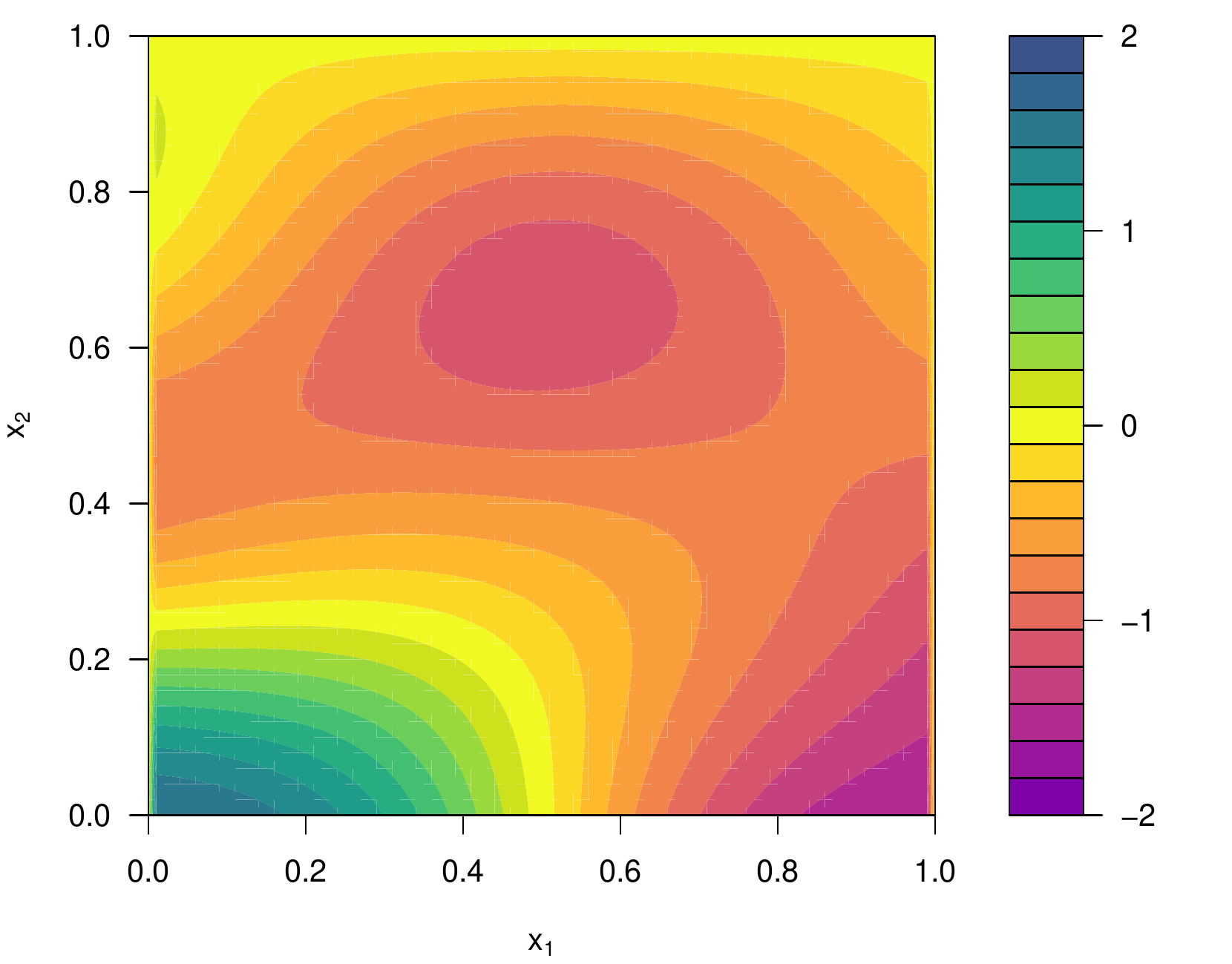}
	\vspace{-0.6cm}
	\caption{\footnotesize{$K\parallel L$: Diagnostics $S_{L\cup K}(x)$}}
	\label{fig_toymod2_par3}
\end{subfigure}
\vspace{-0.4cm}
\caption{\footnotesize{Emulators updated by two boundaries $\mk$ and $\ml$. Top row: perpendicular boundaries, with $\mk: x_1=0$ and $\ml: x_2=0$. Bottom row: parallel boundaries, with $\mk: x_1=0$ and $\ml: x_1=1$.}}
\label{fig_toymod2}
\vspace{-0.4cm}
\end{figure}

An example of an emulator updated by two perpendicular known boundaries is shown in Figures~\ref{fig_toymod2_perp1}--
\ref{fig_toymod2_perp3}, which give $\ed{L \cup K}{f(x)}$, $\sqrt{\vard{L \cup K}{f(x)}}$ and $S_{L\cup K}(x)$ respectively, for the simple function $f(x)$ introduced in section~\ref{sssec_2dmodel}. A second known boundary $\ml$ is now 
located at 
$x_2=0$, where we know that $f(x^L) = f(x_1,0) = -0.9\sin\left(2 \pi x_1\right)$.
%\todo{This needs more explanation - what is the second boundary? how does it behave?} (This is implicit from the definition of f(x).)
%Figure~\ref{fig_toymod2_perp1} displays the emulator expectation updated by the perpendicular boundaries at $\mk: x_1=0$ and $\ml: x_2=0$. 
As expected, we see the emulator expectation agrees exactly with the behaviour of the simulator $f(x)$ on $\mk$ and $\ml$ (as given in Figure~\ref{fig_toymod1_b}). We note also the intuitive property that the variance of the emulator reduces to zero as we approach the boundary, but remains at $\sigma^2=1$ when we are sufficiently distant. This sensibly represents the increase in knowledge about the simulator behaviour the closer we are to $\mk$ or $\ml$\red{, with the scale of the increase governed by the size of the correlation length parameter $\theta$}. Diagnostics $S_{L\cup K}(x)$ are again acceptable. 

%\begin{figure}[t]
%\centering
%\begin{subfigure}{0.325\columnwidth}
%	\includegraphics[scale=0.38]{plots/TwoPerpKBemul_1.pdf}
%	\caption{$K\perp L$: $\ed{L \cup K}{f(x)}$}
%	\label{fig_toymod2_perp1}
%\end{subfigure}
%\begin{subfigure}{0.325\columnwidth}
%	\includegraphics[scale=0.38]{plots/TwoPerpKBemul_2.pdf}
%	\caption{$K\perp L$: $\sqrt{\vard{L \cup K}{f(x)}}$}
%	\label{fig_toymod2_perp2}
%\end{subfigure}
%\begin{subfigure}{0.325\columnwidth}
%	\includegraphics[scale=0.38]{plots/TwoPerpKBemul_3.pdf}
%	\caption{$K\perp L$: Diagnostics $S_{L\cup K}(x)$}
%	\label{fig_toymod2_perp3}
%\end{subfigure}
%
%\begin{subfigure}{0.325\columnwidth}
%	\includegraphics[scale=0.38]{plots/TwoParaKBemul_1.pdf}
%	\caption{$K\parallel L$: $\ed{L \cup K}{f(x)}$}
%	\label{fig_toymod2_par1}
%\end{subfigure}
%\begin{subfigure}{0.325\columnwidth}
%	\includegraphics[scale=0.38]{plots/TwoParaKBemul_2.pdf}
%	\caption{$K\parallel L$: $\sqrt{\vard{L \cup K}{f(x)}}$}
%	\label{fig_toymod2_par2}
%\end{subfigure}
%\begin{subfigure}{0.325\columnwidth}
%	\includegraphics[scale=0.38]{plots/TwoParaKBemul_3.pdf}
%	\caption{$K\parallel L$: Diagnostics $S_{L\cup K}(x)$}
%	\label{fig_toymod2_par3}
%\end{subfigure}
%
%\caption{\footnotesize{Emulators updated by two boundaries $\mk$ and $\ml$. Top row: perpendicular boundaries, with $\mk: x_1=0$ and $\ml: x_2=0$. Bottom row: parallel boundaries, with $\mk: x_1=0$ and $\ml: x_1=1$.}}
%\label{fig_toymod2}
%\end{figure}

%\begin{figure} 
%\begin{tabular}{ccc}
%\hspace{-0.5cm} \includegraphics[scale=0.4]{plots/SingleKBemul_1.pdf} &
%\hspace{-1.5cm}  \includegraphics[scale=0.4]{plots/SingleKBemul_2.pdf} &
%\hspace{-1.5cm}  \includegraphics[scale=0.4]{plots/SingleKBemul_3.pdf} \\
%\hspace{-0.5cm} \includegraphics[scale=0.4]{plots/TwoPerpKBemul_1.pdf} &
%\hspace{-1.5cm}  \includegraphics[scale=0.4]{plots/TwoPerpKBemul_2.pdf} &
%\hspace{-1.5cm}  \includegraphics[scale=0.4]{plots/TwoPerpKBemul_3.pdf} \\
%\hspace{-0.5cm} \includegraphics[scale=0.4]{plots/TwoParaKBemul_1.pdf} &
%\hspace{-1.5cm}  \includegraphics[scale=0.4]{plots/TwoParaKBemul_2.pdf} &
%\hspace{-1.5cm}  \includegraphics[scale=0.4]{plots/TwoParaKBemul_3.pdf} \\
%\end{tabular}
%\caption{stuff}
%\label{fig_toymod}
%\end{figure}

\subsection{Updating by Two Parallel Known Boundaries}\label{ssec_two_para_bound}
Consider now a second boundary $\ml$ located at $x_1=c$, that is therefore parallel to the original boundary $\mk$ at $x_1=0$. 
As updating by $\mk$ leaves the correlation structure $\covd{K}{f(x)}{f(x')}$ still in product form, critically with respect to the $x_1$ term, we can still perform a subsequent analytic update by $\ml$.
%As the product structure of the correlation function is preserved under a boundary update, we can adjust our emulator by $\ml$ as its effects are confined to a further modification of the correlation in the $x_1$ direction. 
We define $L$ as before by \eqref{eq_defL}, and denote the distance from point $x$ to its perpendicular projection, $x^L$, onto $\ml$ as $b$, (see Figure~ \ref{fig_double_perp}, right panel), where now $a+b = c$. See appendix~\ref{app_two_para} for more details of the following derivations.
%\begin{figure}
%\begin{center}
%\includegraphics[scale=0.65]{plots/doubleKB_para_2x.pdf}
%\end{center}
%\caption{The two parallel boundaries case. $x$ and $x'$ are the points of interest for the emulation calculation, while $x^K$ and $x'^K$ are their orthogonal projection onto the known boundary $\mk$, and $x^L$ and $x'^L$ their orthogonal projection onto the known boundary $\ml$. The $y^{(i)}$ and $z^{(i)}$ represent a large number of points on the boundaries $\mk$ and $\ml$ respectively for which we can evaluate $f(y^{(i)})$ and $f(z^{(i)})$ analytically (or at least very quickly).}
%\label{fig_double_bound_para}
%\end{figure}

In this parallel case, analogous results to \eqref{eq_rcovxL} and \eqref{eq_covvar_K} can be derived, which combine to give
\be
\covd{K}{f(x)}{L}\vard{K}{L}^{-1} \;=\; R_1^{(2)}(a,c) (1,0, \cdots, 0)
\ee
%where $R_1^{(2)}(a,c) = R_1(a,c) /R_1(c,c)$ and $R_1(\cdot,\cdot)$ is as in \eqref{eq:UpdCorrComp}. Introducing this result into the sequential adjusted expectation equation \eqref{eq_BLmDK} gives:
where $R_1^{(2)}(a,c) = R_1(a,c) /R_1(c,c)$ and $R_1(\cdot,\cdot)$ is as in \eqref{eq:UpdCorrComp}. Inserting into equation \eqref{eq_BLmDK}, 
the emulator expectation adjusted by both $L$ and $K$ can be shown to be (see appendix~\ref{app_two_para}):
%Introducing this result into the sequential adjusted expectation equation \eqref{eq_BLmDK} gives:
\ba
 \ed{L \cup K}{f(x)} \;&=&\;  \ed{K}{f(x)} + R_1^{(2)}(a,c) (1,0, \cdots, 0) (L - \ed{K}{L})  \nonumber  \\
\;&=&\; \e{f(x)} + \left[ \frac{r_1(a)- r_1(b) r_1(c)}{1-r^2_1(c) } \right] \Delta f(x^K) + 
		\left[\frac{r_1(b) - r_1(a) r_1(c)}{1-r^2_1(c) }\right] \Delta f(x^L) 	\label{eq_expLK1_para} 
\ea
where we have exploited the fact that the projection of $x^L$ onto $\mk$ is just $x^K$. 
Similarly, the covariance adjusted by $L$ and $K$ can be shown to be (see appendix~\ref{app_two_para}):
\begin{align}
\covd{L \cup K}{f(x)}{f(x')} % &=& \covd{K}{f(x)}{f(x')} - \covd{K}{f(x)}{L} \vard{K}{L}^{-1}\covd{K}{L}{f(x')}  \nonumber \\
     &\;=\; \covd{K}{f(x)}{f(x')} - R_1^{(2)}(a,c) (1,0, \cdots, 0)\covd{K}{L}{f(x')}   \nonumber \\
%     &=& \covd{K}{f(x)}{f(x')} - R_1^{(2)}(a,c) \covd{K}{f(x^L)}{f(x')}  \nonumber \\
%     &=&  \covd{K}{f(x)}{f(x')} - R_1^{(2)}(a,c) \covd{K}{f(x^L)}{f(x'^L)} R_1^{(2)}(c,a')  \nonumber \\
%     &=& \sigma^2 R_1(a,a') \, r_{-1}(x^K-x'^K) \nonumber \\
%     && - R_1^{(2)}(a,c) \sigma^2 R_1(c,c) \, r_{-1}(x^K-x'^K) R_1^{(2)}(c,a')  \nonumber \\
%    &=&  \sigma^2 \, r_{-1}(x^K-x'^K) \left\{  R_1(a,a') -  R_1^{(2)}(a,c)  R_1(c,c) R_1^{(2)}(c,a') \right\} \nonumber \\
    &\;=\;  \sigma^2 \, r_{-1}(x^K-x'^K) \left\{  R_1(a,a') -  \frac{R_1(a,c) R_1(c,a')}{R_1(c,c)} \right\}, \nonumber 
\intertext{which is just a generalised form of \eqref{eq_covK2}. We can expand the functions $R_1$ to obtain}
%an expression in terms of the original correlation function components:}
 \covd{L \cup K}{f(x)}{f(x')}   %&=  \sigma^2 \, \frac{r_{-1}(x^K-x'^K)}{R_1(c,c)} \left\{  
    	%(r_1(a-a')-   r_1(a) r_1(a') ) (  1 -   r^2_1(c) ) \; -   \right. \nonumber \\
    %&   \quad\quad \quad \quad \quad \quad \quad \quad \left. (r_1(b)-   r_1(a) r_1(c) )( r_1(b')-   r_1(c) r_1(a') )
     %\right\} \nonumber \\
     &\;=\; \sigma^2 \, \frac{r_{-1}(x^K-x'^K)}{1 -   r^2_1(c)} \Big\{  r_1(a-a')(1 -   r^2_1(c)) -  r_1(a)r_1(a') - r_1(b)r_1(b')   \nonumber \\
&  \phantom{\; \sigma^2 \, \frac{r_{-1}(x^K-x'^K)}{1 -   r^2_1(c)} \Big\{}\quad + \; r_1(c) \big[ r_1(a) r_1(b') +  r_1(b) r_1(a') \big]  \Big\}, \label{eq_covLK1_para}
\end{align}
which we note is now explicitly invariant under interchange of $\mk$ and $\ml$ (with $a\leftrightarrow b$ etc.) as expected. % $\mk \leftrightarrow \ml$ (as $c=a+b$ is invariant under $a\leftrightarrow b, a' \leftrightarrow b'$, as is $a-a' = b-b'$).
Again, the adjusted variance is obtained by setting $x=x'$ 
\ba
\vard{L \cup K}{f(x)}    &\;=\;& \sigma^2 \, \frac{1}{1 -   r^2_1(c)} \Big\{ 1 -   r^2_1(c) -  r^2_1(a) - r^2_1(b)  
 + \; 2 r_1(c) r_1(a) r_1(b)  \Big\} \label{eq_varLK1_para}
\ea
%Again, the form of the analytic expressions for $\ed{L \cup K}{f(x)}$, $\covd{L \cup K}{f(x)}{f(x')}$ and $\vard{L \cup K}{f(x)}$, given now by equations~(\ref{eq_expLK1_para}), (\ref{eq_covLK1_para}) and (\ref{eq_varLK1_para}) show the following:
%\begin{enumerate}[(a)] 
%\item All expressions are invariant under interchange of boundaries $\mk \leftrightarrow \ml$ as before (as of course, they must be). 
%\item Limiting behaviour all seems as expected and is given by equations~(\ref{eq_lim2pKB}) and (\ref{eq_limCov2pKB}), and the 
%equivalent expressions for $a\rightarrow 0$. \ian{We could also examine $\ed{L \cup K}{f(x)}$ and $\vard{L \cup K}{f(x)}$ at the
%halfway point $a=b=c/2$, between the two boundaries, to gain some intuition as to their behaviour (the later will give the maximum variance now possible)}.
%\item Sufficiency: as before, but now just the $2n$ points that are the projection of $D$ onto $\mk$ and $\ml$ are sufficient for $K$ and $L$, as the intersection $\ml \cap \mk$ is empty.
%\en
Again, by inspection of these results we see that the only relevant information for our updated emulator at a general point $x$ are the projections of $x$ onto $\mk$ and $\ml$. 
Thus to update the emulator sequentially by $K$, $L$ then $D$, we only need to include the additional $2(n+1)$ points of the projections of $D$ and $x$ onto $\mk$ and $\ml$, noting that unlike in the perpendicular case, the intersection $\ml \cap \mk$ is now empty.

%Thus to evaluate the emulator at the $n$ points in $D$, required for the sequential update, we just require the $2(n+1)$ points of projections of $D$ and $x$ onto $\mk$ and $\ml$, noting that unlike the perpendicular case the intersection $\ml \cap \mk$ is empty.

%We summarise the results of this section and the previous two sections, for one, two perpendicular and two parallel known boundaries in table~\ref{tab_sum_results}, for ease of comparison.
\begin{table}
\begin{mdframed}
\begin{small}
%\begin{center}
When updating by one boundary $\mk$, with $R_1(a,a') \;=\; r_1(a-a')-r_1(a) r_1(a') $ and $\Delta f(.) \equiv f(.) - \e{f(.)}$:
\vspace{-0.1cm}
\ba
%\ed{K}{f(x)} &=& \e{f(x)} + r_1(a) (f(x^K) -  \e{f(x^K)})  \nonumber \\
\ed{K}{f(x)} \;&=&\; \e{f(x)} + r_1(a)  \Delta f(x^K)  \nonumber \\
\covd{K}{f(x)}{f(x')} \;&=&\; \sigma^2 R_1(a,a') \, r_{-1}(x^K-x'^K)  \nonumber \\
\vard{K}{f(x)} \;&=&\; \sigma^2(1 - r_1(a)^2)    \nonumber
\ea
\vspace{0.3cm}
\end{small}
\end{mdframed}
\vspace{-1cm}
\begin{mdframed}
\begin{small}
When updating by two perpendicular boundaries $\mk$ and $\ml$:
\vspace{-0.1cm}
\ba
 \ed{L \cup K}{f(x)} &=& \e{f(x)} + r_1(a) \Delta f(x^K)  +  r_2(b) \Delta f(x^L) -  r_1(a) r_2(b) \Delta f(x^{LK})   \nonumber \\
\covd{L \cup K}{f(x)}{f(x')} &=& \sigma^2 R_1(a,a') \, R_2(b,b') \, r_{-1,-2}(x^{LK}-x'^{LK})  \nonumber \\
\vard{L \cup K}{f(x)}  & = & \sigma^2 (1- r^2_1(a))(1-r^2_2(b))  \nonumber
\ea
\vspace{0.3cm}
\end{small}
\end{mdframed}
\vspace{-1cm}
\begin{mdframed}
\begin{small}
When updating by two parallel boundaries $\mk$ and $\ml$:
\ba
 \ed{L \cup K}{f(x)} &=& \e{f(x)} + \left[ \frac{r_1(a)- r_1(b) r_1(c)}{1-r^2_1(c) } \right] \Delta f(x^K) + 
		\left[\frac{r_1(b) - r_1(a) r_1(c)}{1-r^2_1(c) }\right] \Delta f(x^L)  \nonumber \\
		\covd{L \cup K}{f(x)}{f(x')} &=&  %\sigma^2 \, r_{-1}(x^K-x'^K) \left\{  R_1(a,a') - 
%								 \frac{R_1(a,c) R_1(c,a')}{R_1(c,c)} \right\}  \nonumber \ian{give \; full \;exp?} \\
 \sigma^2 \, \frac{r_{-1}(x^K-x'^K)}{1 -   r^2_1(c)} \Big\{  r_1(a-a')(1 -   r^2_1(c)) -  r_1(a)r_1(a') - r_1(b)r_1(b')   \nonumber \\
&&  \phantom{\sigma^2 \, \frac{r_{-1}(x^K-x'^K)}{1 -   r^2_1(c)} \Big\{}\quad + \; r_1(c) \big[ r_1(a) r_1(b') +  r_1(b) r_1(a') \big]  \Big\}\nonumber\\
		\vard{L \cup K}{f(x)}    &=& \sigma^2 \, \frac{1}{1 -   r^2_1(c)} \Big\{ 1 -   r^2_1(c) -  r^2_1(a) - r^2_1(b) 
								+ \; 2 r_1(c) r_1(a) r_1(b)  \Big\}    \nonumber
\ea
\vspace{-0.4cm}
%\end{center}
\end{small}
\end{mdframed}
\caption{\footnotesize{A summary for comparison of the updated emulator results found for the three main cases: a single boundary, two perpendicular boundaries and two parallel boundaries. \red{The variance results, which are of course special cases of the covariance results, are included for ease of interpretation.}}}
\label{tab_sum_results}
\vspace{-0.5cm}
\end{table}

\subsection{Continuous Known Boundaries}\label{ssec_continuous_case}
%\todo{JAC: Needs to be its own section or with the single boundary case}

%\ian{While it may be too confusing to do this continuous case, it may elevate the above which is arguably a little simple, to being a bit more new and a bit more cool. I don't think anyone has ever generalised the Bayes Linear update equations to a continuum of points before, so we could claim real originality. Any thoughts on this welcome.}

We now show how to generalise the above calculations from using a slightly artificial discrete and finite set of $m$ known points on each boundary, which only requires a standard Bayes linear update, to a more natural continuum of known points on a continuous boundary, which requires a generalised Bayes linear update.
%In general, when considering the problem of emulation of a complex computer simulation we are (necessarily) limited to performing a finite collection of simulator evaluations as our training set for the emulator. However, as the simulator's behaviour is known precisely for all points along the continuous boundary $\mk$, in principle, we have access to a continuum  of known points along $\mk$ for use in the  boundary update. We now show how the above calculations can be generalised from the traditional case of  updating via a discrete and finite set of $m$ known points on each boundary using the standard Bayes linear update, to updating by a continuum of known points on a continuous boundary.
Let the points on the boundary $\mk$ perpendicular to $x_1$ be denoted $K = \{f(y): y \in \mk \}$. The Bayes linear update can be 
generalised from the case of finite points to that of a continuum of points in the following way (we are unaware of this generalisation having been performed previously, but note that it follows from the foundational position that views the Bayes linear update as a projection~\cite{Goldstein07_BayesLinearBook}). 
The adjusted expectation changes from the matrix equation, 
\[
\ed{K}{f(x)} \;=\; \e{f(x)} + \cov{f(x)}{K} \var{K}^{-1}(K- \e{K}), 
\]
to the integral equation
\be
\ed{K}{f(x)} \;=\; \e{f(x)} + \int_{y \in \mk} \int_{y' \in \mk} \cov{f(x)}{f(y)} \; s(y,y') \; (f(y')- \e{f(y')}) dy dy',  \label{eq_BLmK2}
\ee
and the covariance update becomes
\[
\covd{K}{f(x)}{f(x')} \; =\;  \cov{f(x)}{f(x')} - \int_{y \in \mk} \int_{y' \in \mk} \cov{f(x)}{f(y)} \; s(y,y') \;  \cov{f(y')}{f(x')} \;  dy dy'.
\label{eq_BLcK2} \nonumber
\]
%\ba
%\vard{K}{f(x)} &=& \var{f(x)} - \int_{y \in \mk} \int_{y' \in \mk} \cov{f(x)}{f(y)} \; s(y,y') \;  \cov{f(y')}{f(x)} \;  dy dy'  \label{eq_BLvK2} \\
%\covd{K}{f(x)}{f(x')} &=& \cov{f(x)}{f(x')} - \int_{y \in \mk} \int_{y' \in \mk} \cov{f(x)}{f(y)} \; s(y,y') \;  \cov{f(y')}{f(x')} \;  dy dy' 
%\label{eq_BLcK2} \nonumber
%\ea
Here $s(x,x')$ represents the infinite dimensional generalisation of $\var{K}^{-1}$, and satisfies the equivalent inverse property to that of equation \eqref{eq_covvar} giving
\[
\int_{y' \in \mk} \cov{f(y)}{f(y')} s(y',y'') \; dy' \;=\; \delta(y-y''), \quad \quad {\rm for} \; y,y'' \in \mk \label{eq_infinv}
\]
where $ \delta(y-y'')$ is the Dirac delta function, the generalisation of the identity matrix.
Again, if we denote the projection of a general point $x \in \mx$ onto $\mk$ as $x^K$, we have for $y\in \mk$ that 
\be
\cov{f(x)}{f(y)} = r_1(a) \, \cov{f(x^K)}{f(y)},
\ee
which on substitution into (\ref{eq_BLmK2}) yields
\ba
\ed{K}{f(x)} &=& \e{f(x)} + \int_{y \in \mk} \int_{y' \in \mk} r_1(a) \, \cov{f(x^K)}{f(y)} \; s(y,y') \; (f(y')- \e{f(y')}) dy dy'  \nonumber \\
  &=& \e{f(x)} + r_1(a)  \int_{y' \in \mk} \, \delta(x^K - y') \; (f(y')- \e{f(y')}) dy'  \nonumber \\
  &=& \e{f(x)} + r_1(a) \; (f(x^K)- \e{f(x^K)})  \nonumber 
\ea
in agreement with \eqref{eq_EK1}. 
Similarly the updated covariance becomes
\ba
\covd{K}{f(x)}{f(x')} & =& \cov{f(x)}{f(x')}  -  r_1(a) \int_{y' \in \mk} \delta(x^K - y') \;  \cov{f(y')}{f(x'^K)} r_1(a') \; dy'  \nonumber \\
& =& \cov{f(x)}{f(x')}  -  r_1(a) \cov{f(x^K)}{f(x'^K)} r_1(a')   \nonumber 
\ea
in agreement with equation~(\ref{eq_covK1}), and the derivation of equation~(\ref{eq_covK2}) then follows in exactly the same way 
as shown in section~\ref{ssec_singleKB}.
The continuous versions of the two perpendicular and parallel boundary cases follow similarly, and are given in appendix~\ref{app_contKBE}, and
also agree with the discrete results.

%\subsection{Links to Full Bayesian Analysis}

\section{Design of Known Boundary Emulation Experiments}\label{sec_design_KB}
%\section{Design of KBE Experiments}\label{sec_design_KB}
%\subsection{Space Filling Designs with Known Boundaries}
%A key impact of the presence of known boundaries is on the choice of simulator evaluations, $D$, sited away from the boundaries for emulator construction.  Standard methods typically involve space-filling designs, such as Latin hypercubes or low-discrepancy sequences~\cite{SWMW89_DACE}, all of which typically seek to provide some degree of uniform coverage over the parameter space as a consequence of a prior assumption of a constant level of uncertainty in the behaviour of $f(x)$. However, given a simulator with known boundary behaviour our uncertainty (as quantified by the emulator variance) is no longer constant, and instead exhibits obvious structure as highlighted in Figure~\ref{fig_toymod2}. A simple space-filling design over such problems would needlessly design evaluations close to the boundary where there is little uncertainty and little value in performing expensive simulator evaluations. The design problem for emulators with known boundaries is therefore more complex and simple space-filling design techniques are of limited value. We therefore investigate some simple methods of optimal design, and introduce a mechanism for transforming a uniformly-space-filling design into a more appropriate configuration for bounded problems.
%\ian{could cut here}.

The existence of known boundaries allows us to design a more efficient set of runs over $\mx$ to exploit this additional information. Standard computer model designs involving Latin hypercubes or low-discrepancy sequences~\cite{SWMW89_DACE} are of limited value here, as they seek uniform coverage over $\mx$. As highlighted by Figure~\ref{fig_toymod2}, after the boundary update 
the emulator variance will now exhibit clear (non-uniform) structure, which the design should now reflect. 
%Here we present some simple designs for use in this context that posses various desirable properties.
We therefore investigate some simple methods of optimal design, and introduce a mechanism for transforming a uniformly-space-filling design into a more appropriate configuration for KBE problems.
The general design problem is as follows (see for example~\cite{Johnson:1990aa}): 

\vspace{0.2cm}
\begin{itemize}[leftmargin=0.8cm]
\item Given a simulator and corresponding emulator updated by known boundary $\mk$, select input points $X_D\in \mx$, 
that will give evaluations $D= f(X_D)$, chosen to optimise some criterion $c(X_D)$.
\end{itemize}
\vspace{0.2cm}
%\begin{enumerate}[leftmargin=0.8cm]
%\item Given a simulator and corresponding emulator with known boundary $\mk$, 
%\item Select input points $X_D\in \mx$, that will give evaluations $D= f(X_D)$, chosen to optimise some criterion $c(X_D)$ over the input space.
%%\item Choose these inputs to optimise some criterion $c(X_D)$ over the input space.
%\en
Typically, the criterion is such that we seek to maximise the information content of the chosen design $X_D$, which in computer models typically translates to minimising a function of the emulator variance $\vard{D\cup K}{f(x)}$ over $\mx$~\cite{Santner03_DACE}. Due to the discrete nature of computer experiments, the criterion over $\mx$ is typically approximated by some discrete grid $X$ over $\mx$. The optimisation problem then becomes one of a search over a collection of candidate designs for the `best' candidate under the specified approximate criterion, yielding a locally (not globally) optimal design. This is usually sufficient, as the identification of the global optimum would only be warranted if all the assumptions used in the emulator construction process were thought to be highly accurate, which is rarely the case.
\red{A sensible choice for the design criterion, $c(X_D)$, suitable for our purposes is:}
\vspace{0.2cm}
\begin{itemize}[leftmargin=0.8cm]
\item \red{{\it V-optimality} : $c(X_D) = {\rm trace}(\vard{D \cup K}{f(X)})$, the trace of the adjusted emulator variance matrix over the finite grid of points $X$, given the known boundary and the design $X_D$.}
\end{itemize}
\vspace{0.2cm}

\red{V-optimality just seeks to minimise the sum (and hence also mean) of the point variances across $X$, calculated in the presence of known boundaries, and is 
a discrete approximation of the integrated mean squared prediction error criterion (IMSPE)~\cite{Santner03_DACE,Johnson:1990aa}. To} improve efficiency we exploit the fact that 
%To avoid these potential problems, let us consider an approach based on V-optimality which just seeks to minimise the average point variances across $X$, calculated in the presence of known boundaries. Again, to improve efficiency we exploit the fact that 
\be
 {\rm trace}(\vard{D \cup K}{f(X)}) \;=\; {\rm trace}(\vard{K}{f(X)}) - {\rm trace}(\rvard{D \cup K}{f(X)}), \label{eq_trace_RVar1}
\ee
(see equation~\eqref{eq_BLvDK}) and so we simply need to seek designs that maximise ${\rm trace}(\rvard{D \cup K}{f(X)})$.

\begin{figure}
\begin{center}
\begin{tabular}{ccc}
\hspace{-0.3cm} \includegraphics[scale=0.5]{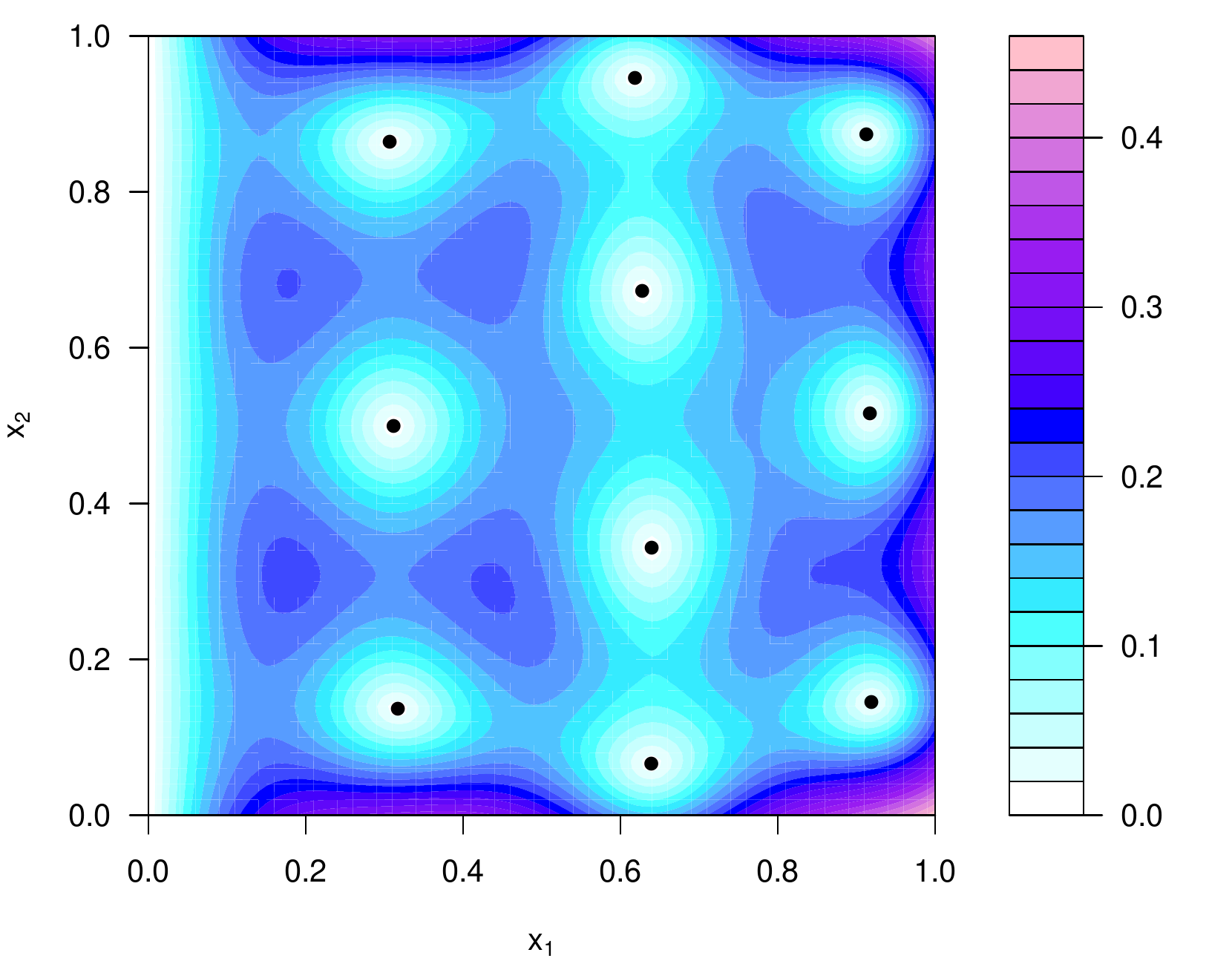} &
\hspace{-0.5cm}  \includegraphics[scale=0.5]{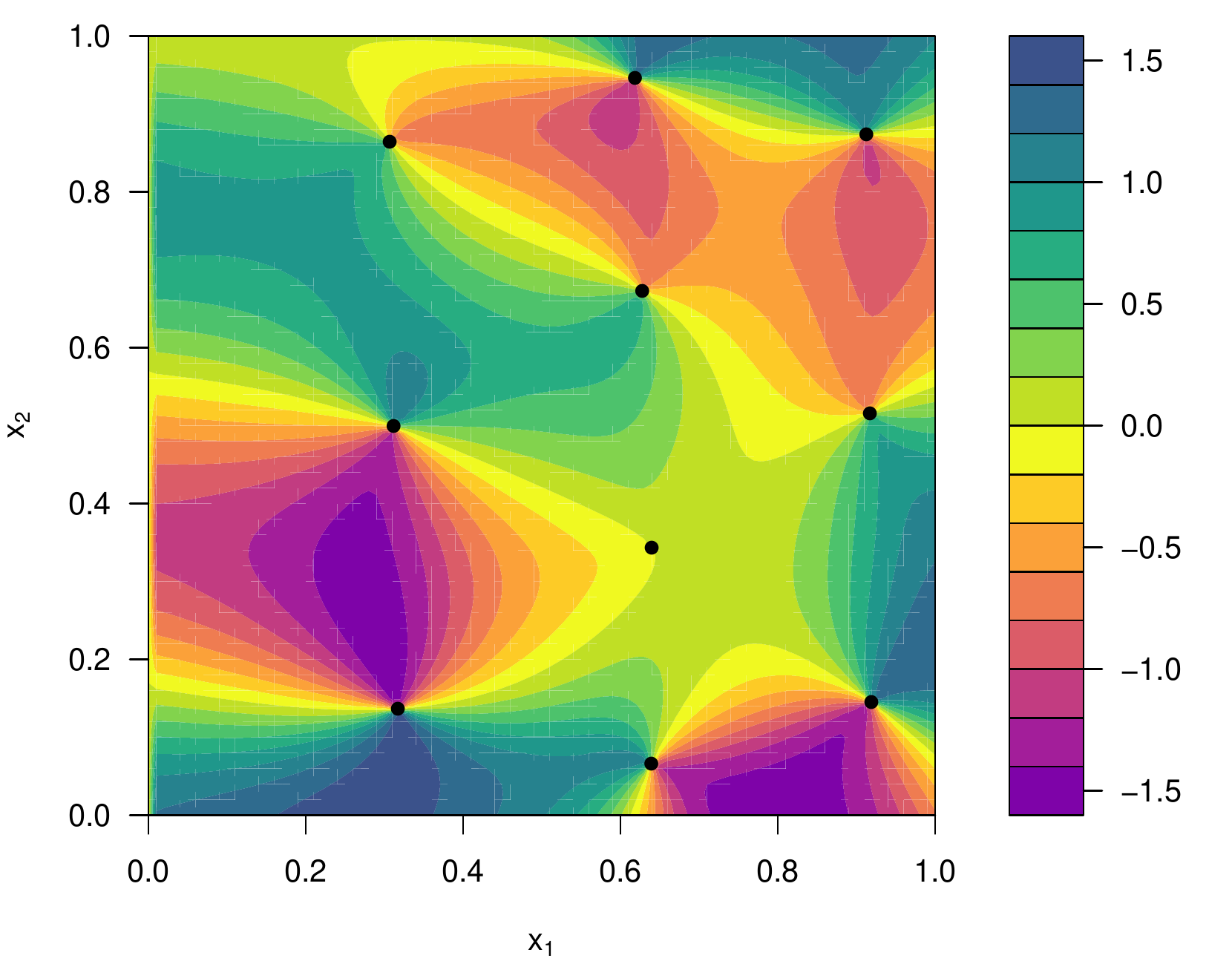} \\
\hspace{-0.3cm} \includegraphics[scale=0.5]{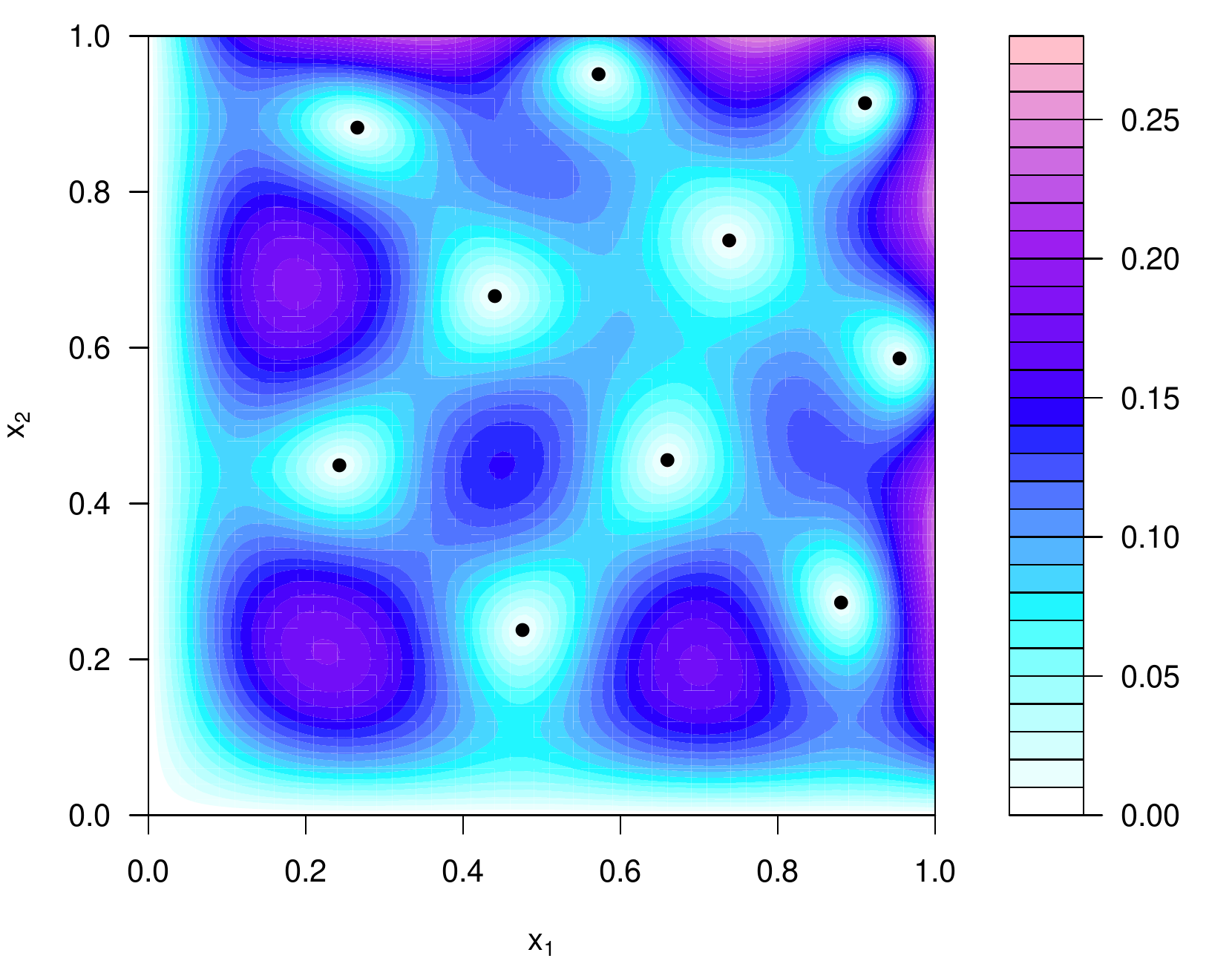} &
\hspace{-0.5cm}  \includegraphics[scale=0.5]{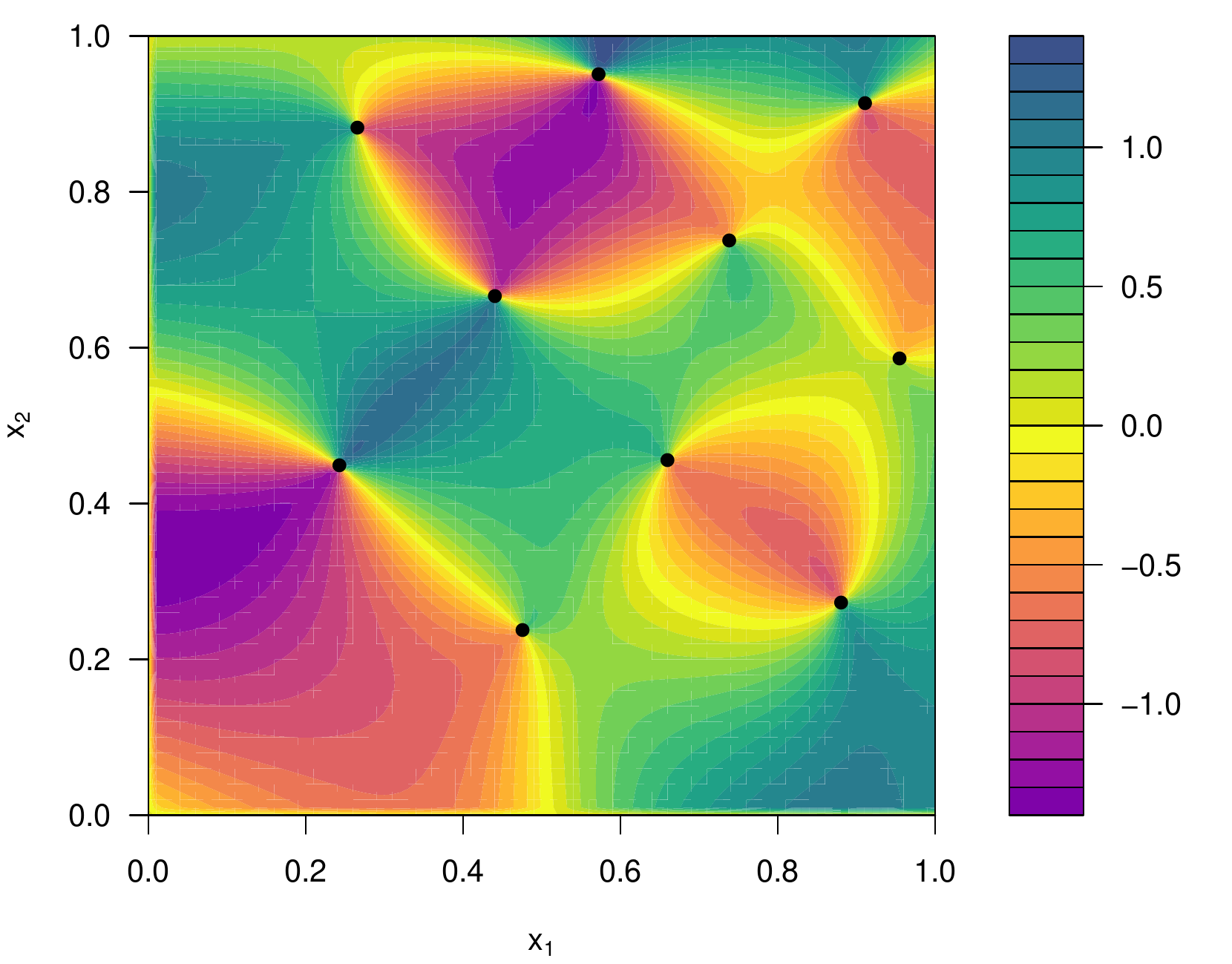} \\
\hspace{-0.3cm} \includegraphics[scale=0.5]{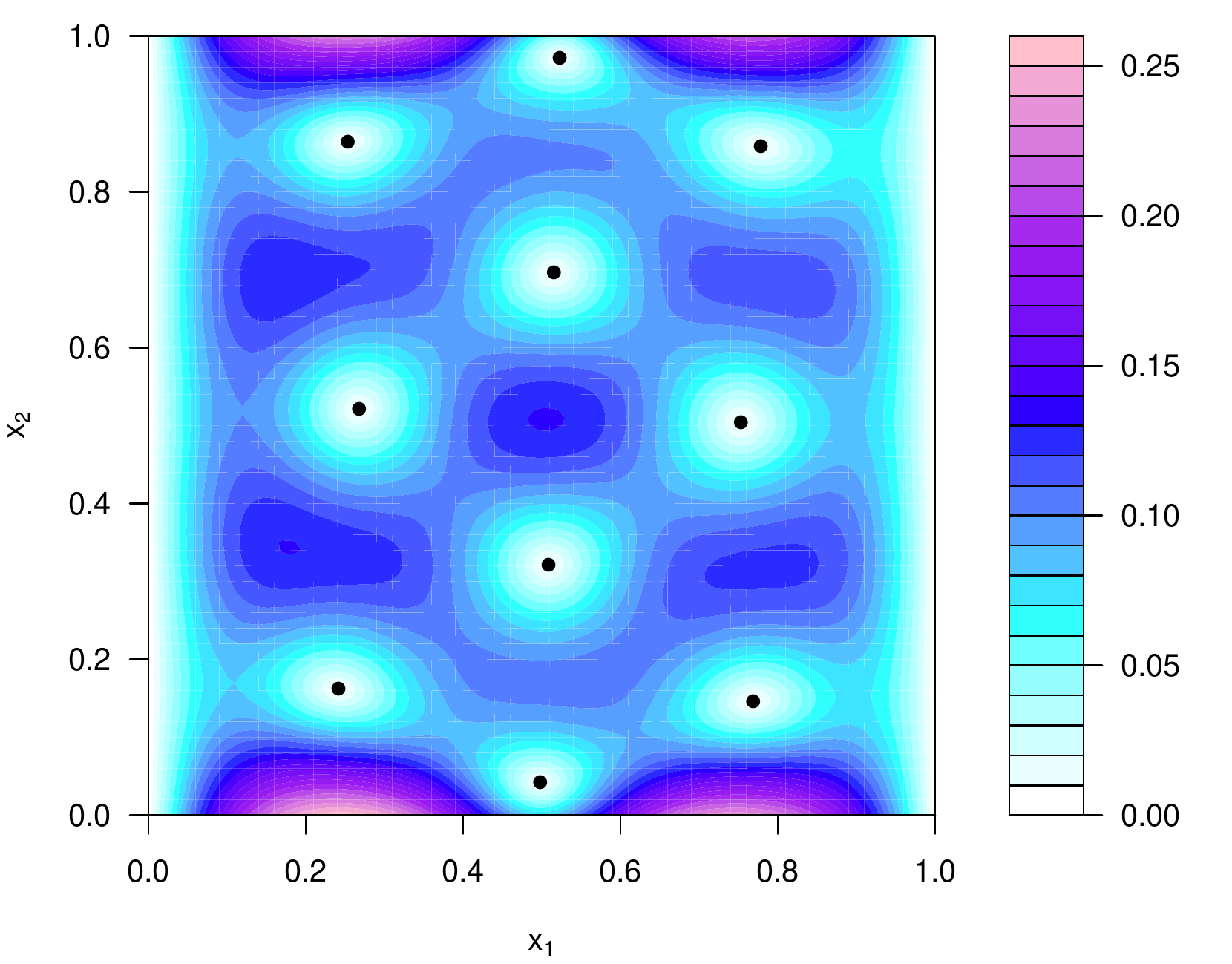} &
\hspace{-0.5cm}  \includegraphics[scale=0.5]{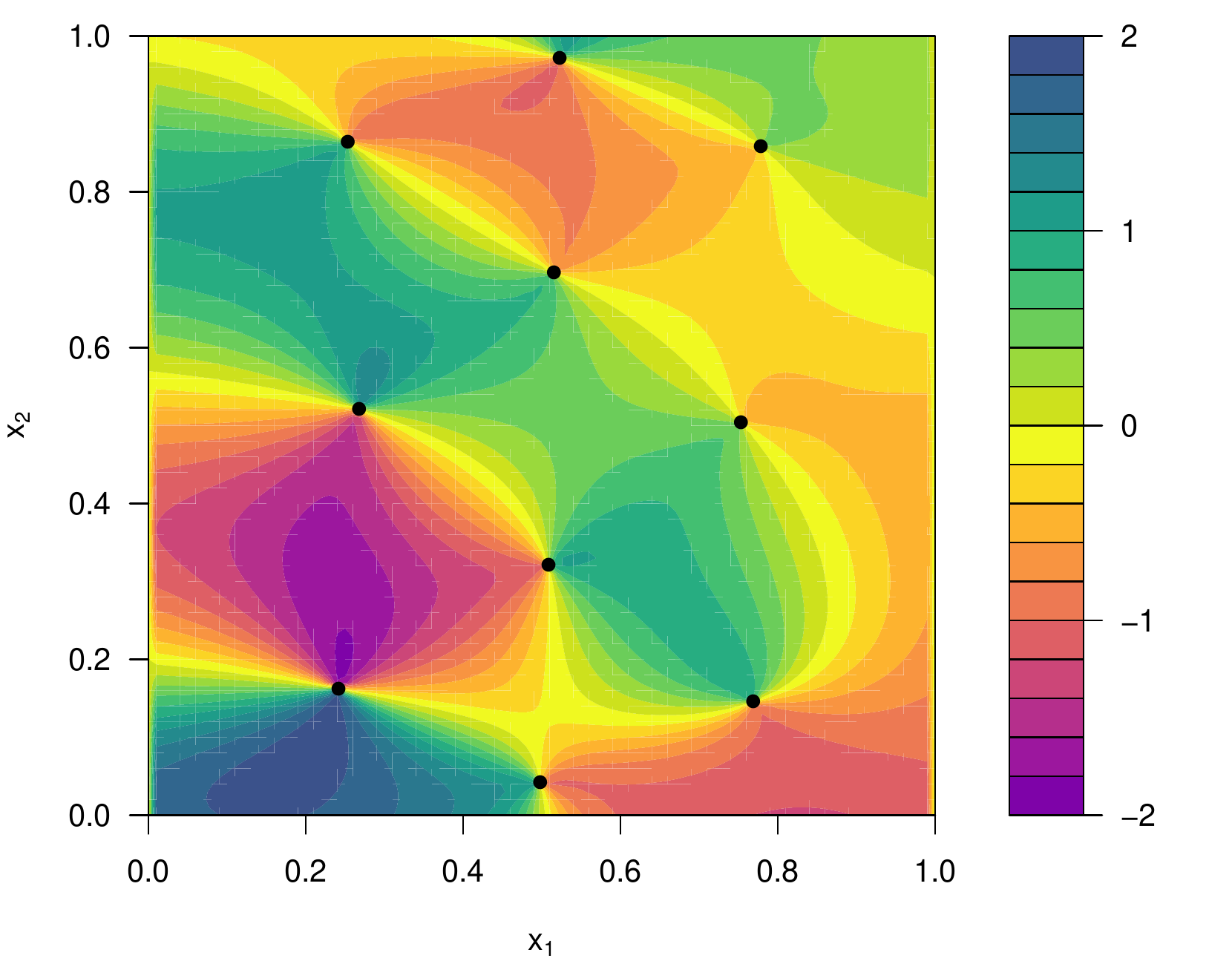} \\
\end{tabular}
\end{center}
\vspace{-0.5cm}
\caption{\footnotesize{Left column: ten point V-optimal designs $X_D$ (black points) and the corresponding emulator standard 
deviation $\sqrt{\vard{D \cup K}{f(X)})}$ defined over $\mx$ (coloured contours) for the single known boundary case (top left), two perpendicular 
known boundaries (middle left) and two parallel 
known boundaries (bottom left) respectively. Emulator diagnostics of the form 
$S_{D\cup K}(x) = (\ed{D\cup K}{f(x)} - f(x))/\sqrt{\vard{D\cup K}{f(x)}}$ are given over $\mx$ in the right column, for each of the three cases.}}
\label{fig_toymod_design1}
\end{figure}

Figure~\ref{fig_toymod_design1} shows ten point V-optimal designs $X_D$ (black points) and the corresponding emulator standard 
deviation $\sqrt{\vard{D \cup K}{f(X)})}$ defined over $\mx$ (coloured contours) for the single known boundary case (top left), two perpendicular 
known boundaries (middle left) and two parallel 
known boundaries (bottom left). 
\red{The V-optimality criteria $c(X_D)$ was assessed using a grid $X$ of size $30 \times 30$, and calculated for any particular candidate design $X_D$ using 
equations~(\ref{eq_trace_RVar1}), (\ref{eq_BLcDK}) and, for the single known boundary case, (\ref{eq_covK2}). The designs in Figure~\ref{fig_toymod_design1} were generated 
using the standard optim() function in R, using 10 point space filling latin hypercube designs as initial conditions.}
Note how the V-optimality criteria 
automatically moves the design points away from the known boundaries toward the less explored regions of $\mx$, while still maintaining 
excellent space filling properties. The toy model was subsequently evaluated at $X$ and the corresponding emulator diagnostics $S_{D\cup K}(x)$ given in the right column of Figure~\ref{fig_toymod_design1}.
%of the form $(\ed{D\cup K}{f(x)} - f(x))/\sqrt{\vard{D\cup K}{f(x)}}$ are given over $\mx$ in the right column, for each of the three cases. 

%
%\begin{figure}
%\begin{center}
%\begin{tabular}{ccc}
%\hspace{-0.8cm} \includegraphics[scale=0.4]{plots/SingleKB_D_m10aopt_emul_1.pdf} &
%\hspace{-0.8cm} \includegraphics[scale=0.4]{plots/SingleKB_D_m10aopt_emul_2.pdf} &
%\hspace{-0.5cm}  \includegraphics[scale=0.4]{plots/SingleKB_D_m10aopt_emul_3.pdf} \\
%\hspace{-0.8cm} \includegraphics[scale=0.4]{plots/TwoPerpKB_D_m10aopt_emul_1.pdf} &
%\hspace{-0.8cm} \includegraphics[scale=0.4]{plots/TwoPerpKB_D_m10aopt_emul_2.pdf} &
%\hspace{-0.5cm}  \includegraphics[scale=0.4]{plots/TwoPerpKB_D_m10aopt_emul_3.pdf} \\
%\hspace{-0.8cm} \includegraphics[scale=0.4]{plots/TwoParaKB_D_m10aopt_emul_1.pdf} &
%\hspace{-0.8cm} \includegraphics[scale=0.4]{plots/TwoParaKB_D_m10aopt_emul_2.pdf} &
%\hspace{-0.5cm}  \includegraphics[scale=0.4]{plots/TwoParaKB_D_m10aopt_emul_3.pdf} \\
%\end{tabular}
%\end{center}
%\caption{\footnotesize{stuff}}
%\label{fig_toymod_design1}
%\end{figure}
%

Despite the desirable properties of such V-optimal designs, the projections onto lower dimensional subspaces of $\mx$ can be less than satisfactory. For example, in the single and also two parallel known boundary cases as illustrated in the top-left and bottom-left panels of Figure~ \ref{fig_toymod_design1}, the projection of $X_D$ onto $x_1$ only covers three distinct values of $x_1$. This could be very inefficient if it was found that $x_1$ was highly influential (and hence deemed an active input) while $x_2$ was found to be inactive. 
%This is important as the use of active and inactive variables is an extremely useful approach in taming high dimensional simulators~\cite{Vernon10_CS}.
This is important as active variables are extremely useful in taming high dimensional simulators~\cite{Vernon10_CS}.

%Such projection concerns may promote the use of a more general purpose design. As mentioned above, a common choice in the computer model literature is the maximin Latin hypercube. To account for the fact that the emulator variance surface is heavily structured, we consider adapting the method to be more appropriate for a known boundary setting via ``warped latin hypercube designs''. These do not optimise any particular criteria, but have good space filling and projection properties. 

Such projection concerns may promote the use of a more general purpose design. As mentioned above, in the computer model literature 
the maximin Latin hypercube is the standard choice~\cite{SWMW89_DACE}. 
Therefore, to account for the non-uniformity of the boundary updated emulator variance $\vard{D \cup K}{f(X)}$, here we explore the use of simple ``warped latin hypercube designs'', that share the useful properties of 
standard Latin hypercubes, but which are adapted to be more appropriate for a known boundary setting. These do not optimise 
any particular criteria, but have good space filling and projection properties.

The warped designs are created by taking a maximin latin hypercube design and warping it so that the marginal density of the design matches the marginal form of the new emulator variance adjusted by the known boundaries, which in the single or two perpendicular boundary cases is proportional to $(1-r_i^2(a))$, as shown by equations \eqref{eq_VK1} and \eqref{eq_varKL1}. For example, in the two perpendicular known boundary case, each point $x^{(i)}$ in a maximin Latin hypercube design is warped via the transformation
\ba
x_1^{(i)}  &\rightarrow&   g_1(x_1^{(i)})/g_1(1)  \quad \quad {\rm where} \quad g^{-1}_1(a) = \int_0^a (1-r_1^2(a')) da' \label{eq_warplh1}\\
x_2^{(i)}  &\rightarrow&   g_2(x_2^{(i)})/g_2(1)  \quad \quad {\rm where} \quad g^{-1}_2(b) = \int_0^b (1-r_2^2(b')) db' \label{eq_warplh2}\\
x_j^{(i)}   &\rightarrow&   x_j^{(i)}  \quad \quad j \ne 1,2 \label{eq_warplh3}
\ea
which ensures the marginal distributions $\pi(x_j^{(i)}) \propto (1-r_j^2(x_j^{(i)}))$, for $j=1,2$, as required.

\begin{figure}
\begin{center}
\begin{tabular}{ccc}
\hspace{-0.3cm} \includegraphics[scale=0.45]{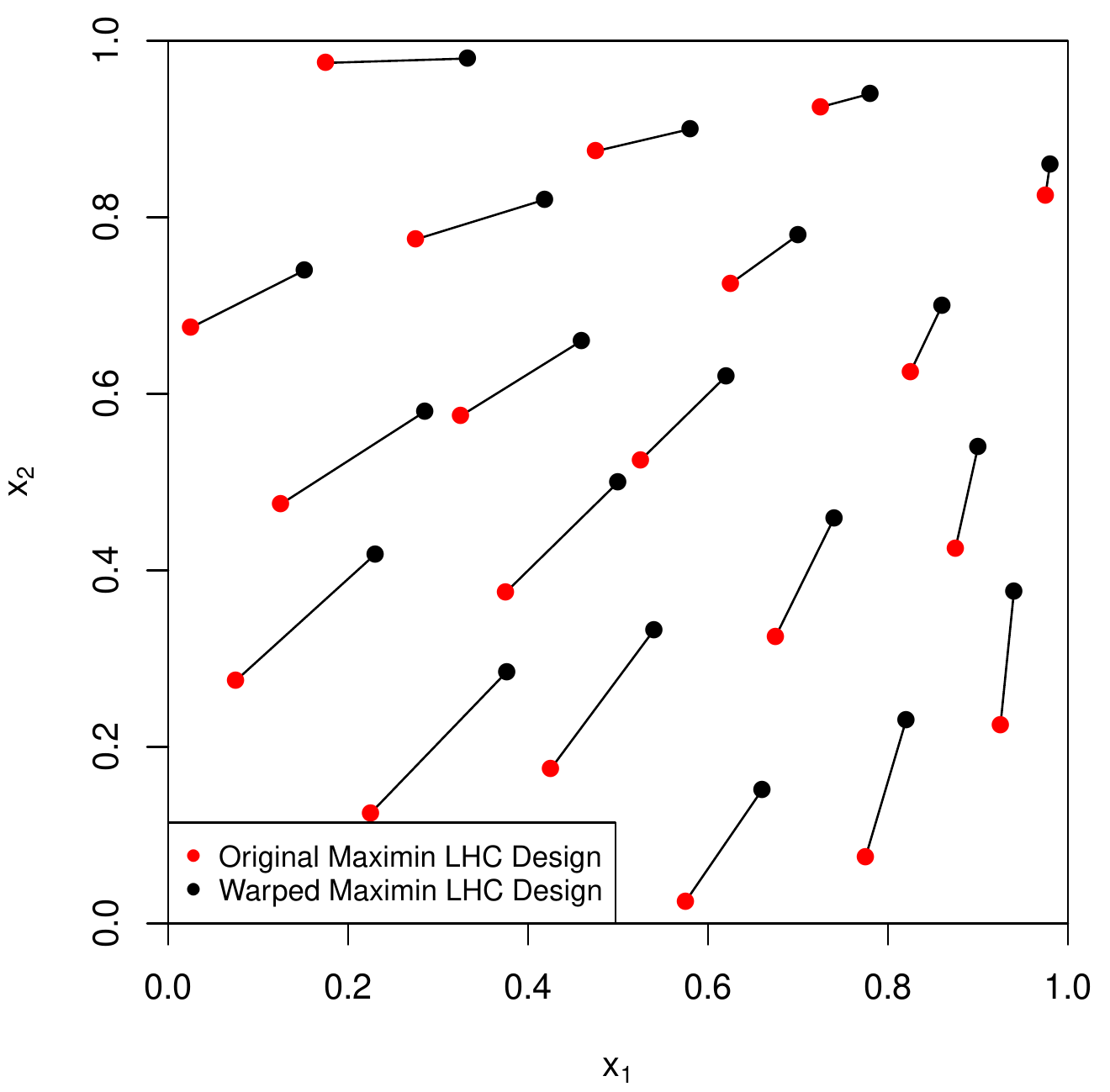} &
\hspace{-0.5cm}  \includegraphics[scale=0.45]{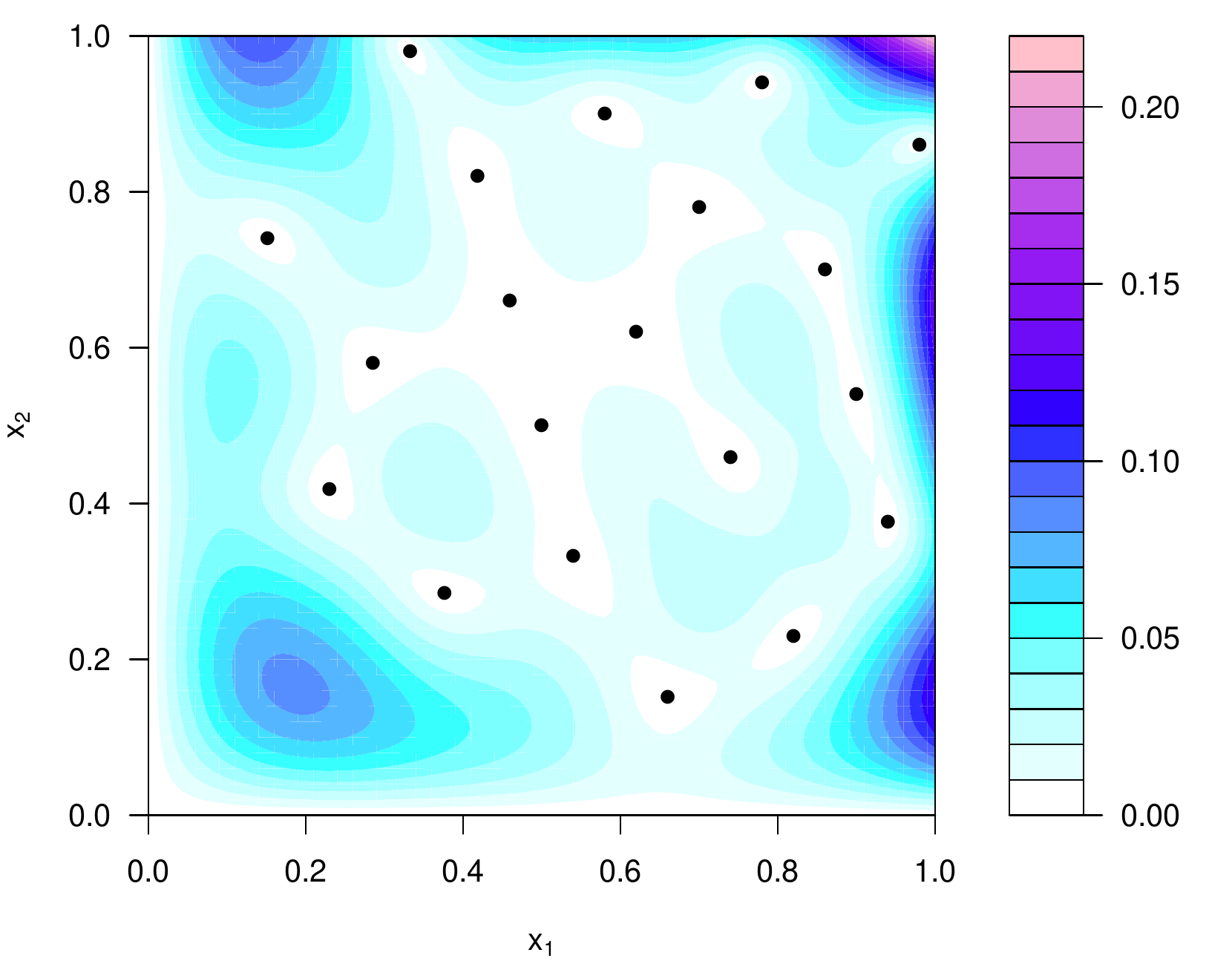} \\
\end{tabular}
\end{center}
\vspace{-0.5cm}
\caption{\footnotesize{Left panel: a 20 point maximin Latin hypercube (red points), and the corresponding warped 
Latin hypercube (black points). The black lines link the pre- and post-warped points to highlight the effect of the warping. Right panel: the emulator standard deviation $\sqrt{\vard{D \cup L \cup K}{f(X)}}$ updated by both boundaries $L$ and $K$ and the warped design $D$. }}
\label{fig_toymod_warped_design}
\vspace{-0.6cm}
\end{figure}

Figure~\ref{fig_toymod_warped_design} (left panel) shows a 20 point maximin Latin hypercube as the red points, and the warped 
Latin hypercube as the black points. The black lines link the pre- and post-warped points to highlight the effect of the warping. The right panel shows the emulator standard deviation $\sqrt{\vard{D \cup L \cup K}{f(X)}}$ updated by both boundaries $L$ and $K$ and the warped design $D$. Such designs are space filling, while they also maintain good projection properties. We illustrate these designs further in the next section to explore improvements to the known boundary emulation of a model of Arabidopsis Thanliana.

%%%%% Sam's corrections %%%%%%

\section{Application to a Systems Biology Model of Arabidopsis}\label{sec_arabid}

In the previous sections of this article we have presented methodology for utilising knowledge of the behaviour of a complex scientific model along particular boundaries of the input parameter space to aid emulation of the model across the whole input space, exploiting both emulation and design procedures. 
%Many models exhibit such known boundaries, particularly if the model is represented as a set of differential equations, in which case particular settings of certain parameters may reduce aspects of the model into components that can be analytically solved.  
In this section we apply this methodology to 
%the benefit of utilising known boundaries for emulation of a 
a model of hormonal crosstalk in the root of an Arabidopsis plant.

%Sections \ref{ATE} and \ref{MHCAT} give a brief introduction to Arabidopsis Thaliana and the associated hormonal crosstalk model of interest.  Section \ref{KBEAM} compares emulation of this model with and without the use of known boundaries within the emulation procedure.  Section \ref{SSKBED} compares emulators constructed using different designs obtained by various design construction methods, and demonstrates how utilising knowledge of boundary model behaviour should affect one's choice of design.

%\subsection{Arabidopsis Thaliana \label{ATE} }

\subsection{Model of Hormonal Crosstalk in Arabidopsis Thaliana}\label{MHCAT}

Arabidopsis Thaliana is a small flowering plant that is widely used as a model organism in plant biology. 
%It is a member of the mustard (Brassicaceae) family, which includes cultivated species such as cabbage and radish. 
Arabidopsis offers important advantages for basic research in genetics and molecular biology for many reasons, including the facts that it has a short life cycle, changes in it are easy to observe and it is genetically relatively simple. Arabidopsis was therefore the first plant to have its genome fully sequenced~\cite{Initiative:2000aa}.
Understanding the genetic structure of Arabidopsis may lead to increased understanding of crop plants such as wheat, and 
hence facilitate the future development of crop plants that are robust to adverse climate conditions.
%Arabidopsis is not an agriculturally important plant, however, there is a strong relationship between the genetics of Arabidopsis and the genetics of more complicated and agriculturally more useful plants such as wheat and other crops.  It is important for scientists to understand how to mutate crop plants to allow them to withstand increasingly adverse climate conditions. 

We demonstrate our known boundary emulation techniques on a model of hormonal crosstalk in the root of an Arabidopsis plant that was constructed by Liu et al. \cite{10.3389/fpls.2013.00075}.  This Arabidopsis model represents the hormonal crosstalk of auxin, ethylene and cytokinin in Arabidopsis root development as a set of 18 differential equations, given in Table \ref{DE}, which must be solved numerically.  The full model takes an input vector of 45 rate parameters $ (k_1, k_{1a}, k_2, ...) $, produces an output vector of 18 chemical concentrations $ ([Auxin], [X], [PLSp], ... )$, \red{and has a typical runtime of approximately 0.1 to 1 seconds, depending on parameter settings and target model time \cite{Vernon_sysbio_hm_2016}}.  This Arabidopsis model has been successfully emulated in the literature in the context of history matching \cite{Jackson1,Vernon_sysbio_hm_2016}.  For the purposes of this article, we are interested in modelling the important output $ [PLSp] $, which represents the concentration of POLARIS peptide \cite{10.3389/fpls.2013.00075} at early time, $ t=2 $.
%for our chosen solver and step size.  
We choose to explore 6 input rate parameters $ \{ k_4, k_6, k_{6a}, k_7, k_8, k_9 \} $ of primary interest, although it is important to note that the benefits of using known boundaries would scale to larger numbers of inputs.  The ranges over which we allowed these 6 inputs to vary is given in Table \ref{IRR}, in appendix~\ref{app_arab_details}.  These ranges were square rooted and mapped to a $ [-1,1] $ scale. The remaining input rate parameters were fixed at values deemed reasonable by biological experts \cite{10.3389/fpls.2013.00075}.

\begin{table}[t]
\scriptsize
\singlespacing
%\begin{mdframed}
\begin{align*}
\frac{d[Auxin]}{dt}  \;\;=\;\; &  \frac{k_{1a}}{\displaystyle 1 + \frac{[X]}{k_1}} + k_2 + k_{2a}   
\frac{[ET]}{\displaystyle 1 + \frac{[CK]}{k_{2b}}} \frac{[PLSp]}{k_{2c} + [PLSp]}   
& \frac{d[Re]}{dt}  \;\;=\;\; & k_{11}[Re^\ast][ET] - (k_{10} + k_{10a}[PLSp])[Re]  
\\ 
&  + \frac{V_{IAA}[IAA]}{Km_{IAA} + [IAA]}   
& \frac{d[Re^\ast]}{dt} \;\;=\;\; & -k_{11}[Re^\ast][ET] + (k_{10} + k_{10a}[PLSp])[Re]   \\
&  - \left( k_3 + \frac{k_{3a}[PIN1pm]}{k3auxin + [Auxin]} \right) [Auxin]    
&\frac{d[CTR1]}{dt}  \;\;=\;\; & -k_{14}[Re^\ast][CTR1] + k_{15}[CTR1^\ast]   
\\
\frac{d[X]}{dt}  \;\;=\;\; & k_{16} - k_{16a}[CTR1^\ast] - k_{17}[X]  
&\frac{d[CTR1^\ast]}{dt}  \;\;=\;\; & k_{14}[Re^\ast][CTR1] - k_{15}[CTR1^\ast]   \\
\frac{d[PLSp]}{dt}  \;\;=\;\; & k_8[PLSm] - k_9[PLSp]   
&\frac{d[PIN1m]}{dt}  \;\;=\;\; & \frac{k_{20a}}{k_{20b} + [CK]} [X] \frac{[Auxin]}{k_{20c} + [Auxin]}     \\ 
\frac{d[Ra]}{dt}  \;\;=\;\; & -k_4 [Auxin] [Ra] + k_5 [Ra^\ast]   
& &- k_{1_v21}[PIN1m]  \\
\frac{d[Ra^\ast]}{dt}  \;\;=\;\; & k_4 [Auxin] [Ra] - k_5 [Ra^\ast]  
&\frac{d[PIN1pi]}{dt}  \;\;=\;\; & k_{22a}[PIN1m] - k_{1_v23}[PIN1pi]   \\ 
\frac{d[CK]}{dt}  \;\;=\;\; & \frac{k_{18a}}{\displaystyle 1 + \frac{[Auxin]}{k_{18}}} - k_{19} [CK]   
& & - k_{1_v24}[PIN1pi] + \frac{k_{25a}[PIN1pm]}{\displaystyle 1 + \frac{[Auxin]}{k_{25b}}}   \\
& + \frac{V_{CK}[cytokinin]}{Km_{CK} + [cytokinin]}   
&\frac{d[PIN1pm]}{dt}  \;\;=\;\; & k_{1_v24}[PIN1pi] - \frac{k_{25a}[PIN1pm]}{\displaystyle 1 + \frac{[Auxin]}{k_{25b}}}   \\
\frac{d[ET]}{dt}  \;\;=\;\; & k_{12} + k_{12a}[Auxin][CK] - k_{13}[ET]   
&\frac{d[IAA]}{dt}  \;\;=\;\; & 0   \\
& + \frac{V_{ACC}[ACC]}{Km_{ACC} + [ACC]}   
&\frac{d[cytokinin]}{dt}  \;\;=\;\; & 0   \\
\frac{d[PLSm]}{dt}  \;\;=\;\; & \frac{k_6[Ra^\ast]}{\displaystyle 1 + \frac{[ET]}{k_{6a}}} - k_7[PLSm]   
&\frac{d[ACC]}{dt}  \;\;=\;\; & 0   
\end{align*}
%\end{mdframed}
\vspace{-0.4cm}
\caption[Arabdiopsis Equations]{\footnotesize{Arabidopsis model differential equations.} \label{DE}}
\vspace{-0.6cm}
\end{table}

\subsection{Known Boundary Emulation of the Arabidopsis Model}\label{KBEAM}

%In this section we apply our known boundary emulation techniques to the Arabidopsis model.  
%Section \ref{EKB} establishes the known boundaries relevant for the chosen output $ [PLSp] $, 
%Section \ref{ESPS} outlines our emulation strategy, and Section \ref{Res} compares the results of emulation with and without incorporating the knowledge of the model output's behaviour along the known boundaries.

\subsubsection{Establishing Known Boundaries}\label{EKB}

Establishing known boundaries requires some understanding of the scientific model. 
%For the purposes of a scientific model represented as a set of differential equations, boundaries of the input parameter space for which the model output is analytically tractable can be considered known.  This is because solving the system for this output for any input combination along these boundaries is trivial.  
It is not uncommon for one or more known boundaries to occur in a model system for some outputs. Often, setting certain parameters to
specific values will decouple smaller subsections of the system, which may allow subsets of the equations to be solved analytically, for particular outputs. This is the case for the Arabidopsis model.

We establish the known boundaries for output $ [PLSp] $ by considering its rate equation:
\be
\frac{d[PLSp]}{dt} \;=\; k_8[PLSm] - k_9[PLSp]
\ee
A known boundary exists when rate parameter $ k_8 = 0 $, since in this case:
\ba
\frac{d[PLSp]}{dt} \;&=&\; - k_9[PLSp]   \\
\Rightarrow \;\;\;\; [PLSp] \;&=&\; [PLSp^0] e^{-k_9t}
\ea
where $ [PLSp^0] $ is the initial condition of the $ [PLSp] $ output, and we see that $[PLSp]$ has been entirely decoupled from the rest of the system. The output $ [PLSp] $ can now be obtained along the boundary $ k_8 = 0 $ with negligible computational cost.  

The second (perpendicular) known boundary for output $ [PLSp] $ occurs when $ k_6 = 0 $. This decouples the combined system of 
$[PLSm]$ and $[PLSp]$. We can solve for $[PLSm]$ first using:
\ba
\frac{d[PLSm]}{dt} \;&=&\; - k_7[PLSm]   \\
\Rightarrow \;\;\;\; [PLSm] \;&=&\; [PLSm^0] e^{-k_7t}
\ea
Inserting this solution for $ [PLSm] $ into the rate equation for $ [PLSp] $ then yields:
\be
[PLSp] \;=\; [PLSp^0] e^{-k_9t} + \frac{k_8 [PLSm^0]}{k_9 - k_7} (e^{-k_7t} - e^{-k_9t})
\ee
which again requires negligible computational cost to evaluate for any given input combination.
We now use these known boundaries to aid emulation of $ [PLSp] $ in the Arabidopsis model.

\subsubsection{Emulator Structure and Parameter Specification}\label{ESPS}

%In this section we outline the general emulation strategy and parameter choices used throughout this article.  This general emulator structure will then be used to investigate the use of known boundaries in the construction of an emulator, and how such use may affect design choices.

The emulation strategy used is as follows.
As discussed in section \ref{sec_emulCCM}, we restrict the form of our emulator to a pure Gaussian process, as given by equation~(\ref{eq_GPem}).  We used a product Gaussian correlation function of the form given in equation~(\ref{eq_prodgausscor}), as we assumed the solution to the Arabidopsis model 
would most likely be smooth and that many orders of derivatives would exist.
The prior emulator expectation and variance were taken to be constant, that is $ \e{f(x)} = \beta $ and $\var{f(x)} = \sigma^2$, where $ \beta $ and $\sigma^2$ were estimated to be the sample mean and variance of a set of previously evaluated scoping runs. The correlation length parameter $ \theta $ was set to $ \theta = 0.7 $ for each input, a choice consistent with the argument for approximately assessing correlation lengths presented in~\cite{Vernon10_CS}. This value for $\theta$ was also checked for adequacy using standard emulator diagnostics~\cite{Tony_EmDiag}.
We have made this relatively simple emulator specification for illustrative purposes, so that we can focus on the effect of the inclusion of known boundaries. 

%Prior emulator expectation was taken to be constant, that is $ \e{f(x)} = \beta $.  $ \beta $ was estimated to be the sample mean of a set of 100 scoping runs previously evaluated, the design for which was taken to be a maximin Latin hypercube across the 6 dimensional input space.  The variance scaling parameter $ \sigma^2 $ was taken to be the sample variance of the same set of 100 model evaluations.  The correlation length parameter $ \theta $ was chosen to be the same for each input with a value of $ \theta = 0.7 $, this value being assessed for adequacy using diagnostic tests. 
%\sam{Do we need to say that this simple specification is taken for illustrative purposes etc. or justify it any other way?} \ian{Yes.}

%In the next section we compare emulators of this general form constructed with and without use of the known boundaries $ k_6 = 0 $ and $ k_8 = 0 $.  All emulators are compared using a fixed set of 2000 diagnostic points, evaluated across a maximin Latin hypercube.

 % Linear models without a correlated residual process added can form suitable emulators themselves depending on what the emulator is to be used for, and they are inherently much faster than an emulator with a correlated second-order stationary residual process \cite{}.  We include the linear model fitted without correlated residual process in our comparisons along with the emulators involving a correlated residual process based on further training points, and additionally the known boundaries.

\subsubsection{Results}\label{Res}

We now compare emulators of the above form constructed both with and without use of the known boundaries $\mk: k_6 = 0 $ and $\ml: k_8 = 0 $, and with and without the addition of training points. In this section we fix the design for the training points as a maximin Latin hypercube design of size 60 across the 6 dimensional input space, and explore the effects of more tailored designs in section~\ref{SSKBED}.  Bayes linear updates by one and two known boundaries were carried out using the single and two perpendicular boundary updates given by equations~(\ref{eq_EK1}), (\ref{eq_VK1}), (\ref{eq_covK2}) and (\ref{eq_expKL1}), (\ref{eq_covKL1}), (\ref{eq_varKL1}) respectively. 
Additional updating using the set of training points $D$ was then performed using the sequential update formula given by equations~(\ref{eq_BLmDK})-(\ref{eq_BLcDK}).

We use visual representations of the emulators and various diagnostics in order to compare emulators built under the six scenarios of interest.  These will be referred to using numerical labelling as follows, with the data used to update the emulators given in parenthesis:
\smallskip
\begin{enumerate}
\item Prior emulator beliefs only, no training points and no known boundaries: ($\emptyset$)
\item Single known boundary $ k_6 = 0 $, no training points: ($K$)
\item Two perpendicular known boundaries $ k_6 = 0 $ and $ k_8 = 0 $, no training points: ($L \cup K$)
\item Training points only: ($D$)
\item Single known boundary and training points: ($D\cup K$)
\item Two perpendicular known boundaries $ k_6 = 0 $ and $ k_8 = 0 $, and training points: ($D\cup L \cup K$)
\end{enumerate}
\smallskip
\begin{figure}[t]
\center
\includegraphics[width=10.5cm]{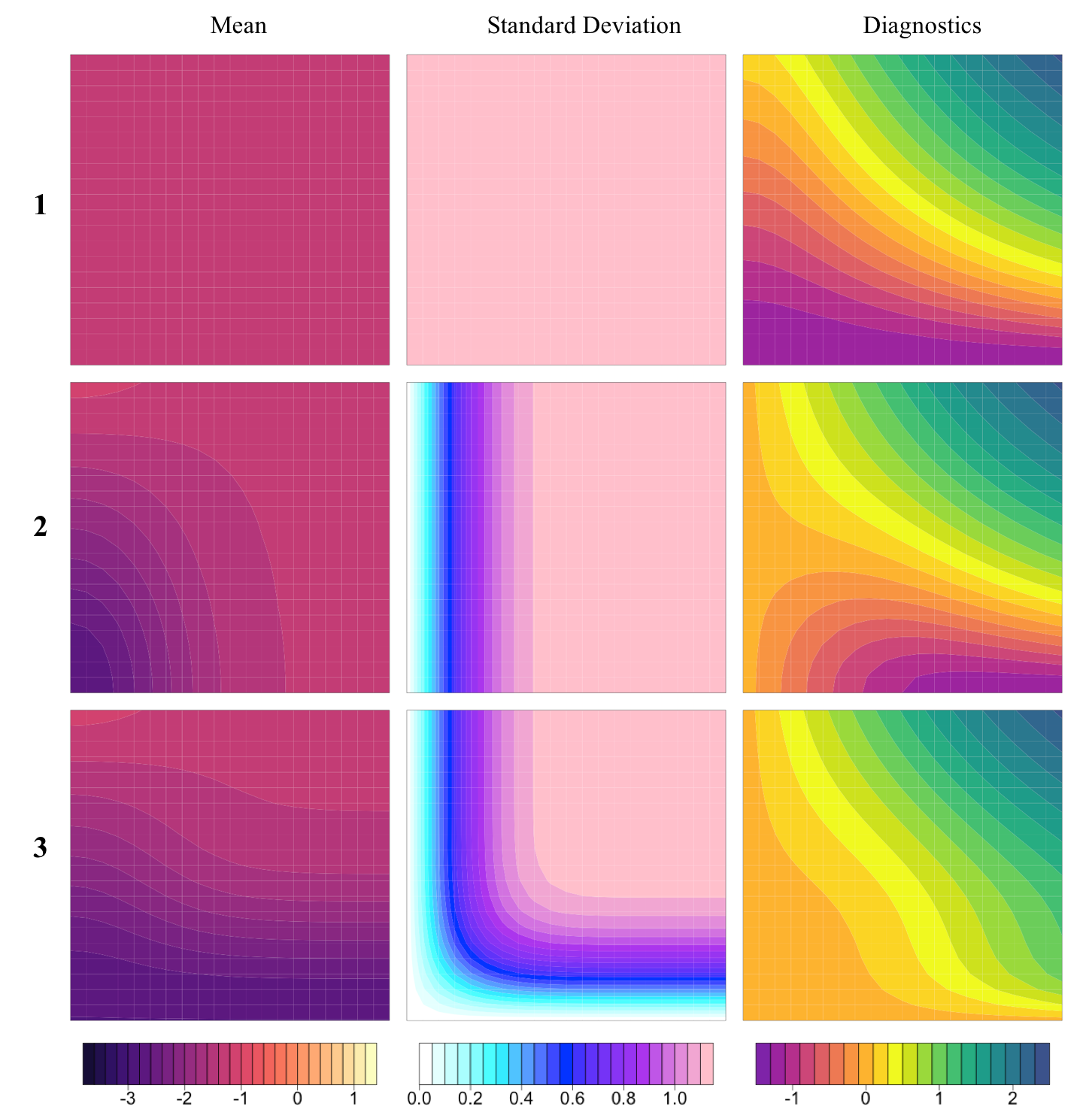}
\caption[2D Slice Emulation Results]{\footnotesize{Results of emulating, without training points, a 2-dimensional $ k_6 $ (x-axes) by $ k_8 $ (y-axes) slice of the 6-dimensional input space, with each of the inputs $ \{ k_4, k_{6a}, k_7, k_9 \} $ set to the mid-values of their square root ranges.  The first row shows the results when using prior emulator beliefs only, the second row shows the results when updating by the boundary $\mk: k_6 = 0 $ only, and the third row shows the results when updating using both boundaries $\mk: k_6 = 0 $ and $\ml: k_8 = 0 $.  Each column from left to right shows emulator mean, 
standard deviation and diagnostics respectively.}  \label{CP1}}
\vspace{-0.6cm}
\efi
\begin{figure}[t]
\center
\includegraphics[width=10.5cm]{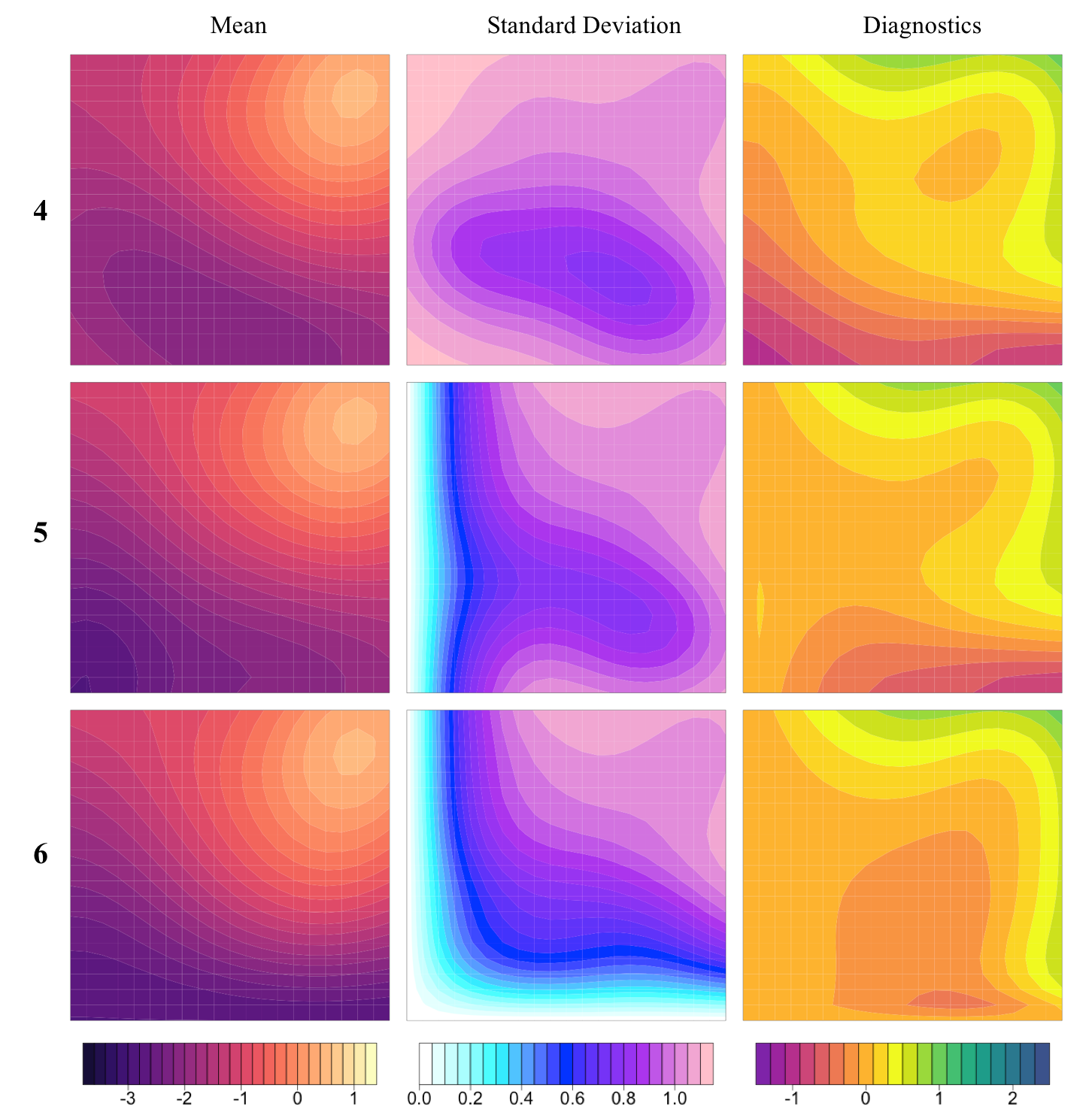}
\caption[2D Slice Emulation Results]{\footnotesize{Results of emulating, with training points, a 2-dimensional $ k_6 $ (x-axes) by $ k_8 $ (y-axes) slice of the 6-dimensional input space, with each of the inputs $ \{ k_4, k_{6a}, k_7, k_9 \} $ set to the mid-values of their square root ranges.  The first row shows the results when updating by the training points only, the second row shows the results when updating by the training points and the known boundary $\mk:  k_6 = 0 $, and the third row shows the results when updating by the training points and the two known boundaries 
$\mk:  k_6 = 0 $ and $\ml:  k_8 = 0 $.  Each column from left to right shows emulator means, variances and diagnostics respectively.}  \label{CP2}}
\vspace{-0.9cm}
\efi
\begin{figure}[t]
\center
\includegraphics[width=6.0cm]{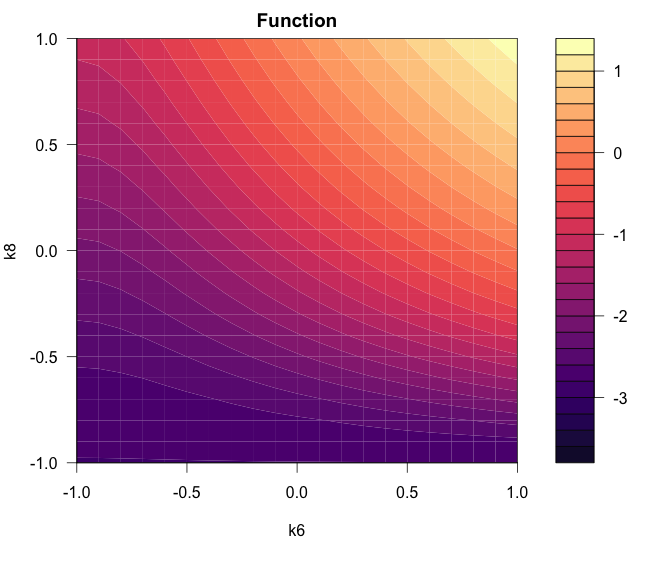}
\vspace{-0.4cm}
\caption[2D Slice Emulation Results]{\footnotesize{A $ k_6 \times k_8 $ cross section of the simulator output with each of the inputs $ \{ k_4, k_{6a}, k_7, k_9 \} $ set to the mid-values of their square root ranges. This should be compared with the left column of Figures~\ref{CP1} and \ref{CP2}.} \label{func}}
\vspace{-0.6cm}
\efi
Equivalent plots to those shown in figures~\ref{fig_toymod1} and \ref{fig_toymod2} are substantially more difficult to visualise across all dimensions of a high-dimensional space. Instead, to show intuitively the effect of the various known boundaries, we first examine a slice of the full 6-dimensional space.
Figures \ref{CP1} and \ref{CP2} show the results of emulating, with and without training points using no, one and two boundaries respectively, a 2-dimensional $ k_6 $ (x-axes) by $ k_8 $ (y-axes) slice of the 6-dimensional input space, with each of the inputs $ \{ k_4, k_{6a}, k_7, k_9 \} $ set to the mid-values of their square root ranges. The rows are labelled in terms of the above six scenarios, and the columns give the emulator mean, standard deviation and diagnostics, defined as in section~\ref{sssec_2dmodel}. These figures can be compared to the true function, shown in Figure \ref{func}.

Figure \ref{CP1} shows that updating using the boundary $ k_6 = 0 $ results in an updated mean near to the boundary which closely reflects the true function, whilst further away from the boundary it tends back towards the prior mean.  The standard deviation tends to zero at the boundary and increases further away from it, tending back towards the prior standard deviation.  The diagnostic plots show that the emulator gives acceptable predictions across the input space, tending to zero at the boundary.  Introducing the second boundary results in accurate predictions close to both boundaries and acceptable diagnostic plots.  Behaviour of the mean and variance tends to the prior specification in the sections of the input space far from both boundaries.

Figure \ref{CP2} shows that emulator variance modestly decreases when the 60 training points are incorporated, a result which is sensitive to how close
any of the training runs are to this particular slice. The emulator mean does show noticeable improvement, but note that the inclusion of the two boundaries  $\mk$ and $\ml$ still has a far more significant effect on the emulator than that of the 60 runs. 
%This is because these boundaries are both 5-dimensional objects and hence carry a lot more information than 60 single runs.
%into the construction of emulators in comparison to when they aren't, with small sections of greatly improved accuracy occurring near training point locations.  
The diagnostic plots are comparable to those in Figure \ref{CP1}, the most notable difference being the diagnostic values at the top right corner of the input space, which have now been reduced.  Diagnostic plots such as these have been compared at several combinations of the other input values with similarly adequate results, and we examine more comprehensive diagnostics below.  Since the correlation structure is more heavily influential for updating our beliefs of simulator behaviour when known boundaries are utilised, it is even more important to ensure that parameters of the correlation function have been adequately specified, in particular the correlation lengths.  Large amounts of poor diagnostics for points near the boundary may indicate that the correlation length has been overestimated: an easy mistake if the function rapidly changes its behaviour as it moves away from the boundary. 

Figures \ref{CP1} and \ref{CP2} demonstrate a major advantage of being able to update simulator beliefs using known boundaries over just using individual points.  Individual points are usually large distances away from each other in high dimensions.  However, as can be seen from these variance plots (and would similarly be shown by any other slice with different values of the four fixed inputs), the known boundaries here are $ d-1 $ dimensional objects and hence carry far more information than individual runs (which are 0-dimensional objects), which results in significant variance resolution across substantial amounts of the input space for very little computational cost.

\begin{figure}[t]
\center
\includegraphics[width=14cm]{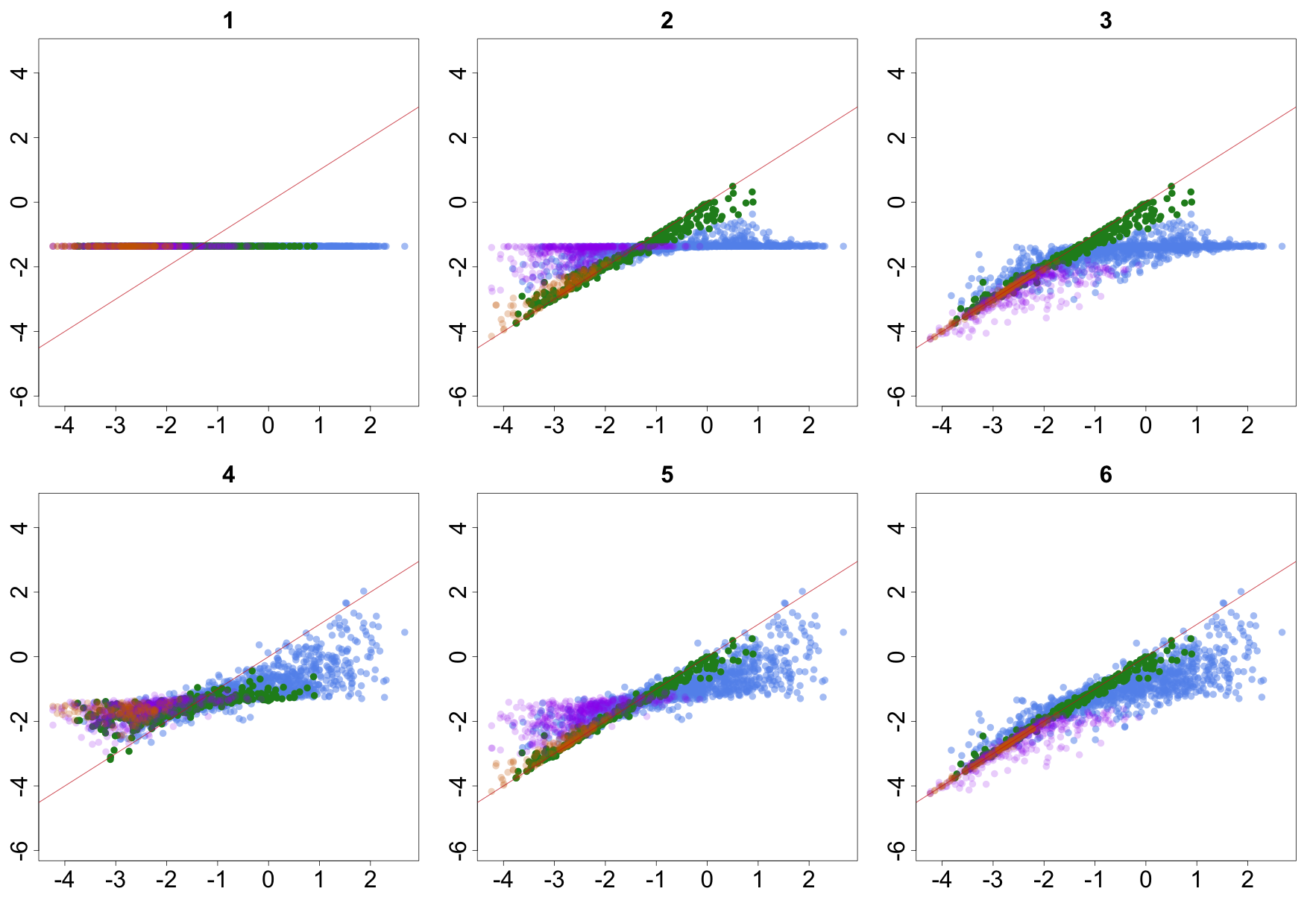}
\caption[Diagnostic Plots]{\footnotesize{$ f(x) $ against $ \ed{D}{f(x)} $ for the diagnostic set of 2000 points.  Blue points are such that $ k_6 > -0.5 $ and $ k_8 > -0.5 $, green points are such that $ k_6 < -0.5 $ and $ k_8 > -0.5 $, purple points are such that $ k_6 > -0.5 $ and $ k_8 < -0.5 $, and orange points are such that $ k_6 < -0.5 $ and $ k_8 < -0.5 $.  The red line is the function $ y=x $.   The columns (from left to right) show the results of emulating without boundaries, with one boundary and with two boundaries.  The rows show the results of emulating without training points on the top and with training points on the bottom.} \label{MO} }
\vspace{-0.4cm}
\efi

We now perform a more detailed comparison by evaluating the emulators over a fixed set of 2000 diagnostic points, which form a maximin Latin hypercube.
Figure \ref{MO} shows $ f(x) $ against $ \ed{D}{f(x)} $ for the set of 2000 diagnostic points, for the six scenarios outlined above.  We divide the 
points according to their $(k_6,k_8)$ coordinates (each scaled to $[-1,1]$) as follows: blue points are such that $ k_6 > -0.5 $ and $ k_8 > -0.5 $, green points have $ k_6 < -0.5 $ and $ k_8 > -0.5 $, purple points have $ k_6 > -0.5 $ and $ k_8 < -0.5 $, and orange points have $ k_6 < -0.5 $ and $ k_8 < -0.5 $.  The red line is the function $ y=x $.  Panel 2 shows that updating the emulator mean by the single boundary $\mk: k_6 = 0 $ results in larger changes in the mean prediction towards the true value for (green and orange) points close to that boundary.  We notice that, although they are affected, there are relatively large numbers of blue and purple points for which the prediction is largely unchanged from the prior specification.  Panel 3 shows that incorporating the second boundary into the emulation process results in (purple) points close to that boundary having greatly altered emulator mean values towards the true simulator values.  Orange points, which are close to both boundaries, have their accuracy increased even further, with many of them lying very close to the line $ y=x $.  Panels 4 to 6 show the effect of using training points in the construction of the emulators.  The effect of updating our beliefs about the simulator output at any particular point in the input space depends on its location relative to the training points. In the case when beliefs have been updated using both boundaries, subsequent updating using training points informs us most about the blue points, namely those which are far from the boundaries.  This suggests that training point design should be affected by knowledge of boundary behaviour such that a greater increase in accuracy in those areas largely unaffected by (that is far from) the boundaries is obtained. We now explore these design issues in section~\ref{SSKBED}. We also give further diagnostic analysis of the 2000 points in appendix~\ref{app_arab_details}.

\subsection{Simulation Study of KBE Design}\label{SSKBED}

We now compare emulators constructed using various training point designs, introduced in section~\ref{sec_design_KB}, that exploit the known boundaries. We wish to explore the improvements to the emulators due to such designs, compared to the improvements 
seen from just using the known boundaries directly, as were examined in the previous section. 
%We propose that the optimal design when known boundaries are utilised is different to the optimal design without the known boundaries.  
%Many standard techniques of design proposal aim to achieve a good (or nearly optimal) design, accepting that obtaining the absolute optimal design under any given criteria is usually too computationally expensive, particularly in high dimensions.  
We do this by comparing the use of several designs using the root mean square error (RMSE) of the 2000 diagnostic training points obtained in Section \ref{ESPS} under knowledge of the boundaries $\mk: k_6 = 0 $ and $\ml: k_8 = 0 $.
Firstly, we demonstrate that a warped maximin Latin hypercube is preferable to a standard maximin Latin hypercube.  We then compare the chosen design of three V-optimality design procedures; two which take account of the known boundaries and one which doesn't. 
%\ian{It is obvious that warped designs will be better. What we are trying to show here is how much better they are.}

\subsubsection{Warped Maximin Latin Hypercube Designs}

We generated 1000000 Latin hypercubes of size 60 over the 6-dimensional input space and chose the one with maximal minimum distance between any two of its points.  We compare the emulator constructed using this design with that constructed using the warped version of this design, constructed using equations~(\ref{eq_warplh1}), (\ref{eq_warplh2}) and (\ref{eq_warplh3}). Table \ref{MLHR} shows the RMSEs of the 2000 diagnostic points for emulators constructed with and without both the known boundaries for each of these two designs, for two choices of correlation length parameter $\theta$.

%\begin{table}[h]
%	\center
%	\begin{tabular}{|c|cc|} \hline
%		 & Maximin LH & Warped Maximin LH \\ \hline
%		Without KBs  & 0.9247 & 0.9489  \\ 
%		With KBs  & 0.6763 & 0.5886 \\ \hline
%	\end{tabular}
%	\caption[Input Ranges for Arabidopsis model]{A table of RMSEs of the 2000 diagnostic points using emulators constructed with and without the known boundaries for a maximin Latin hypercube design and the warped version of this design.  \label{MLHR}}
%\end{table} 

\begin{table}[t]
	\center
	\begin{tabular}{|c|c|cc|} \hline
	$\theta$	 & Known Boundaries & Maximin LH & Warped Maximin LH \\ \hline
\multirow{2}{*}{0.7} &	 Without   & {\bf 0.9247} & 0.9489  \\ 
	&	With   & 0.6763 & {\bf 0.5886} \\ \hline
\multirow{2}{*}{1.2} & Without   & {\bf 0.4427} & 0.6601  \\ 
		&With   & 0.2986 & {\bf 0.2530} \\ \hline
	\end{tabular}
	\caption[Input Ranges for Arabidopsis model]{\footnotesize{A table of RMSEs of the 2000 diagnostic points using emulators constructed with and without both the known boundaries $\mk$ and $\ml$ for a maximin Latin hypercube design and the warped version of this design, for two choices of 
	correlation length $ \theta$. The numbers in bold correspond to the preferred strategy, for the given knowledge of the boundaries.} \label{MLHR}}
	\vspace{-0.6cm}
\end{table} 

%\begin{table}[h]
%	\center
%	\begin{tabular}{|c|cc|} \hline
%		 & Maximin LH & Warped Maximin LH \\ \hline
%		Without KBs  & 0.4427 & 0.6601  \\ 
%		With KBs  & 0.2986 & 0.2530 \\ \hline
%	\end{tabular}
%	\caption[Input Ranges for Arabidopsis model]{\footnotesize{A table of RMSEs of the 2000 diagnostic points using emulators constructed with and without the known boundaries for a maximin Latin hypercube design and the warped version of this design.   $ \theta = 1.2 $.} \label{MLHR}}
%\end{table} 

We observe that the RMSE is greatly improved when the known boundaries are incorporated into the construction process of the emulator, relative to when they are not included, as expected.  It is also the case that the RMSE shows noticeable improvement for the emulators constructed using the warped design, compared to the standard design, when the known boundaries are utilised, for both choices of correlation length. This suggests that a warped maximin Latin hypercube is indeed a reasonable general purpose design to use if known boundaries are present which  maintains the good projection properties of a Latin hypercube, as discussed in section~\ref{sec_design_KB}.

\subsubsection{Approximate V-optimality Designs}

Finding global V-optimal designs such as shown in figures~\ref{fig_toymod_design1} is extremely computationally demanding, and impractical 
for moderate to high run number. We instead investigate three approximately V-optimal 60 point designs, constructed as follows, the first being constructed without including the known boundaries and the second two including them. 

For the first design, which ignores the known boundaries, we iteratively chose individual design points that optimise the current V-optimal criteria, that is, the $i$th point was chosen to optimise ${\rm trace}(\vard{D_i \cup D_{i-1}}{f(X)})$, given that the previous $(i-1)$ points, as represented by $D_{i-1}$, had already been chosen in the previous iterations. This is highly unlikely to lead to a global solution, but should result in designs which are quick to generate and that have high V-optimality criteria that are good enough for our purposes. 
The second design was created by warping the first design, using equations~(\ref{eq_warplh1}), (\ref{eq_warplh2}) and (\ref{eq_warplh3}) as in the previous section. 
The third design used the known boundaries $\mk$ and $\ml$, and was generated in a similar iterative manner to the first design, 
but now the $i$th point was chosen to optimise ${\rm trace}(\vard{D_i \cup D_{i-1} \cup K \cup L}{f(X)})$. In this case we expect the points 
to land further away from the two boundaries, similar to the designs shown in figure~\ref{fig_toymod_design1}. In all three cases $X$ was chosen to be a $ 6^6 $ grid across the 6 dimensional input space, which represents a pragmatic approximation to limit the design calculation time.

%We constructed two designs of size 60 using sequential V-optimality criteria.
% \ian{Need to say it is hard to find the V-optimal design for 60 points analogous to that of figure~\ref{fig_toymod_design1} as the optimisation is too hard, hence we do it sequentially. Need to be careful in our wording as this is different from full sequential design where we would run the model again at each new point, before designing the next point.} 
% The first design involved starting with prior expectation and variance, and subsequently adding points one at a time to minimise the sum of the updated emulator variances at a set of points forming an equally spaced $ 6^6 $ grid across the input space.  At each iteration, 5000 uniformly sampled points across the input space were proposed as candidates to be selected 
%\ian{Should we then optimise the best point, to ensure it really is good? Doing things sequentially should allow us to do this.}. 
%The second design was constructed similarly, although starting from the emulator expectation and variance obtained from updating by the known boundaries.

Table~\ref{tab_RMSE_Aopt} shows the RMSEs of the 2000 diagnostic points for emulators constructed with and without the known boundaries $\mk$ and $\ml$ for each of the above three designs: iterative V-optimal, warped iterative V-optimal and iterative V-optimal with known boundaries. We give the results for two values of the correlation length $\theta$. The RMSE numbers in bold correspond to the appropriate design for that scenario, with the other numbers provided for a fair comparison, for example, if we are not aware of the known boundaries we would use the first design, but if we include them 
we would use either the second or third designs. 
We observe that there is a substantial drop in RMSE when known boundaries are incorporated into the construction of the emulator, as expected. For example, when using the standard iterative V-optimal design the RMSE drops from 0.8166 to 0.5815 when known boundaries are included. 
We also see a further drop in RMSE when the existence of the known boundaries are used in the design process, for example the RMSE drops from 0.5815 (iterative V-optimal) to 0.5091 (warped iterative V-optimal) and to a similar 0.5101 for the full iterative V-optimal with known boundaries design. We note that the second and third designs give similar RMSEs in the bold cases, up to the noise resulting from the finite size of the 2000 diagnostic runs 
(table~\ref{tab_Aopt} gives the calculated V-optimality criteria $c(X_D)$ for each of the cases in table~\ref{tab_RMSE_Aopt}, and shows that this 
criteria is very similar for the second and third design in these cases). 
Comparing tables~\ref{MLHR} and \ref{tab_RMSE_Aopt} we can see that the approximate V-optimal designs have lower RMSEs than their Latin hypercube counterparts, which is mainly due to their better space filling properties, justifying their use, provided we are not too concerned about their projection properties, as discussed in section~\ref{sec_design_KB}. This improvement is less for the larger value of $\theta=1.2$.

%The warped V-optimal sequential design constructed without considering the boundaries results in an improved RMSE relative to the unwarped design.  However, a sequential V-optimal design which incorporated the known boundaries from the beginning resulted in an even lower RMSE.  In particular, this design placed no points within a distance of 0.74 from either of the known boundaries.

The results of this design simulation study suggest that knowledge of known boundaries should affect our choice of training point design, which can lead to substantial benefits in addition to those obtained by the direct incorporation of the boundaries into the emulator.

\begin{table}[t]
	\center
	\begin{tabular}{|c|c|ccc|} \hline
\multirow{ 2}{*}{$\theta$}   & Known & Iterative  & Warped  & Iter. V-Opt.  \\
	 & Boundaries & V-Opt. & Iter. V-Opt. & with KBs \\ \hline
\multirow{2}{*}{0.7}	   &Without  & {\bf 0.8166} & 0.9013 & 0.9700 \\ 
	 &With   & 0.5815 &{\bf  0.5091} & {\bf 0.5101}  \\  \hline
\multirow{2}{*}{1.2}	   &Without   & {\bf 0.4476} & 0.6687 & 0.9028 \\ 
		 &With   & 0.2830 & {\bf 0.2340} & {\bf 0.2414 } \\  \hline
	\end{tabular}
	\caption{\footnotesize{A table of RMSEs of the 2000 diagnostic points using emulators constructed with and without the known 
	boundaries $\mk$ and $\ml$ for three designs, namely a standard iterative V-optimal design without the known boundaries, the warped version of this design, and an iterative V-optimal design which takes account of the known boundaries.} \label{tab_RMSE_Aopt}}
	\vspace{-0.6cm}
\end{table}
\begin{table}[t]
	\center
	\begin{tabular}{|c|c|ccc|} \hline
\multirow{ 2}{*}{$\theta$}   & Known & Iterative  & Warped  & Iter. V-Opt.  \\
	 & Boundaries & V-Opt. & Iter. V-Opt. & with KBs \\ \hline
\multirow{2}{*}{0.7}	  &Without  & {\bf 55605} & 56018 & 56464 \\ 
	 &With   & 28673 & {\bf 27707} & {\bf 27675}  \\  \hline
\multirow{ 2}{*}{1.2}	   &Without   & {\bf 33668} & 36839 & 40015 \\ 
		 &With   & 7993 & {\bf 6627} & {\bf 6607}  \\  \hline
	\end{tabular}
	\caption{\footnotesize{A table of the V-optimality criterion values $c(X_D)$ of the $ 6^6 $ grid of points, using emulators constructed with and without the known boundaries for three designs, namely a standard iterative V-optimal design without the known boundaries, the warped version of this design, and an iterative V-optimal design which takes account for the known boundaries.} \label{tab_Aopt}}
	\vspace{-0.6cm}
\end{table}

\section{Conclusion}\label{sec_conc}

We have discussed how improved emulation strategies have the potential to benefit multiple scientific areas, allowing more accurate analyses with lower computational cost, and therefore if additional prior insight into the physical structure of the model is available, it is of real importance that emulator structures capable of incorporating such insights are indeed developed.  

Here it was shown that if a simulator has boundaries or hyperplanes in its input space where it can be either analytically solved 
or just evaluated far more efficiently, these known boundaries can be formally incorporated into the emulation process by Bayesian updating of 
the emulators with respect to the information contained on the boundaries. It was also shown that this is possible for a large class of emulators, for multiple boundaries of various forms\footnote{We note that although for a general product correlation structure we require the boundary to be
a hyperplane perpendicular to one (or more for lower dimensional boundaries) input directions, for the Gaussian correlation structure the boundary can be a hyperplane in any orientation. Our analysis also naturally extends to lower dimensional boundaries.}, and most importantly, for trivial extra computational cost. 
This analysis also demonstrated how to include known boundaries when using standard black box Gaussian Process software (for users that do not have access to 
alter the code), by simply incorporating all the projections of the input points of interest and the simulator runs into the emulator update.
This method is simple to implement, but is of course 
less powerful than direct implementation of the fully updated emulator equations that we have developed here, especially if one needs to evaluate 
the emulator at a large number of points.

The design problem of how to choose an efficient 
set of runs of the full simulator, given that we are aware of the existence of one or more known boundaries was then examined. V-optimal and warped latin hypercube designs were suggested as reasonable choices in this context, and their relative strengths and weaknesses explored. 
Finally we applied this approach to a model of hormonal crosstalk in Arabidopsis, an important model in systems biology, which possesses two perpendicular known boundaries, and analysed the improvements to the emulator of the $[PLSp]$ output due to first the known boundaries, and then due to the use of more careful designs of simulator runs. 

Obviously, the applicability of this approach depends on whether any such boundaries can be found for the complex model in question. We note that in some scenarios the input space of interest $\mathcal{X}\subset \mathbb{R}^d$ may be defined such that $\mathcal{X}$ does not contain a known boundary $\mk$, however, a boundary may exist just outside of $\mathcal{X}$ (for example when some physical parameter was set to zero, but the lower limit of $\mathcal{X}$ for that parameter is just above zero), such that were $\mk$ to be included in the emulation process, the resulting emulator would still be improved over a significant proportion of $\mathcal{X}$\red{, with the extent of the improvement dependent on the correlation parameters and on the distance from $\mk$ to $\mathcal{X}$}. This may in fact be fairly common as when specifying $\mathcal{X}$ the domain expert may be aware that the boundary $\mk$ is not of primary physical interest, as
the more complex model that is employed away from $\mk$ has been constructed for a reason, and hence they may not have originally included $\mk$ within $\mathcal{X}$. As the benefits of using $\mk$ in the emulation process come with trivial computational cost, all such boundaries should be included.

A further point we would make is that, as was seen in the application to the Arabidopsis model, known boundaries may exist only for a subset of the simulator outputs. This can still be useful, especially in applications of emulation such as history matching \cite{Vernon10_CS,Vernon10_CS_rej} whereby the input parameter space $\mathcal{X}$ is searched to find acceptable matches between model and observed data, by iteratively discarding regions of $\mathcal{X}$ that seem unlikely to lead to good matches based only on subsets of the outputs that are easy to emulate. Further outputs are included as the history match progresses, and are usually found to be easier to emulate in later iterations, after $\mathcal{X}$ has been substantially reduced (see \cite{Vernon10_CS,Vernon10_CS_rej,Jackson1} for extended discussions and examples of this). Therefore the iterative structure of the history matching 
approach may still allow substantial exploitation of known boundaries that improve the emulation of only a subset of the outputs. 

The results presented here can of course be extended in several directions, for example \red{to collections of multiple parallel or perpendicular boundaries of varying dimension, or to} multivariate emulators providing suitable 
product correlation structures are used~\cite{Rougier:2008aa}. Extensions to the case of uncertain regression 
parameters (the $\beta_j$ in equation~(\ref{eq_fullem})) are also possible, although the formal update would now depend on the specific form of 
the correlation function $r_1(a)$ which may not be tractable for many choices. Curved boundaries of various geometries could of course be incorporated, provided both that suitable transformations were found to convert them to hyperplanes and that we were happy to adopt the induced transformed product correlation structure as our prior beliefs. We leave these considerations to future work.

%\section*{Acknowledgments}
%We would like to acknowledge the assistance of volunteers in putting
%together this example manuscript and supplement.
\vspace{-0.6cm}

\bibliographystyle{siamplain}
\bibliography{../../../Master_Bibliography_and_Related_Files/master_bib}

\newpage

\appendix

\section{Two Perpendicular Boundary Emulator Derivations}\label{app_two_perp}

Here we provide the full derivation of the expectation and covariance of $f(x)$, adjusted by perpendicular boundaries $L$ and $K$ as discussed in section~\ref{ssec_two_perp_bound}. 

We perform the update by $K$ using the results of section~\ref{ssec_singleKB}. We then use an analogous proof to that 
of equation~(\ref{eq_covvar}), but now applied to the vector $L$ after performing the update\footnote{We are assuming there are no problems here due to the non-empty $\mk \cap \ml$. In fact the full Bayes linear update would instead use the generalised inverse if $L$ contains points on $\mk$ (which would possess zero variance), but equation~(\ref{eq_covvar_Ka}) will remain the same.} for $K$:
\ba
 \vard{K}{L}  \vard{K}{L}^{-1} &=& I_{(m+1)} \nonumber \\
\Rightarrow \quad \covd{K}{f(x^L)}{L} \vard{K}{L}^{-1} &=& (1,0, \cdots, 0) \label{eq_covvar_Ka}
\ea

We can also use equation~(\ref{eq_rcovxK}), which is a direct consequence of the product correlation structure, which still holds after the update by $K$, now with $K$ replaced by $L$ to give
\be
\covd{K}{f(x)}{L} \;=\; r_2(b) \, \covd{K}{f(x^L)}{L}  \label{eq_rcovxLa} 
\ee
where $b$ is the perpendicular distance from $x$ to $\ml$, as shown in figure~\ref{fig_double_perp} (left panel).
The expectation as given by equation~(\ref{eq_expKL1}) can now be calculated using the sequential update equation~(\ref{eq_BLmDK}) as 
\ba
 \ed{L \cup K}{f(x)} &=& \ed{K}{f(x)} + \covd{K}{f(x)}{L} \vard{K}{L}^{-1}(L- \ed{K}{L})  \nonumber \\
  &=&  \ed{K}{f(x)} +  r_2(b) (1,0, \cdots, 0) (L- \ed{K}{L}) \nonumber \\
  &=&  \ed{K}{f(x)} +  r_2(b) (f(x^L) - \ed{K}{f(x^L)} )  \label{eq_app_twoperp1}\\
 &=&   \e{f(x)} + r_1(a) (f(x^K) -  \e{f(x^K)})  \;+\; r_2(b) f(x^L)  \nonumber \\
 && \quad \;-\;  r_2(b) ( \e{f(x^L)} + r_1(a) (f(x^{LK}) -  \e{f(x^{LK})}) )  \nonumber \\
 &=&  \e{f(x)} + r_1(a) (f(x^K) -  \e{f(x^K)})  +  r_2(b) (f(x^L)  -  \e{f(x^L)}) \nonumber \\ 
 && \quad \; - r_1(a) r_2(b) (f(x^{LK}) -  \e{f(x^{LK})}) \nonumber \\
 &=&  \e{f(x)} + r_1(a) \Delta f(x^K)  +  r_2(b) \Delta f(x^L) -  r_1(a) r_2(b) \Delta f(x^{LK})    \label{eq_app_twoperp2}
\ea
where we have also used equation~(\ref{eq_EK1}) for $ \ed{K}{f(x)}$, defined $\Delta f(.) \equiv f(.) - \e{f(.)}$ and denoted the projection of
$x^L$ onto $\mk$ as $x^{LK}$, which is just the perpendicular projection of $x$ onto $\ml \cap \mk$.
%\vard{D \cup K}{f(x)} &=& \vard{K}{f(x)} - \covd{K}{f(x)}{D} \vard{K}{D}^{-1}\covd{K}{D}{f(x)} \label{eq_BLvDK}
The corresponding expression for the covariance as shown in equation~(\ref{eq_covKL1}), adjusted by $L$ and $K$ is
\ba
\covd{L \cup K}{f(x)}{f(x')} &=& \covd{K}{f(x)}{f(x')} - \covd{K}{f(x)}{L} \vard{K}{L}^{-1}\covd{K}{L}{f(x')}  \nonumber \\
     &=& \covd{K}{f(x)}{f(x')} - r_2(b) (1,0, \cdots, 0)\covd{K}{L}{f(x')}   \nonumber \\
     &=& \covd{K}{f(x)}{f(x')} - r_2(b) \covd{K}{f(x^L)}{f(x')}  \nonumber \\
%     &=& \covd{K}{f(x)}{f(x')} - r_2(b) \covd{K}{f(x^L)}{f(x'^L)} r_2(b')  \nonumber \\
     &=& r_2(b-b') \covd{K}{f(x^L)}{f(x'^L)}  - r_2(b) \covd{K}{f(x^L)}{f(x'^L)} r_2(b')  \label{eq_app_twoperpcov1} \\
     &=&  (r_2(b-b')- r_2(b) r_2(b') ) \covd{K}{f(x^L)}{f(x'^L)} \nonumber \\
     &=& \sigma^2 R_1(a,a') \, R_2(b,b') \, r_{-1,-2}(x^{LK}-x'^{LK}) \nonumber
\ea
where we have defined the correlation function of the projection of $x$ and $x'$ onto $\ml \cap \mk$ as
\be
r_{-1,-2}(x^{LK}-x'^{LK})   \;=\;  \prod_{i=3}^d r_i(x^{LK}_i - x'^{LK}_i)  \;=\; \cov{f(x^{LK})}{f(x'^{LK})} \nonumber
\ee
The limiting behaviour looks to be as expected in the large and small $a,a',b,b'$ limits e.g. 
\ba
\lim_{b\to 0}  \ed{L \cup K}{f(x)} \;=& f(x^L), \qq \qq  \lim_{b\to 0}  \vard{L \cup K}{f(x)} &=\; 0, \label{eq_lim2pKB}\nonumber \\
\lim_{b \to \infty} \ed{L \cup K}{f(x)} \;=& \ed{K}{f(x)}, \qq \quad  \lim_{b\to \infty}  \vard{L \cup K}{f(x)} &=\; \vard{K}{f(x)} ,\nonumber 
\ea
and similarly for the covariances 
\ba
\lim_{b\to 0} \covd{L \cup K}{f(x)}{f(x')} &=&  \lim_{b'\to 0} \covd{L \cup K}{f(x)}{f(x')} \;=\; 0  \label{eq_limCov2pKB} \nonumber \\
\lim_{b,b' \to \infty}  \covd{L \cup K}{f(x)}{f(x')} &=& \covd{K}{f(x)}{f(x')} , \quad  b-b' \text{ finite} \quad\nonumber 
\ea

\section{Two Parallel Boundary Emulator Derivations}\label{app_two_para}

Here we now provide the full derivation of the expectation and covariance of $f(x)$, adjusted by parallel boundaries $L$ and $K$, 
as discussed in section~\ref{ssec_two_para_bound}.

First we need to find the analogous version of equation~(\ref{eq_rcovxL}), which relates $\covd{K}{f(x)}{L}$ to $\covd{K}{f(x^L)}{L}$. Noting 
that 
\be
\covd{K}{f(x^L)}{f(z^{(j)})}  \;=\;  \sigma^2 R_1(c,c) \, r_{-1}(x^K-y^{(j)})
\ee
it follows that
\ba
\covd{K}{f(x)}{f(z^{(j)})}  &=& \sigma^2 R_1(a,c) \, r_{-1}(x^K-y^{(j)})  \nonumber \\
&=&  \frac{R_1(a,c)}{R_1(c,c)} \, \sigma^2  R_1(c,c) \, r_{-1}(x^K-y^{(j)}) \nonumber \\
&=&  \frac{ R_1(a,c)}{R_1(c,c)} \, \covd{K}{f(x^L)}{f(z^{(j)})} \nonumber\\
&=& R_1^{(2)}(a,c) \, \covd{K}{f(x^L)}{f(z^{(j)})}   \label{eq_app_para_prod}
\ea
where we have defined $R_1^{(2)}(a,c) = R_1(a,c) /R_1(c,c)$. Therefore we have
\be
\covd{K}{f(x)}{L} \;=\; R_1^{(2)}(a,c) \covd{K}{f(x^L)}{L}
\ee
Here equation~(\ref{eq_covvar_K}) holds as before, implying we can again avoid explicit evaluation of the intractable 
$ \vard{K}{L}^{-1} $ term. Hence the adjusted expectation, as given by equation~(\ref{eq_expLK1_para}) can be calculated using the 
sequential update equation~(\ref{eq_BLmDK}) as
\ba
 \ed{L \cup K}{f(x)} &=& \ed{K}{f(x)} + \covd{K}{f(x)}{L} \vard{K}{L}^{-1}(L- \ed{K}{L})  \nonumber \\
 &=&  \ed{K}{f(x)} + R_1^{(2)}(a,c) \covd{K}{f(x^L)}{L} \vard{K}{L}^{-1}(L- \ed{K}{L})  \nonumber \\
 &=&  \ed{K}{f(x)} + R_1^{(2)}(a,c) (1,0, \cdots, 0) (L - \ed{K}{L})  \nonumber  \\
  &=&  \ed{K}{f(x)} + R_1^{(2)}(a,c) (f(x^L) - \ed{K}{f(x^L)} ) \label{eq_app_para1h}  \\
    &=&   \e{f(x)} + r_1(a) (f(x^K) -  \e{f(x^K)})  +  \nonumber  \\ 
    && \frac{r_1(b)-   r_1(a) r_1(c)}{1 -  r^2_1(c) }  \left\{ f(x^L) - \left(\e{f(x^L)} + r_1(c) (f(x^{K}) -  \e{f(x^K)})  \right) \right\}  \nonumber  \\
&=& \e{f(x)} + \left[ \frac{r_1(a)- r_1(b) r_1(c)}{1-r^2_1(c) } \right] \Delta f(x^K) + 
		\left[\frac{r_1(b) - r_1(a) r_1(c)}{1-r^2_1(c) }\right] \Delta f(x^L)  \label{eq_app_finalpara}
\ea
where we have used the fact that for parallel boundaries the projection of $x^L$ onto $\mk$ is just $x^K$.   
Similarly we find the covariance adjusted by $L$ and $K$ to be
\ba
\covd{L \cup K}{f(x)}{f(x')} &=& \covd{K}{f(x)}{f(x')} - \covd{K}{f(x)}{L} \vard{K}{L}^{-1}\covd{K}{L}{f(x')}  \nonumber \\
     &=& \covd{K}{f(x)}{f(x')} - R_1^{(2)}(a,c) (1,0, \cdots, 0)\covd{K}{L}{f(x')}   \nonumber \\
     &=& \covd{K}{f(x)}{f(x')} - R_1^{(2)}(a,c) \covd{K}{f(x^L)}{f(x')}   \nonumber  \\
     &=&  \covd{K}{f(x)}{f(x')} - R_1^{(2)}(a,c) \covd{K}{f(x^L)}{f(x'^L)} R_1^{(2)}(c,a')  \label{eq_app_par1}\\
     &=& \sigma^2 R_1(a,a') \, r_{-1}(x^K-x'^K) \nonumber \\
     && - R_1^{(2)}(a,c) \sigma^2 R_1(c,c) \, r_{-1}(x^K-x'^K) R_1^{(2)}(c,a')  \nonumber \\
    &=&  \sigma^2 \, r_{-1}(x^K-x'^K) \left\{  R_1(a,a') -  R_1^{(2)}(a,c)  R_1(c,c) R_1^{(2)}(c,a') \right\} \nonumber \\
    &=&  \sigma^2 \, r_{-1}(x^K-x'^K) \left\{  R_1(a,a') -  \frac{R_1(a,c) R_1(c,a')}{R_1(c,c)} \right\} \nonumber 
 \ea
 which is just a generalised form of equation~(\ref{eq_covK2}). To make the invariance under interchange of boundaries explicit, we expand out the $R_1(.,.)$ terms giving 
 \ba   
 \covd{L \cup K}{f(x)}{f(x')}   &=&  \sigma^2 \, \frac{r_{-1}(x^K-x'^K)}{R_1(c,c)} \left\{  
    	(r_1(a-a')-   r_1(a) r_1(a') ) (  1 -   r^2_1(c) ) \; -   \right. \nonumber \\
    &&   \quad\quad \quad \quad \quad \quad \quad \quad \left. (r_1(b)-   r_1(a) r_1(c) )( r_1(b')-   r_1(c) r_1(a') )
     \right\} \nonumber \\
     &=& \sigma^2 \, \frac{r_{-1}(x^K-x'^K)}{1 -   r^2_1(c)} \Big\{  r_1(a-a')(1 -   r^2_1(c)) -  r_1(a)r_1(a') - r_1(b)r_1(b')   \label{eq_app_par_cov1} \\
&&  \quad \quad\quad\quad\quad \quad \quad \quad \quad + \; r_1(c) \big[ r_1(a) r_1(b') +  r_1(b) r_1(a') \big]  \Big\}  \nonumber
\ea
which is now explicitly invariant under the interchange of the two boundaries $\mk \leftrightarrow \ml$ (as $c=a+b$ is invariant under $a\leftrightarrow b, a' \leftrightarrow b'$, as is $a-a' = b-b'$). 

The emulator variance is given by equation~\eqref{eq_varLK1_para}, but we note that it is now bounded from above, and will attain its maximum at the midpoint between the two parallel boundaries. Setting $a=b=c/2$ in equation~\ref{eq_varLK1_para} shows that this maximum bound is given by:
\be
\vard{L \cup K}{f(x)}  \;\; \le \;\;  \sigma^2 \left\{1 - \frac{2 r_1^2(\frac{c}{2})}{1+r_1(c)} \right\}
\ee

In addition, we note that the emulator expectation, covariance and variance update calculations for a single boundary will easily generalise to a boundary of lower dimension than $d-1$. However, for multiple boundaries there will be some additional constraints, specifically on the dimension and intersection of the boundaries in the perpendicular case (the parallel case is easier), a full discussion of which we leave to future work.

\section{Continuous Known Boundary Proofs}\label{app_contKBE}

\subsection{Two perpendicular continuous boundaries}
For the two perpendicular boundary continuous case, after the update by boundary $\mk$, we use $s_K(z,z')$ to represent the infinite dimensional generalisation of $\vard{K}{L}^{-1}$, which satisfies the corresponding inverse property:
\be
\int_{z' \in \ml} \covd{K}{f(z)}{f(z')} s_K(z',z'') \; dz' \;=\; \delta(z-z''), \quad \quad {\rm for} \; z,z'' \in \ml  \label{eq_app_delta}
\ee
Then, noting that $\covd{K}{f(x)}{f(z)} \;=\; r_2(b) \, \covd{K}{f(x^L)}{f(z)} $, the emulator expectation adjusted sequentially by first $K$ and then $L$, becomes
\ba
\ed{L \cup K}{f(x)} &=& \ed{K}{f(x)} + \int_{z \in \ml} \int_{z' \in \ml} \covd{K}{f(x)}{f(z)} \; s_K(z,z') \; (f(z')- \ed{K}{f(z')}) dz dz'   \nonumber \\
&=& \ed{K}{f(x)} + \int_{z \in \ml} \int_{z' \in \ml} r_2(b) \covd{K}{f(x^L)}{f(z)} \; s_K(z,z') \; (f(z')- \ed{K}{f(z')}) dz dz'   \nonumber \\
&=& \ed{K}{f(x)} + r_2(b) \int_{z' \in \ml} \delta(x^L-z') \; (f(z')- \ed{K}{f(z')}) dz'   \nonumber \\
&=& \ed{K}{f(x)} + r_2(b) \; (f(x^L)- \ed{K}{f(x^L)})    \nonumber 
\ea
which is identical to equation~(\ref{eq_app_twoperp1}), and the rest of the proof of equation~(\ref{eq_app_twoperp2}) follows as before. 
Similarly for the covariance we have:
\ba
\covd{L \cup K}{f(x)}{f(x')} & =& \covd{K}{f(x)}{f(x')}  -  r_2(b) \int_{z' \in \ml} \delta(x^L-z') \;  \covd{K}{f(z')}{f(x'^L)} r_2(b') \; dz'  \nonumber \\
& =& r_2(b-b') \covd{K}{f(x^L)}{f(x'^L)} -  r_2(b) \covd{K}{f(x^L)}{f(x'^L)} r_2(b')   \nonumber 
\ea
which agrees with equation~(\ref{eq_app_twoperpcov1}), and the rest of the proof follows as before.

\subsection{Two parallel continuous boundaries}

The proof for continuous parallel boundaries follows a similar form to the perpendicular case. We use $s_K(z,z')$ as before, which 
still satisfies equation~(\ref{eq_app_delta}). However, here we have instead from equation~(\ref{eq_app_para_prod}) that
\be
\covd{K}{f(x)}{f(z)} \;=\;  R_1^{(2)}(a,c) \, \covd{K}{f(x^L)}{f(z)}, \quad \quad z\in \ml  \nonumber
\ee
Therefore the emulator expectation adjusted sequentially by first $K$ and then $L$ becomes
\ba
\ed{L \cup K}{f(x)} &=& \ed{K}{f(x)} + \int_{z \in \ml} \int_{z' \in \ml} R_1^{(2)}(a,c) \covd{K}{f(x^L)}{f(z)} \; s_K(z,z') \; (f(z')- \ed{K}{f(z')}) dz dz'   \nonumber \\
&=& \ed{K}{f(x)} + R_1^{(2)}(a,c) \int_{z' \in \ml} \delta(x^L-z') \; (f(z')- \ed{K}{f(z')}) dz'   \nonumber \\
&=& \ed{K}{f(x)} + R_1^{(2)}(a,c)\; (f(x^L)- \ed{K}{f(x^L)})    \nonumber 
\ea
which is identical to equation~(\ref{eq_app_para1h}), and the rest of the proof of equation~(\ref{eq_app_finalpara}) follows as before. 
Similarly for the covariance we have:
\ba
\covd{L \cup K}{f(x)}{f(x')} & =& \covd{K}{f(x)}{f(x')}  -  R_1^{(2)}(a,c) \int_{z' \in \ml} \delta(x^L-z') \;  \covd{K}{f(z')}{f(x'^L)}  R_1^{(2)}(c,a') \; dz'  \nonumber \\
& =&  \covd{K}{f(x)}{f(x')}  -  R_1^{(2)}(a,c) \covd{K}{f(x^L)}{f(x'^L)}  R_1^{(2)}(c,a')  \nonumber 
\ea
which agrees with equation~(\ref{eq_app_par1}), and the rest of the proof of equation~(\ref{eq_app_par_cov1}) again follows as before.

%\section{D-optimality in KBE Design Discussion}\label{app_des_dopt}
%Continuing the discussion in section~\ref{sec_design_KB}, while D-optimality is in some sense more sophisticated 
%as it also accounts for the covariances across $X$, we see that in the context of the emulation of deterministic computer models it can be of limited value. The primary issue is that locating a single point of $D$ at one of the grid points in $X$ will cause the emulator variance to equal zero there. Consequently, this introduces a zero eigenvalue in $\vard{D \cup K}{f(X)}$ and hence ${\rm det} (\vard{D \cup K}{f(X)})$ will attain its lower bound of 0.
%% JAC: Not sure about this bit 
%%  
%%It is possible to remove the offending point from $X$ and evaluate $c(X_D)$ on the remaining points (and keep doing this for all other 
%%points in $X_D$ which will similarly seek out points in $X$). This will lead to a design made 
%%up entirely of elements of X which may not be ideal. \ian{would this still work?}.
%
%Alternatively, we can note that
%\ba
%\vard{D \cup K}{f(X)} &=& \vard{K}{f(X)} - \covd{K}{f(X)}{D} \vard{K}{D}^{-1}\covd{K}{D}{f(X)} \nonumber \\
% &=& \vard{K}{f(X)} - \rvard{D \cup K}{f(X)}
%\ea
%and for fixed $X$ and $K$, the prior variance $\vard{K}{f(X)}$ is unaffected by $D$. So we could choose instead to  maximise the determinant of the resolved variance ${\rm det}(\rvard{D \cup K}{f(X)})$, although this will not be equivalent to full D-optimality.

\section{Arabidopsis model details and emulator diagnostics}\label{app_arab_details}

Table~\ref{IRR} gives the parameter ranges for the Arabidopsis Thaliana model, used throughout section~\ref{sec_arabid}.

\begin{table}[h]
	\center
	\begin{tabular}{|c|c|c|} \hline
		Input Rate & Minimum & Maximum \\ 
		Parameter & & \\ \hline
		$ k_4 $  & 0 & 10 \\
		$ k_{6} $  & 0 & 1 \\ 
		$ k_{6a} $  & 0 & 20 \\
		$ k_{7} $ & 0 & 10 \\
		$ k_{8} $ & 0 & 10 \\
		$ k_{9} $ & 0 & 1  \\ \hline
	\end{tabular}
	\caption[Input Ranges for Arabidopsis model]{\footnotesize{A table of parameter ranges for the Arabidopsis Thaliana, (which were square rooted and converted to $ [-1,1] $ for the analysis)}}\label{IRR}
\vspace{-0.6cm}
\end{table}

\begin{figure}[t]
\center
\includegraphics[width=14cm]{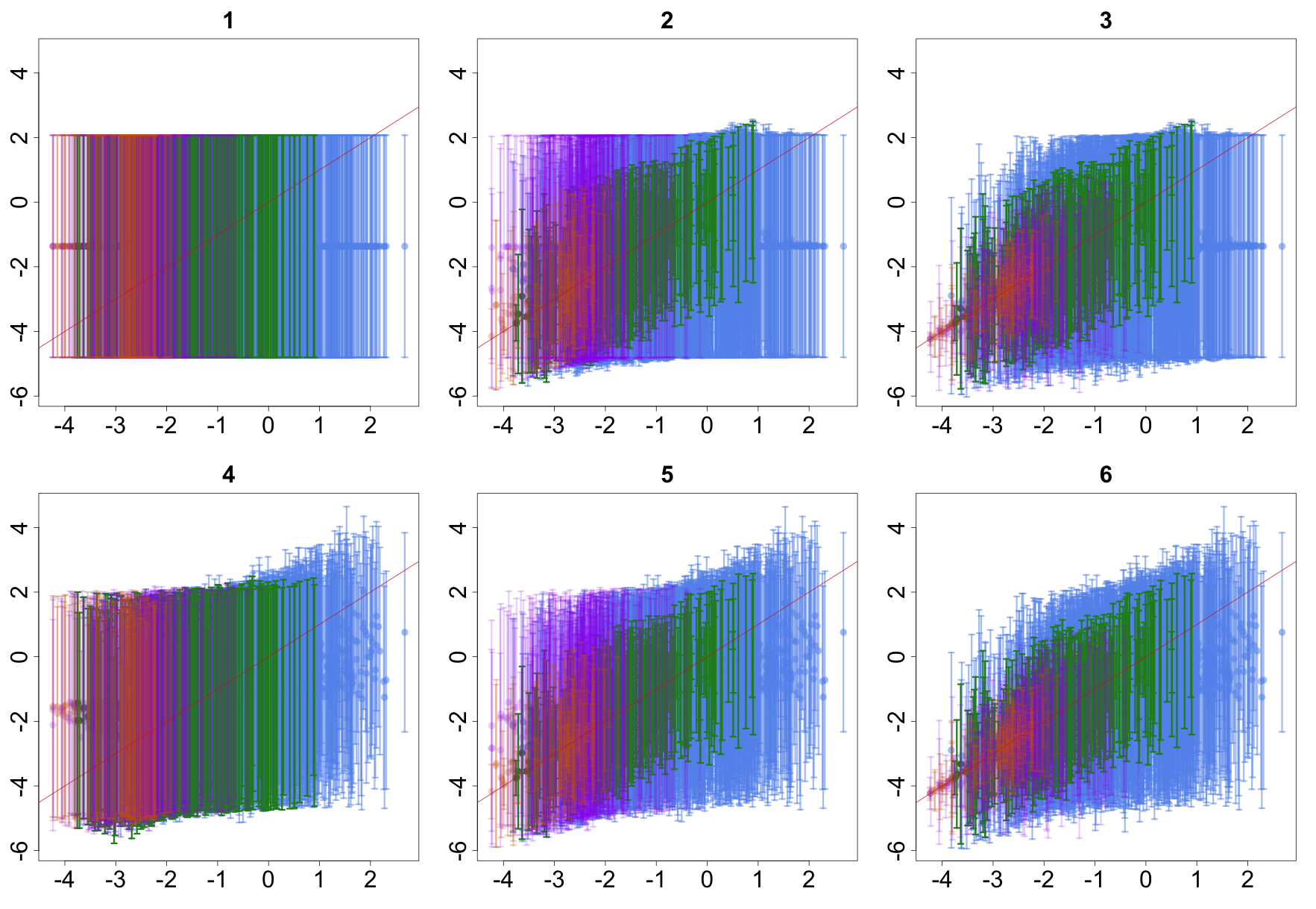}
\caption[Diagnostic Plots]{\footnotesize{$ f(x) $ against $ \ed{D}{f(x)} \pm 3 \sqrt{\vard{D}{f(x)}} $ for the diagnostic set of 2000 points.  Blue points are such that $ k_6 > -0.5 $ and $ k_8 > -0.5 $, green points are such that $ k_6 < -0.5 $ and $ k_8 > -0.5 $, purple points are such that $ k_6 > -0.5 $ and $ k_8 < -0.5 $, and orange points are such that $ k_6 < -0.5 $ and $ k_8 < -0.5 $.  The red line is the function $ y=x $.  The columns (from left to right) show the results of emulating without boundaries, with one boundary and with two boundaries.  The rows show the results of emulating without training points on the top and with training points on the bottom.} \label{Trans} }
\vspace{-0.6cm}
\efi

Figure \ref{Trans} shows $ f(x) $ against $ \ed{D}{f(x)} \pm 3 \sqrt{\vard{D}{f(x)}} $ for each of the six emulator scenarios for the diagnostic set of 2000 points, with the colour scheme the same as for Figure \ref{MO}.  These plots show how the variance at each point is updated in correspondence to its expectation. We observe that the error bars on some of the points decrease as the boundaries get utilised, particularly for points which lie close to at least one or other of the boundaries.  The majority of the error bars still cross the line $ y=x $, indicating that the emulator expectation lies within three emulator standard deviations of the true simulator value.  

\begin{figure}[t]
\center
\includegraphics[width=14cm]{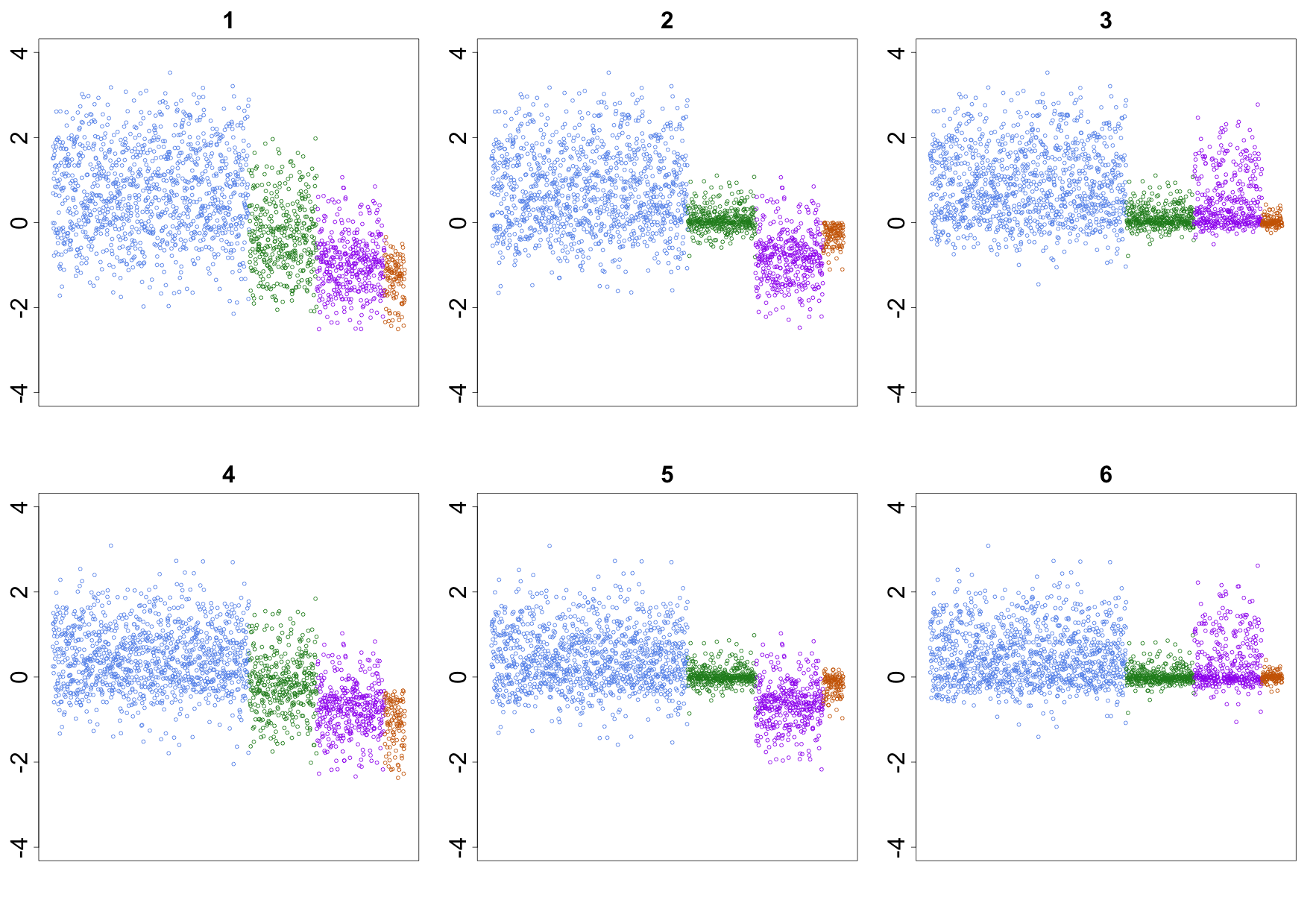}
\vspace{-0.3cm}
\caption[Diagnostic Plots]{\footnotesize{$ \frac{(f(x) - \ed{D}{f(x)})}{\sqrt{\vard{D}{f(x)}}} $ for the diagnostic set of 2000 points.  The columns (from left to right) show the results of emulating without boundaries, with one boundary and with two boundaries.  The top row shows the results of emulating without training points and the bottom row shows the results of emulating with the training points.}  \label{Diag} }
\vspace{-0.6cm}
\efi

Figure \ref{Diag} shows $ \frac{(f(x) - \ed{D}{f(x)})}{\sqrt{\vard{D}{f(x)}}} $ for each of the six emulator scenarios for the diagnostic set of 2000 points.  A value with magnitude greater than 3 is equivalent to the corresponding error bar in Figure \ref{Trans} not containing the true simulator value.  We conclude that the diagostic plots are acceptable for all emulators, with those points far from the boundary $ k_6 = 0 $ having larger values once the boundaries have been utilised.  The small diagnostic values corresponding to points close to the boundary $ k_6 = 0 $ may suggest that a larger correlation length could be appropriate, particularly in certain dimensions of the input space such as $ k_6 $ itself.

\end{document}